\documentclass[fleqn,usenatbib]{mnras}
\usepackage{caption}
\usepackage{lineno}
\usepackage{footnote}
\usepackage{hyperref}
\usepackage{threeparttable}
\usepackage{graphicx}	
\usepackage{amsmath}	% Advanced maths commands
\usepackage{amssymb}	% Extra maths symbols
\usepackage{multicol}        
\usepackage{bm}		
\usepackage{pdflscape}	% Landscape pages

\newcommand{\hi}{{\rm H}{\textsc i}}

\title{CHANG-ES XXV: HI Imaging of Nearby Edge-on Galaxies -- Data Release 4}

\author[Y. Zheng et al.]{
Yun Zheng,$^{1}$
Jing Wang,$^{1}$ \thanks{corresponding author: Jing Wang ~  jwang$\_$astro@pku.edu.cn}
Judith Irwin,$^{2}$
Jayanne English,$^{3}$
Qingchuan Ma,$^{4}$
Ran Wang,$^{1}$
\newauthor{
Ke Wang,$^{1}$
Q. Daniel Wang,$^{5}$
Marita Krause,$^{6}$
Toky H. Randriamampandry,$^{1}$
Jiangtao Li,$^{7,8}$
}
\newauthor{
and Rainer Beck$^{6}$}
\\
$^{1}$Kavli Institute for Astronomy and Astrophysics, Peking University, Beijing 100871, People's Republic of China\\
$^{2}$Department of Physics, Engineering Physics and Astronomy, Queen's University, Kingston, ON, K7L 3N6, Canada\\
$^{3}$Department of Physics and Astronomy, University of Manitoba, Winnipeg, Manitoba, R3T 2N2, Canada\\
$^{4}$Peking University, Beijing 100871, People's Republic of China\\
$^{5}$Astronomy, University of Massachusetts, Amherst, Amhert, MA\\
$^{6}$Max-Planck-Institut für Radioastronomie, Auf dem Hügel 69, 53121 Bonn, Germany\\
$^{7}$Department of Astronomy, University of Michigan, 311 West Hall, 1085 S. University Avenue, Ann Arbor, MI, 48109-1107, USA\\
$^{8}$Purple Mountain Observatory, Chinese Academy of Sciences, 10 Yuanhua Road, Nanjing 210023, China
}

\begin{document}
\label{firstpage}
\pagerange{\pageref{firstpage}--\pageref{lastpage}}
\maketitle

\begin{abstract}
We present the $\hi$ distribution of galaxies from the Continuum Halos in Nearby Galaxies – an EVLA Survey (CHANG-ES).
Though the observational mode was not optimized for detecting $\hi$, we successfully produce $\hi$ cubes for 19 galaxies. The moment-0 maps from this work are available on CHANG-ES data release website, i.e., \url{https://www.queensu.ca/changes}. Our sample is dominated by star-forming, $\hi$-rich galaxies at distances from 6.27 to 34.1 Mpc. $\hi$ interferometric images on two of these galaxies (NGC 5792 and UGC 10288) are presented here for the first time, while 12 of our remaining sample galaxies now have better $\hi$ spatial resolutions and/or sensitivities of intensity maps than those in existing publications. We characterize the average scale heights of the $\hi$ distributions for a subset of most inclined galaxies (inclination $>$ 80 deg), and compare them to the radio continuum intensity scale heights, which have been derived in a similar way. The two types of scale heights are well correlated, with similar dependence on disk radial extension and star formation rate surface density but different dependence on mass surface density. This result indicates that the vertical distribution of the two components may be governed by similar fundamental physics but with subtle differences.
\end{abstract}


\section{Introduction} 

A key component of cosmological galaxy formation models is the correct handling of galaxy outflows, in order to reproduce the correct amount and distribution of mass in stars, gas and metals \citep{2009ApJ...695..292C, 2013MNRAS.436.3031V, 2015ARA&A..53...51S}. But how outflows are launched and affect the interstellar medium (ISM) and circum-galactic medium (CGM) remains poorly understood, leading to a variety of results and large uncertainties in cosmological simulations of galaxy formation \citep{2015ARA&A..53...51S, 2017ARA&A..55...59N}. It is likely that the cosmic rays (CRs) accelerated by supernovae play a key role in driving outflows \citep{1990ApJ...365..544B, 2008ApJ...674..258E, 2010ApJ...711...13E}, while the structure of magnetic fields in and around galaxies seems to be strongly shaped by the outflows \citep{2009RMxAC..36...25K, 2012mfu3.conf..155K}. The non-thermal pressure from magnetic fields and CRs provides significant support for the CGM, and they may affect the recycling and new accretion of gas onto galactic disks, thereby affecting the future evolutionary path of the galaxy \citep{2020MNRAS.497.1712B}. How these processes exactly work, particularly for different types of galaxies, has largely been limited by a lack of observational constraints. It was against this background that the CHANG-ES program was initiated to map galaxy halos at radio wavelengths (1.5 and 6 GHz) for 35 highly inclined galaxies, including the magnetic fields and the CRs \citep{2016ApJ...824...30D, 2019A&A...623A..33S, 2019A&A...622A...9M, 2019A&A...632A..10M}.

So far, fruitful results have been obtained in the CHANG-ES project characterizing the radio properties of the halo. For example, high latitude radio continuum emission seems to be prevalent around star-forming galaxies \citep{2012AJ....144...44I}. Depending on the star formation rate (SFR), the galaxies can display global radio continuum halos, or discrete high latitude radio continuum features with strong local magnetic fields \citep{2013AJ....146..164I}. A budget for the energy injected by supernovae into different ISM/CGM components can be derived by investigating the distribution above a galaxy's midplane of spectral index, star formation rate, and luminosity (e.g., radio continuum and X-ray) \citep{2016ApJ...824...30D, 2019A&A...632A..10M}. The results seem to depend on the level of global SFR \citep{2016MNRAS.456.1723L}.

The vertical structures of the non-thermal radio halo depend on the whether CR electron transportation is mainly advective or diffusive \citep{heesen2018halos,2019A&A...632A..12S, 2019A&A...623A..33S, 2020A&A...639A.111S, 2021MNRAS.tmp.2538H}. Modeling these structures reveals outflow velocities close to escape velocities and mass outflow rates on the order of the SFR \citep{2016ApJ...824...30D, 2021MNRAS.tmp.2538H}. Large-scale, coherent magnetic fields expected to constrain the hot gas halo \citep{2012AJ....144...44I} are detected in all non-thermal radio halos \citep{2020A&A...639A.112K}, but can have complex structures including reversals and `ropes' \citep{2019A&A...623A..33S, 2019A&A...632A..10M}. The morphology and pressure in the radio halo are further regulated by pressure from the intracluster medium if the galaxy is in a cluster \citep{2016ApJ...824...30D}.

To further the understanding of the galactic disk/halo interaction, we need to explore the interplay of the cosmic-ray/magnetic field with various gaseous components of the ISM. One expects that diffuse atomic neutral hydrogen gas ($\hi$) should play a major role in this interplay. As the same outflow physics which shapes the radio halos regulates the baryonic flow, it is natural to expect that the properties of galactic $\hi$ gas can provide additional clues. For example, there is already previous observational evidence that the local morphology and kinematics of the $\hi$ are affected by outflows, forming holes, shells, and spurs \citep{1993AJ....105.2098R, rand1994atomic, 2000ApJ...536..645C, 2004ApJ...606..258W}, producing turbulence and increasing velocity dispersion \citep{tamburro2009driving}, and thickening the $\hi$ disk \cite{2002AJ....123.3124F, 2007AJ....134.1019O, 2019A&A...631A..50M, 2021ApJ...916...26R}. To investigate if $\hi$ gas has the same outflow physics as cosmic-ray gas traced in radio continuum, we explore the thickness of $\hi$ component of galaxies following the similar method of \cite{2018AA...611A..72K}, who measured the thickness of radio continuum.

As $\hi$ gas also serves as the raw material for forming interstellar molecular clouds and stars, globally its richness and locally its spatial correlation with star formation sites constrain the balance between fueling and feedback \citep{leroy2008star, 2009ApJ...693..216K, 2012ApJ...745...69K}. $\hi$ has also been recognized as a sensitive probe of external environmental perturbations, both gravitational and hydrodynamic, which may modify CGM properties \citep{2009AJ....138.1741C, 2015MNRAS.448.1715J,  2017ApJ...838...81Y}. Thus, $\hi$ provides powerful additional information to help interpret the observed halo properties in the context of understanding outflow physics. Fortunately, the 21 cm emission line tracing the $\hi$ gas was covered in the bandwidth used by CHANG-ES to observe the radio halos (see below).

$\hi$ images for highly inclined galaxies like those in the CHANG-ES sample have additional more specific scientific value beyond the more general research of galaxy evolution. This is because those images provide opportunities to look into several unique features of $\hi$ disks. \cite{kalberla2009hi} observed $\hi$ disk flaring --- that the scale height of the disk exponentially increases with radius in the `edge-on' Milky Way between 5 and 35 kpc. $\hi$ flares are found to be a common property in spiral galaxies, and reflect a quasi-hydrostatic equilibrium between the ISM pressure and a galaxy's gravity \citep{brinks1984high, bigiel2012universal}. 
\cite{2007AJ....134.1019O} and \cite{2013MNRAS.434.2069K} used deep $\hi$ observations of edge-on galaxies and found a thick $\hi$ layer with a typical scale height of 1$-$2 kpc. The thick $\hi$ disk is also referred to as the extraplanar gas or the $\hi$ halo. It may form due to gas recycling, which is driven by outflows in the galactic fountain model \citep{shapiro1976consequences, bregman1980galactic}. 

$\hi$-warps are also ubiquitous in the outer regions of disk galaxies  where the $\hi$ extends beyond the edge of the optical disk.  Such an effect is most clearly observable when galaxies are edge-on \citep{1957AJ.....62...90B, 1957AJ.....62...93K, 1976A&A....53..159S, 1977MNRAS.181..573N, 1978PhDT.......195B, 1990ApJ...352...15B}. \cite{1976A&A....53..159S} observed $\hi$ warps in four out of five edge-on samples, while \cite{garcia2002neutral} observed warps in 20 out of 26 of edge-on galaxies. The warps provide clues about several processes of galaxy evolution, potentially including halo-galaxy misalignment \citep{toomre1983theories, dekel1983galactic, sparke1988model}, gas accretion \citep{jiang1999warps, shen2006galactic, rovskar2010misaligned}, and environmental perturbation \citep{weinberg1995production, weinberg2006magellanic}.
Thus, an observational census of edge-on systems can provide useful insights into the different physics that shape the vertical extent of $\hi$ disks.

The CHANG-ES observations have a spatial resolution that in most cases is comparable to or finer than previous observations, which aids in revealing fine structure.
With these motivations, we extracted $\hi$ images from the CHANG-ES data cubes in the 1.5 GHz band and present them as a uniform dataset as part of the CHANG-ES project. 
Although this project was not initially designed for $\hi$ observations, in this paper we report the successfully extraction of $\hi$ fluxes from within the wide bands that were employed (section 2).
$\hi$ interferometric images of two of our galaxies have not previously been published. We use standard scaling relations between $\hi$ mass, SFR, and stellar mass, to characterize the type of galaxies included in the sample (section 3).
We measure the exponential $\hi$ scale heights (Appendix A) and compare the $\hi$ scale heights with radio scale heights (section 3.2.1) and $\hi$ radius (section 3.2.2). We compare the normalized $\hi$ scale heights with other physical parameters, i.e., total mass surface density (section 3.2.3). We analyze the consistency and inconsistency between the behaviors of $\hi$ component and radio continuum scale heights and discuss the possible physics scenario (section 4).
We also inspect the $\hi$ moment-0 images of the sample and compare them to optical and H$\alpha$ images (Appendix B and C), confirming the complexities in morphology as discussed above.
We assume a $\Lambda$CDM Cosmology (H$_{0}$=73 km s$^{-1}$ Mpc$^{-1}$) and Kroupa initial mass function \citep{kroupa2001variation} for calculations throughout this paper.

This paper is not meant to be an exhaustive study of the $\hi$ in individual galaxies, but to provide and make public a consistent data set for follow-up work. The major goal of this paper is to present the new $\hi$ image data obtained from the CHANG-ES observation, and investigate drivers for the thickness of the $\hi$ component in comparison with the thickness of the radio continuum which has been investigated thoroughly in \cite{2018AA...611A..72K}.

\section{Sample and Data reduction}
\subsection{Sample}
\begin{figure}
\includegraphics[width=0.45\textwidth]{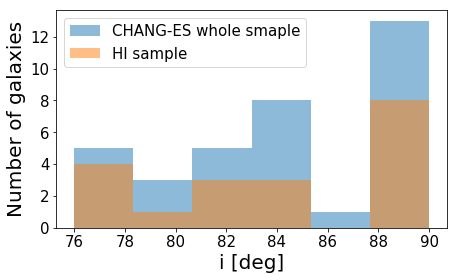}
{\textbf{(a)}}
\includegraphics[width=0.45\textwidth]{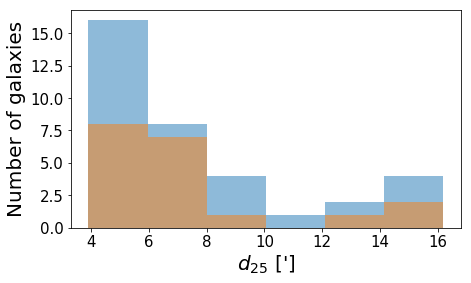}
{\textbf{(b)}}
\includegraphics[width=0.45\textwidth]{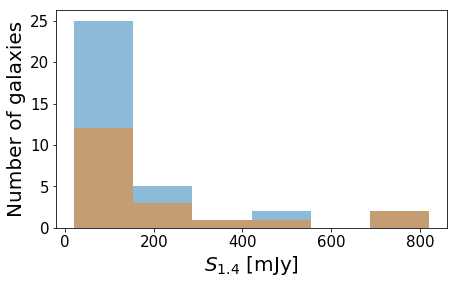}
{\textbf{(c)}}
\caption{The histograms of comparing (a) inclination (i), (b) blue isophotal diameters ($d_{25}$), and (c) flux densities at 1.4 GHz ($S_{1.4}$) between the original CHANG-ES sample (blue) and the current $\hi$ sample (orange). The K-S test probabilities P value of these three distributions are all larger than 0.99. We do not find significant differences between the two samples.}
\label{changes}
\end{figure}

The CHANG-ES sample was selected from the Nearby Galaxies Catalog \citep{tully1988nearby}. The details of sample were described in \cite{2012AJ....144...43I}. In Figure \ref{changes}, we plot histograms of comparing inclination (i), blue isophotal diameters ($d_{25}$), and flux densities at 1.4 GHz ($S_{1.4}$), between the original CHANG-ES sample and the current $\hi$ sample. The major selection criteria of CHANG-ES have been based on these parameters \citep{2012AJ....144...43I}. We derive the P value of K-S test probabilities which are all larger than 0.99. Based on the histograms and K-S test probabilities, we do not find significant differences between the two samples. 

We use the coordinates, local galaxy number densities ($\rho$), and optical diameter ($D_{25}$) presented in CHANG-ES Paper I \citep{2012AJ....144...43I}, and distances from CHANG-ES Paper IV \citep{wiegert2015chang}. The galaxies have distances ranging from 6.27 to 34.1 Mpc, and are in a variety of environments, including interactions, groups, and clusters. We also take the available radio scale heights derived in the L-band for 9 galaxies from \cite{2018AA...611A..72K}.

\subsection{HI data reduction}
The galaxies were observed with the The Karl G. Jansky Very Large Array (VLA) using the B, C, and D configurations, with on-source times per pointing of 2 hr, 30 min, and 20 min, respectively. The observations relevant to this paper were conducted in L band, employing the same data sets that were used to obtain the continuum images. Although CHANG-ES was not designed to obtain $\hi$ images, the fact that CHANG-ES data were obtained as multi-frequency products allowed for the possibility of extracting the $\hi$ spectral line as well.
Details of the observations and continuum data reduction have been previously described in \cite{2012AJ....144...43I} and \cite{wiegert2015chang}.

We start with the calibrated visibilities from the CHANG-ES continuum data reductions. The reduction steps follow the standard routines of the Common Astronomy Software Applications (CASA) \citep{2007ASPC..376..127M}.
We split out the spectral windows that contain the $\hi$ signal and apply self-calibration to the continuum emission to improve the image quality. Then we subtract the continuum emission from the visibilities using the UVCONTSUB task. To do so we first mask channels that clearly contained emission line flux and then fit the continuum of each visibility with a polynomial function of no more than order 1. We produce the cubes using the CLEAN task, with a Briggs robustness weighting of 0.5. The resulting data cubes have an average RMS of $\sim$0.4 mJy/beam, average beam size of $\sim$14.5'' in full width half maximum (FWHM), and a typical channel width of 52.8 km/s. 

We successfully process the $\hi$ signal for 19 galaxies in the L-band C-configuration observations. We refer to these as the $\hi$ sample. We are unable to properly extract the $\hi$ emission from the VLA data of the remaining 16 galaxies, typically because the $\hi$ line coincides with radio interference (RF) channels or edges of broadband windows. The basic information and properties of the data cubes for the $\hi$ sample are shown in Table \ref{sample}. The moment-0 maps from this work are available on the CHANG-ES data release website, i.e., \url{https://www.queensu.ca/changes}.

\begin{table*}
\centering
\caption{The Properties of $\hi$ Sample}
\resizebox{\textwidth}{40mm}{
\begin{tabular}{lcccccccccccc}
 \hline
Galaxies & R.A. (J2000)$^{a}$ & Decl. (J2000)$^{a}$ & Distance$^b$ & $\rho^c$ & SFR$^d_{\rm H_{\alpha}+22\mu m}$ & $M_{*}$ & $D_{25}^e$ & Bmaj & Bmin & RMS & 3-$\sigma$ column density limit\\
 & [h m s] & [$^{\circ}$ ' ''] & [Mpc] & [Mpc$^{-3}$] & [$M_{\odot}\rm yr^{-1}$] & 
$10^9 M_{\odot}$ & ['] & [''] & [''] & [mJy beam$^{-1}$] & [$10^{20}$ cm$^{-2}$]\\
 \hline
NGC 660  & 01 43 02.40 & $+$13 38 42.20 & 12.3 & 0.12 & 3.53 $\pm$ 0.34 & 12.52 & 7.2 & 16.24 & 14.81 & 0.497 & 3.61\\
NGC 2683 & 08 52 41.33 & $+$33 25 18.26 & 6.27 & 0.09 & 0.27 $\pm$ 0.03 & 13.89 & 9.1 & 13.74 & 12.84 & 0.374 & 3.71\\
NGC 3003 & 09 48 36.05 & $+$33 25 17.40 & 25.4 & 0.64 & 1.66 $\pm$ 0.17 & 19.44 & 6.0 & 13.09 & 12.93 & 0.401 & 4.14\\
NGC 3044 & 09 53 40.88 & $+$01 34 46.70 & 20.3 & 0.19 & 1.86 $\pm$ 0.17 & 14.10 & 4.4 & 15.74 & 14.19 & 0.352 & 2.76\\
NGC 3079 & 10 01 57.80 & $+$55 40 47.24 & 20.6 & 0.29 & 5.41 $\pm$ 0.48  & 44.12  & 7.7 & 13.81 & 13.09 & 0.567 & 5.49\\
NGC 3448 & 10 54 39.20 & $+$54 18 17.50 & 24.5 & 0.24 & 1.90 $\pm$ 0.19  & 5.77  & 4.9 & 13.92 & 13.01 & 0.368 & 3.55\\
NGC 3556 & 11 11 30.97 & $+$55 40 26.80 & 14.09 & 0.15 & 3.80 $\pm$ 0.32  & 44.74  & 7.8 & 12.89 & 11.95 & 0.450 & 5.12\\
NGC 3877 & 11 46 07.70 & $+$47 29 39.65 & 17.7 & 1.53 & 1.44 $\pm$ 0.13  & 29.79  & 5.1 & 13.54 & 13.13 & 0.440 & 4.33\\
NGC 4096 & 12 06 01.13 & $+$47 28 42.40 & 10.32 & 0.40 & 0.76 $\pm$ 0.09  & 9.77  & 6.4 & 14.56 & 13.42 & 0.437 & 3.91\\
NGC 4157 & 12 11 04.37 & $+$50 29 04.80 & 15.6 & 1.19 & 1.87 $\pm$ 0.19 & 24.22  & 7.0 & 13.80 & 13.75 & 0.431 & 3.97\\
NGC 4217 & 12 15 50.90 & $+$47 05 30.40 & 20.6 & 0.95 & 2.01 $\pm$ 0.19  & 39.69  & 5.1 & 14.65 & 13.37 & 0.373 & 3.33\\
NGC 4302 & 12 21 42.48 & $+$14 35 53.90 & 19.41 & 3.60 & 0.98 $\pm$ 0.09  & 28.80  & 4.7 & 15.77 & 14.52 & 0.390 & 2.98\\
NGC 4565 & 12 36 20.78 & $+$25 59 15.63 & 11.9 & 1.00 & 1.02 $\pm$ 0.10  & 56.49  & 16.2 & 14.60 & 13.78 & 0.371 & 3.23\\
NGC 4631 & 12 42 08.01 & $+$32 32 29.40 & 7.4  & 0.41 & 2.79 $\pm$ 0.23  & 10.74  & 14.7 & 14.16 & 13.44 & 0.494 & 4.54\\
NGC 4666 & 12 45 08.59 & $-$00 27 42.79 & 27.5 & 0.54 & 11.19 $\pm$ 0.98  & 79.45  & 4.2 & 18.32 & 14.41 & 0.381 & 2.52\\
NGC 5084 & 13 20 16.92 & $-$21 49 39.30 & 23.4 & 0.29 & 0.11 $\pm$ 0.03$^*$ & 110.76 & 12.5 & 25.27 & 13.89 & 0.400 & 1.99\\
NGC 5775 & 14 53 57.60 & $+$03 32 40.05 & 28.9 & 0.67 & 8.05 $\pm$ 0.69  & 64.40  & 3.9 & 21.25 & 14.76 & 0.345 & 1.93\\
NGC 5792 & 14 58 22.71 & $-$01 05 27.90 & 31.7 & 0.52 & 4.70 $\pm$ 0.39 & 100.04   & 7.2 & 17.00 & 14.16 & 0.383 & 2.78\\
UGC 10288& 16 14 24.80 & $-$00 12 27.10 & 34.1 & 0.23 & 0.70 $\pm$ 0.07  & 16.99  & 4.9 & 18.46 & 14.52 & 0.378 & 2.46\\
 \hline
\end{tabular}}
\begin{tablenotes}
\item $^a$ From the NASA Extragalactic Database (NED).
\item $^b$ Presented in CHANG-ES Paper IV \citep{wiegert2015chang}.
\item $^c$ Number density of galaxies brighter than $-$16 mag in the vicinity of the galaxy; presented in CHANG-ES Paper I \citep{2012AJ....144...43I}.
\item $^d$ SFRs were estimated using a combination of H$\alpha$ and 22 $\mu$m data; presented in CHANG-ES Paper XVII \citep{2019ApJ...881...26V}. An exception is NGC 5084, which lacks H$\alpha$ observations. Therefore its SFR is estimated from 22 $\mu$m data alone; see CHANG-ES Paper IV. \citep{wiegert2015chang}.
\item $^e$  Optical blue band diameter at the 25th mag arcsec$^{-2}$ isophote presented in CHANG-ES Paper I \citep{2012AJ....144...43I}.
\end{tablenotes}
\label{sample}
\end{table*}

\subsection{HI mass, radius, and scale height}
We extract $\hi$ emission and generate $\hi$ intensity maps with the software SoFiA \citep{2015MNRAS.448.1922S}, which uses a smooth$+$clip algorithm to detect fluxes, and an algorithm based on the noise distribution to determine detection reliabilities.
We use a threshold of 4-$\sigma$, a maximum smoothing kernel of 6 beams and no spectral smoothing, and a reliability threshold of 0.99. We also inspect the detection masks to ensure reliable products from SoFiA. 

The total $\hi$ mass $M_{\rm HI}$ is calculated as
\begin{equation}
M_{\rm HI}/M_{\odot} = 2.356 \times 10^5 S_{\rm HI} D^2_{\rm Mpc} ,
\end{equation}
where $S_{\rm HI}$ is the total flux in Jy km s$^{-1}$ and $D_{\rm Mpc}$ is the distance in Mpc. The galaxies, NGC 3448 and NGC 5775, are merging with their companion galaxies. For those galaxies, we update the $\hi$ mass after roughly segmenting each galaxy from their companions on the moment-0 map (see red dash lines in Figure \ref{atlas}).

As the galaxies are highly inclined, $\hi$ self-absorption can be significant. However, as there is no reliable way yet to accurately determine the extent of self-absorption without detecting absorption lines along sight-lines to quasars, we leave the data uncorrected at present. It is thus likely that the $\hi$ masses derived are underestimated. It is also possible that the self-absorption is strongest in the mid-plane, causing a systematic overestimate of the $\hi$ scale heights (see below). But comparing the $\hi$ mass or scale heights within this sample of galaxies should still lead to meaningful results since the galaxies have similar inclinations. We note that, the fraction of self-absorbed $\hi$ is roughly $11 \pm 3\%$ for a galaxy with an inclination of 80 degrees \citep{jones2018alfalfa}, which is the typical inclination in our sample.

\begin{figure}
    \centering
	\includegraphics[width=0.4\textwidth]{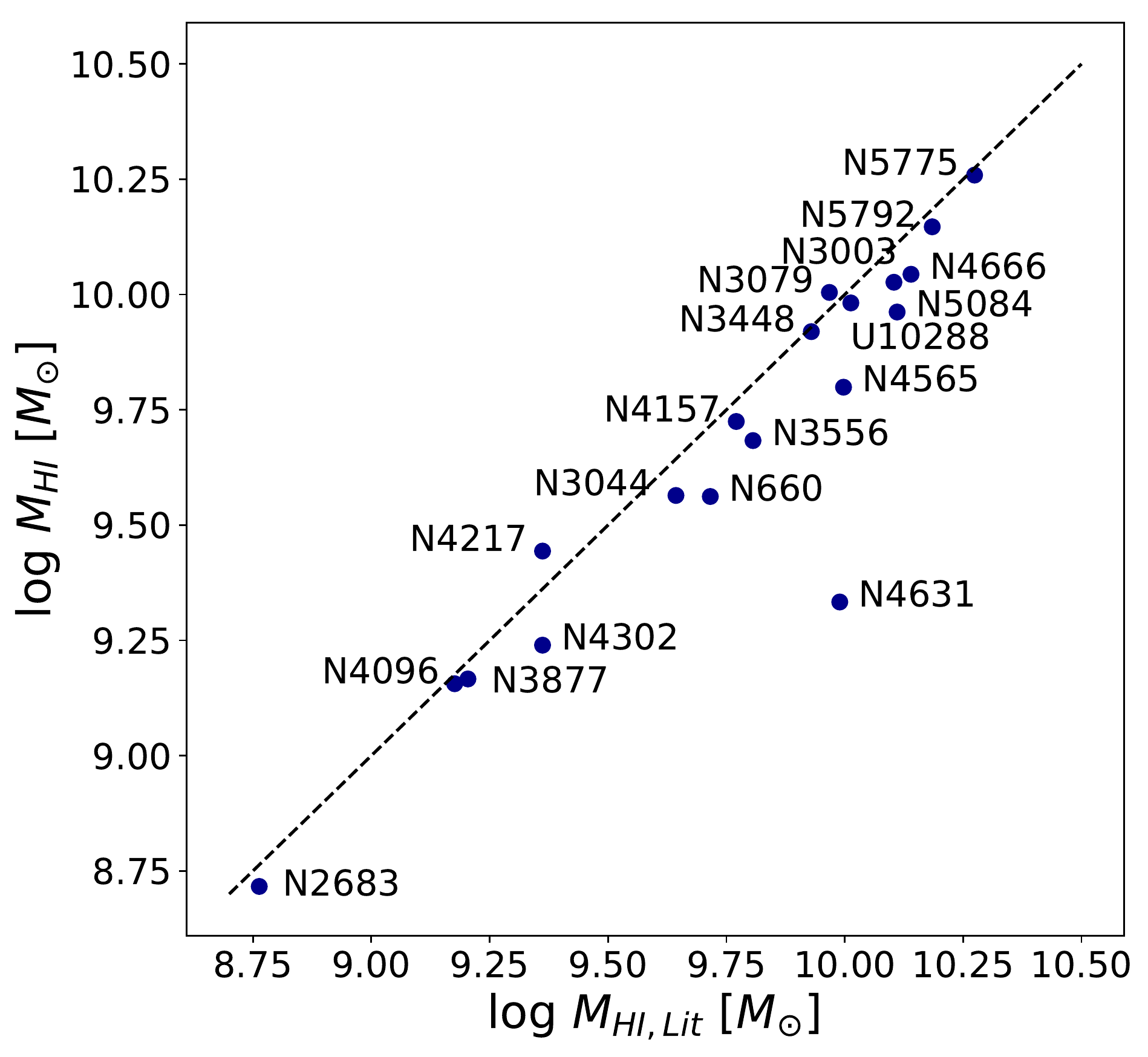}
	\caption{$\hi$ mass comparison with literature ($M_{\rm HI,Lit}$) from single-dish observations. The dash line represents the 1:1 line.}
	\label{himass}
\end{figure}

Figure \ref{himass} shows the comparison between the $\hi$ emission masses obtained from our data and those determined from a variety of single-dish observations in the literature; the references are listed in Table~2 along with these masses. The uncertainties in this paper's $\hi$ mass mainly result from the calibration process and are typically 15\%.
The CHANG-ES $\hi$ masses are consistent with those in literature, but are slightly lower by 0.06 dex (excluding the interacting galaxy NGC 4631). This is likely due to the well-known problem that interferometry misses large angular scale fluxes \citep{rohlfs2013tools}.

NGC 4631 has a significantly lower $M_{\rm HI}$ derived using our data than the literature value, probably because it is a merger system consisting of NGC 4631 and two companions. Our relatively shallow $\hi$ image of NGC 4631 misses most of the extended $\hi$ tail from interaction with NGC 4646 \citep{1993AJ....105.2098R}, but successfully separates the disk from that of NGC 4627 (3' away) which is confused in the single dish data. In addition, because NGC 4631 is the second largest galaxy in $\hi$ angular size in our $\hi$ sample, the missing flux is also significant.

The $\hi$ characteristic radius $R_{\rm HI}$, defined by the 1 $M_{\odot}~{\rm pc}^{-2}$ isophote along the major axis. For our analysis, this radius is estimated based on the $\hi$ size-mass relation from \cite{2016MNRAS.460.2143W}
\begin{equation}
{\rm log} R_{\rm HI} = (0.506 \pm 0.003) ~{\rm log}M_{\rm HI} - (3.593 \pm 0.009).
\end{equation}
We derive each galaxies' $\hi$ radius using our $M_{\rm HI}$ and this $\hi$ size-mass relation instead of directly measuring $R_{\rm HI}$ from the $\hi$ images to avoid uncertainties involved in de-projections of these highly edge-on galaxies. Meanwhile, the scatter of size-mass relation is extremely small ($\sim$0.06 dex) and does not change with galaxy luminosity, $\hi$ richness or morphological type \citep{2016MNRAS.460.2143W}.

We measure the average $\hi$ scale heights of the galaxies which have inclination larger than 80 deg through the task “BoxModels” of the NOD3 program package \citep{2017A&A...606A..41M}. The average $\hi$ scale heights indicate the thickness of the galactic $\hi$ gas, with the thin disk, the thick disk, and the extraplanar (possibly halo like and$\slash$or fountain like) $\hi$ mixed together. Limited by the resolution, the vertical profiles in our sample do not present two or three different components. The same method was used in \cite{2018AA...611A..72K} to measure the scale heights of radio continuum. We largely follow the procedure of \cite{2018AA...611A..72K} but make two additional adjustments for the $\hi$ component. The detailed description of “BoxModels” and our adjustments can be found in the Appendix~A. The major steps include, binning the $\hi$ intensity map into strips perpendicular to the major axis, using a radially dependent correcting equation to account for beam smearing, fitting the surface density vertical profiles of each strip with an exponential function, and calculating the radially averaged exponential scale height within the optical radial range ($R_{25}$) as the final measure of scale height for each galaxy ($\bar{h}_{\rm HI,25}$). 

As the method is simply photometric, $\bar{h}_{\rm HI,25}$ can differ from the real scale heights of galaxies due to several systematic effects. Firstly, the planar projection can cause considerable over-estimation since the disks are not perfectly edge-on.
Secondly, the flaring of a disk expected in a typical galaxy under quasi-static equilibrium is barely measurable when mixed with projection effects. It is undetected by $\bar{h}_{\rm HI,25}$ which is a radially averaged measure.
Thirdly, the warp in the outer $\hi$ disk can further project onto the optical radial range and cause the asymmetry and over-estimation of $\bar{h}_{\rm HI,25}$. This effect should be mitigated by the fact that most of the $\hi$ disks in our sample are not highly extended with respect to the optical disks (i.e. median $R_{\rm HI}/R_{25}\sim 1.18$). Finally, although we have corrected for the beam smearing following the empirical method of \cite{2018AA...611A..72K}, residual systematic uncertainties may remain for $\hi$ disks. Possibly due to these uncertainties, the derived $\bar{h}_{\rm HI,25}$ is systematically larger than a few hundred parsecs typically expected for a $\hi$ disk \citep[e.g.,][]{2019A&A...622A..64B}.

Despite these uncertainties, $\bar{h}_{\rm HI,25}$ provides a first-order characterization of the $\hi$ disk shape in the vertical direction and, within the sample, is expected to be statistically correlated with and thus indicate the galaxies' physical scale heights. Because the radio continuum scale heights from \cite{2018AA...611A..72K} and $\hi$ scale heights here are measured using a consistent method, thus suffering from the similar systematic biases, they can be directly compared (see Section 3.2).
The $\hi$ masses ($M_{\rm HI}$), $\hi$ radius ($R_{\rm HI}$), and average $\hi$ scale height ($\bar{h}_{\rm HI,25}$) are listed in Table \ref{HIsample}.

\begin{table}
\caption{The $\hi$ Properties}
\resizebox{0.48\textwidth}{40mm}{
\begin{tabular}{lccccccc}
  \hline
Galaxies & ${\rm log} M_{\rm HI}$ & $R_{\rm HI}$ & ${\rm log} M_{\rm HI,\rm Lit}^a$ & Ref & $\bar{h}_{\rm HI,25}$\\
 & [$M_{\odot}$] & [kpc] & [$M_{\odot}$] & & [kpc]\\
  \hline
NGC 660 & 9.56 & 17.55 & 9.72 & 1 & -\\
NGC 2683 & 8.72 & 6.55 & 8.76 & 1 & 0.76 $\pm$ 0.28\\
NGC 3003 & 10.03 & 30.16 & 10.10 & 1 & 3.48 $\pm$ 2.85\\
NGC 3044 & 9.56 & 17.60 & 9.64 & 1 & 0.59 $\pm$ 0.20\\
NGC 3079 & 10.00 & 29.40 & 9.97 & 1 & 1.52 $\pm$ 1.02\\
NGC 3448 & 9.92(9.89) & 25.64 & 9.93 & 2 & -\\
NGC 3556 & 9.68 & 20.22 & 9.81 & 1 & 1.19 $\pm$ 0.35\\
NGC 3877 & 9.17 & 11.07 & 9.20 & 1 & 1.33 $\pm$ 0.22\\
NGC 4096 & 9.16 & 10.94 & 9.18 & 1 & 1.05 $\pm$ 0.25\\
NGC 4157 & 9.72 & 21.22 & 9.77 & 1 & 0.75 $\pm$ 0.28\\
NGC 4217 & 9.44 & 15.29 & 9.36 & 1 & 0.58 $\pm$ 0.14\\
NGC 4302 & 9.24 & 12.06 & 9.36 & 3 & 0.61 $\pm$ 0.17\\
NGC 4565 & 9.80 & 23.14 & 9.94 & 3 & 0.73 $\pm$ 0.30\\
NGC 4631 & 9.33 & 13.45 & 9.76 & 4 & 0.73 $\pm$ 0.26\\
NGC 4666 & 10.04 & 30.78 & 10.14 & 1 & -\\
NGC 5084 & 9.96 & 27.98 & 10.11 & 5 & 2.10 $\pm$ 0.30\\
NGC 5775 & 10.26(10.09) & 32.82 & 10.27 & 1 & 1.30 $\pm$ 0.45\\
NGC 5792 & 10.15 & 34.70 & 10.18 & 5 & -\\
UGC 10288 & 9.98 & 28.63 & 10.01 & 1 & 1.17 $\pm$ 0.07\\
  \hline
\end{tabular}}
\begin{tablenotes}
\item The $\hi$ diameters of NGC 3448 and NGC 5775 are calculated by the $\hi$ mass in parentheses.
\item $^a$ The $\hi$ mass in literature from single-dish observation.
\item Ref.--- Reference to published $\hi$ mass.
\item 1. \cite{2005ApJS..160..149S}. 2. \cite{1974AJ.....79..767P}. 3. \cite{2011AJ....142..170H}. 4. \cite{rots1980neutral}. 5. \cite{2014AJ....148..127Y}.
\end{tablenotes}
\label{HIsample}
\end{table}

\subsection{SFR and stellar mass}

\cite{2019ApJ...881...26V} presented new narrow-band H$\alpha$ imaging for CHANG-ES and estimated SFRs in conjunction with the Wide-field Infrared Survey Explorer (WISE) 22 $\mu$m imaging.
We exclude NGC 5084 from our SFR related analysis, because NGC 5084 lacks an H$\alpha$ observation.
The stellar mass $M_*$ of each galaxy is estimated by the i-band luminosity from the Sloan Digital Sky Survey (SDSS) and the g-i color correlated mass-to-light ratio \citep{2003ApJS..149..289B}. We derive the i-band flux with the python package \textit{photutils} \citep{bradley2020astropy}. There might be some concern about uncertainties in the estimates of $M_*$ because of the high level of dust attenuation in these highly inclined galaxies. However, the reddening of the color (which artificially raises the $M_*$-to-light ratio) and the attenuation of the optical luminosity nearly cancel out \citep{2003ApJS..149..289B}, thus dust attenuation does not significantly affect the estimate of $M_*$. 

We test the reliability of our stellar mass estimation by comparing the $M_*$ values with those from the literature. \cite{2019ApJS..244...24L} estimated stellar masses based on the WISE mid-infrared bands, W1 (3.4 $\mu$m) and W2 (4.6 $\mu$m), and covered the whole sample in this study. 
The Two Micron All Sky Survey (2MASS) provides the Ks band luminosity \citep{2006AJ....131.1163S} for all galaxies in our sample, and a mass-to-light ratio of 0.6 $M_{\odot}/L_{\odot}$ \citep{2014AJ....148...77M} has been commonly adopted in the literature. The NASA-Sloan Atlas (NSA) catalog \footnote{\href{http://nsatlas.org}{http://nsatlas.org}} also bases on luminosity from SDSS images, but uses K-correction based spectral energy distribution fitting to derive the stellar mass. It provides stellar masses for 12 galaxies in our sample. 

\begin{figure}
    \centering
    \includegraphics[width=0.4\textwidth]{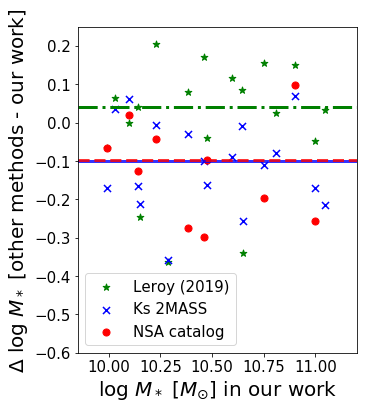}
    \caption{The differences between stellar masses derived in other methods and that in our work ($\Delta \log M_{*}$) vs. stellar mass in our work. The green, blue, and red colors points represent three stellar masses from WISE W1 (3.4 $\mu$m) and W2 (4.6 $\mu$m) \citep{2019ApJS..244...24L}, 2MASS Ks band luminosity \citep{2006AJ....131.1163S} with mass-to-light ratio of 0.6 $M_{\odot}/L_{\odot}$ \citep{2014AJ....148...77M}, and NSA catalog, respectively. The green, blue, and red lines present the median value of $\Delta \log M_{*}$ equal to 0.040, -0.100, and -0.098, respectively.}
    \label{stellarmass}
\end{figure}

Figure \ref{stellarmass} shows the differences between the three types of stellar masses (green, blue, and red points) from the literature and stellar masses in our work ($\Delta \log M_{*}$). The green, blue, and red lines present the median value of $\Delta \log M_{*}$ equaling to 0.040, -0.100, and -0.098, respectively. The offsets are consistent with the uncertainty reported in the literature \citep[e.g.,][]{2006AJ....131.1163S, 2003ApJS..149..289B}.

The results of the SFRs and stellar masses are listed in Table \ref{sample}. We also calculate the $\hi$ mass fraction ($M_{\rm HI}/M_{*}$) as the ratio between $\hi$ mass and stellar mass and the specific SFR (sSFR) as the ratio between SFR and stellar mass.

\section{Results}

\begin{figure}
\includegraphics[width=0.45\textwidth]{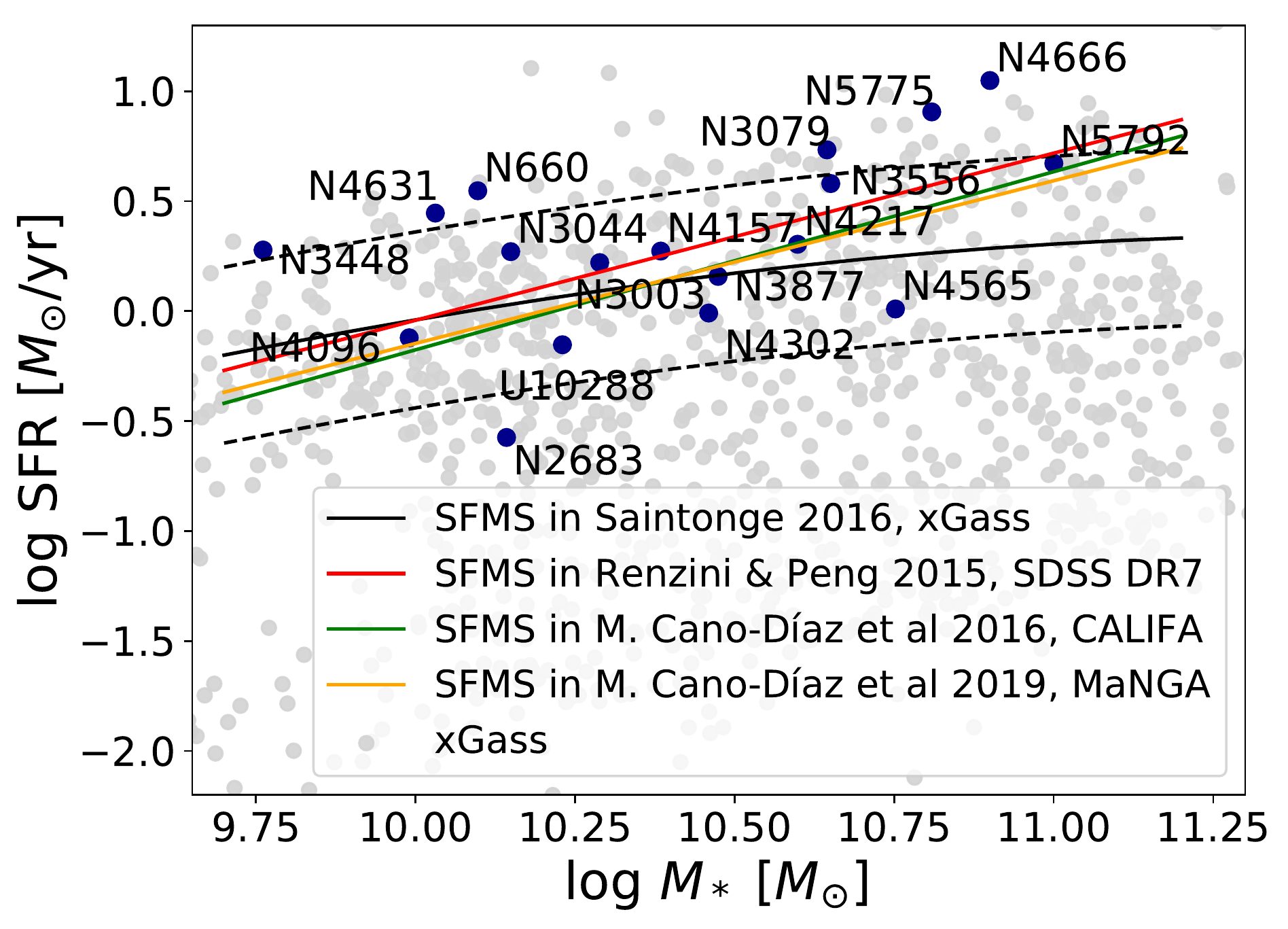}
{\textbf{(a)}}
\includegraphics[width=0.45\textwidth]{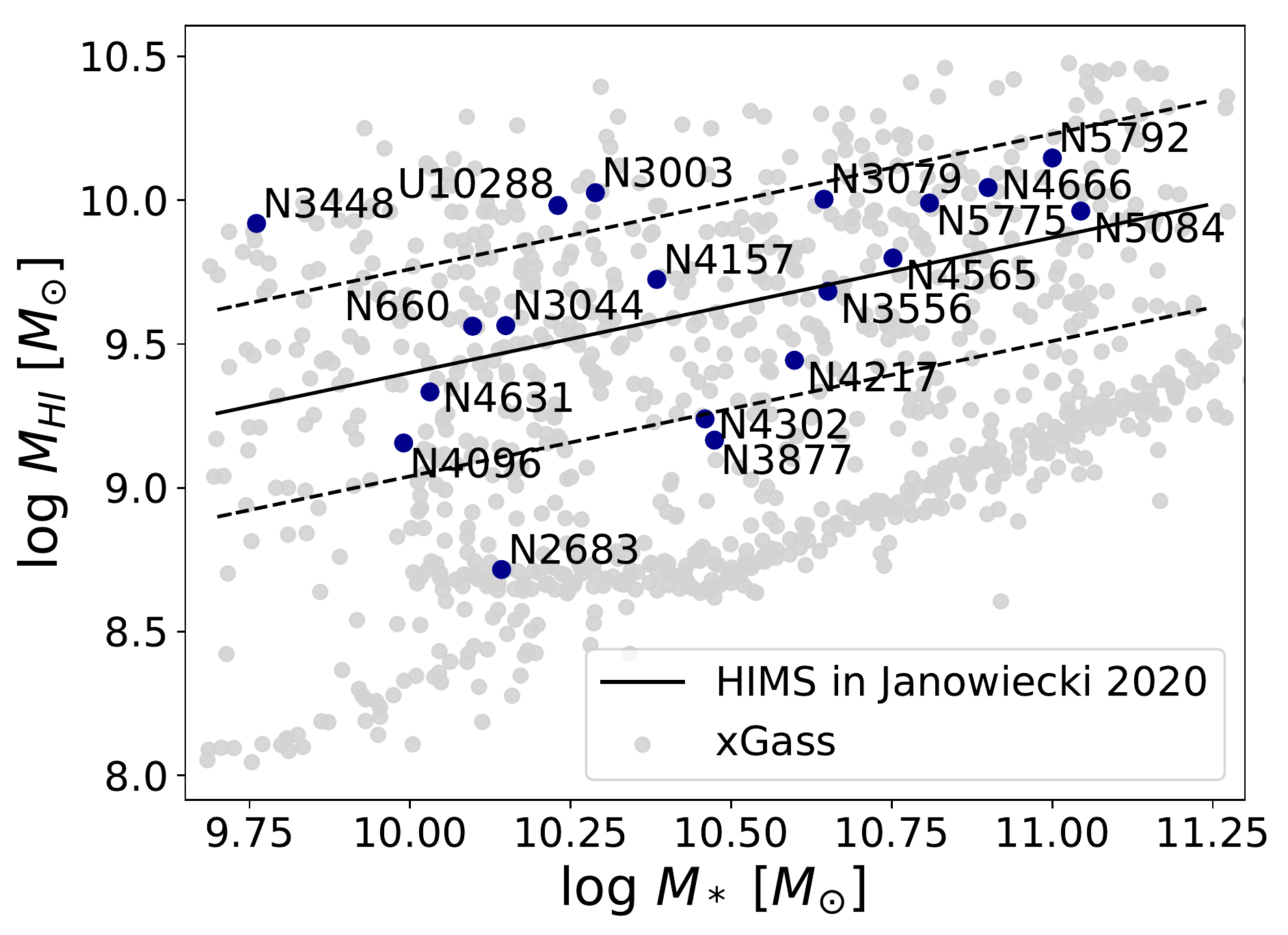}
{\textbf{(b)}}
\includegraphics[width=0.45\textwidth]{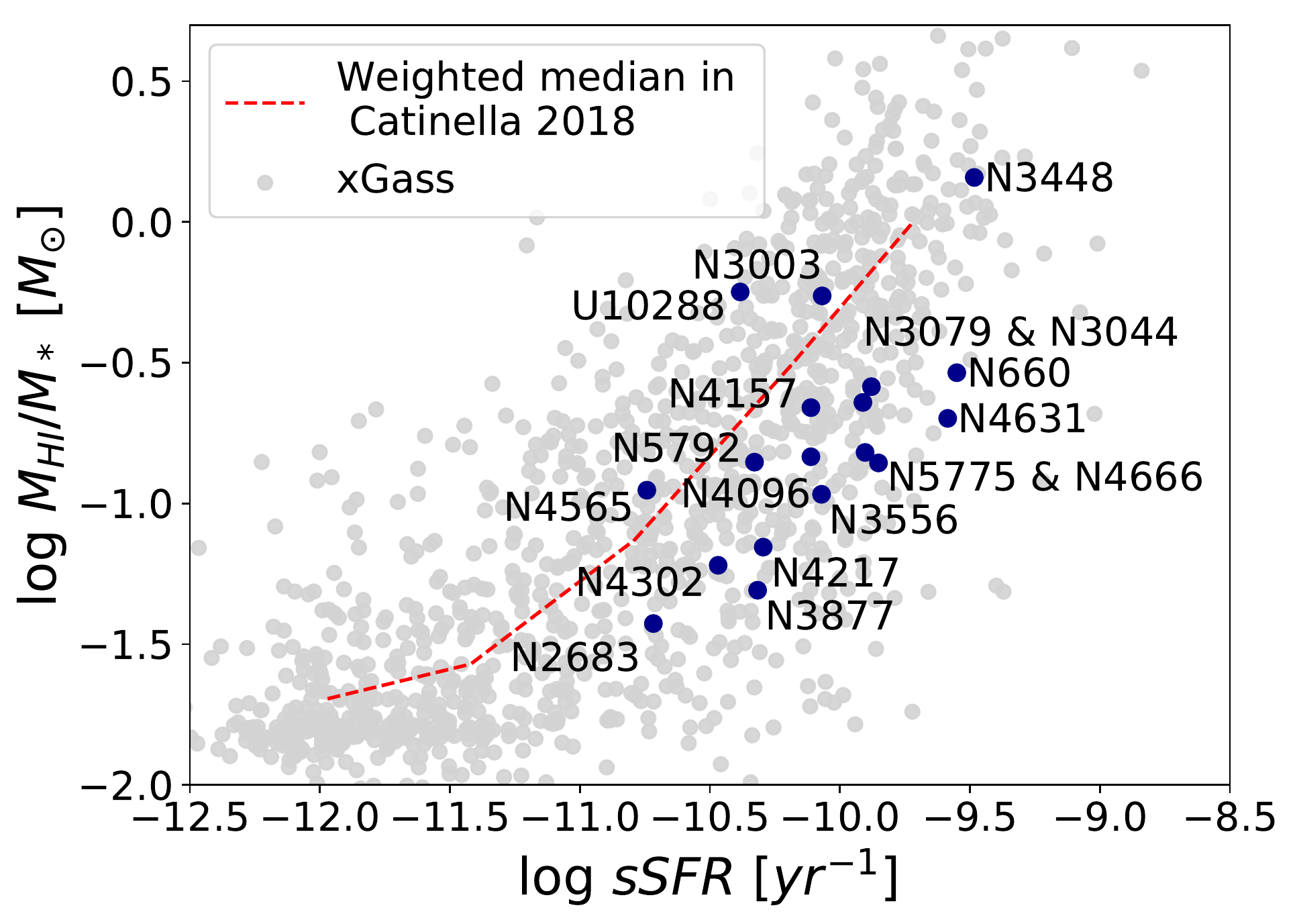}
{\textbf{(c)}}
\caption{The $\hi$ sample and SFR, $M_*$, and $M_{\rm HI}$ scaling relations. (a) The distribution of our $\hi$ sample (dark blue points) in the star formation rate versus stellar mass relation. The black solid and dashed lines represent the median position and 1-$\sigma$ envelope of the star forming main sequence (SFMS) \citep{2016MNRAS.462.1749S}. The color lines represent the SFMS for different surveys \citep{2015ApJ...801L..29R, 2016ApJ...821L..26C, 2019MNRAS.488.3929C}. (b) The distribution of our $\hi$ sample (dark blue points) in the $\hi$ mass versus stellar mass relation. The black solid and dashed lines show the median position and 1-$\sigma$ envelope of the $\hi$ main sequence \citep{2020MNRAS.493.1982J}. The median relation between $M_{\rm HI}$ and $M_*$ is derived from galaxies that are close to the SFMS. (c) The $\hi$ mass fraction is plotted as a function of specific SFR. In all three panels, the background galaxies in grey are from the extended GALEX Arecibo SDSS Survey (xGASS) \citep{catinella2018xgass}.}
\label{xGASS}
\end{figure}

We place the CHANG-ES $\hi$ sample on plots of scaling relations between SFR, $M_*$, and $M_{\rm HI}$ (Figure \ref{xGASS}) to provide an overview of the type of galaxies included in the $\hi$ sample. Most of the galaxies distribute along or above the star forming main sequence (SFMS) except for NGC 2683, but even NGC 2683 is only 1.51-$\sigma$ below the the SFMS. The majority of the galaxies also have $M_{\rm HI}$ along or above the $\hi$ main sequence (HIMS). Our $\hi$ sample is hence dominated by star-forming, $\hi$-rich galaxies. 

In Figure \ref{xGASS}(a), we plot the SFMSs derived with different samples, including xGASS \citep{2016MNRAS.462.1749S}, SDSS-DR7 \citep{2015ApJ...801L..29R}, CALIFA \citep{2016ApJ...821L..26C}, and MaNGA \citep{2019MNRAS.488.3929C}. The difference between the SFMSs can be the result of different measurements of star formation rates. \cite{2015ApJ...801L..29R} used SFR which is firstly derived with the H$\alpha$ luminosity within the 3-arcsec central fiber of SDSS and then extrapolated to represent the whole galaxy based on optical photometry \citep{2004MNRAS.351.1151B}. \cite{2016ApJ...821L..26C, 2019MNRAS.488.3929C} used dust attenuation corrected H$\alpha$ luminosity within the whole integral field unit (IFU) covered area (typically 1.5 to several times the effective radius) to derive the SFR. For xGass survey \citep{2016MNRAS.462.1749S}, the SFR was derived by combining GALEX NUV$\slash$FUV band and WISE 22 $\mu$m luminosity. In our work, the SFR has been derived by combining H$\alpha$ and WISE 22 $\mu$m luminosity  \citep{2019ApJ...881...26V}, which is close to those from xGASS. Although these different SFMSs have different slopes, particularly at the massive end ($M_*>10^{10.5}~M_{\odot}$), most galaxies in our sample tend to have SFR around or above all these SFMSs.

On the other hand, Figure \ref{xGASS}(c) suggests that most galaxies  have an $\hi$ gas fraction below the median value expected for their sSFR. This suggests that the star forming efficiency (out of $\hi$ gas) of the sample galaxies is significantly enhanced, unless the $\hi$ mass has been under-estimated by $\sim$0.3 dex due to $\hi$ self-absorption or missing large-scale flux. The potentially enhanced star forming efficiency may be consistent with the selection criteria of the CHANG-ES sample for high-SFR galaxies.

In the Appendix~B and C, we first investigate the $\hi$ morphologies of the whole $\hi$ sample by comparing the $\hi$ distribution with the distribution of optical light and H$\alpha$ fluxes (whenever data are available). We describe the morphologies of two galaxies which did not have $\hi$ images in the literature before, and then we describe the $\hi$ morphologies of the rest galaxies by comparing to results in the literature. We use the deviations of galaxies from the scaling relations presented in Figure \ref{xGASS} as indicators of the $\hi$-richness and star forming status compared to normal star-forming disk galaxies.

\subsection{HI morphologies}
 
\begin{figure*}
\centering
\includegraphics[height=0.27\textheight]{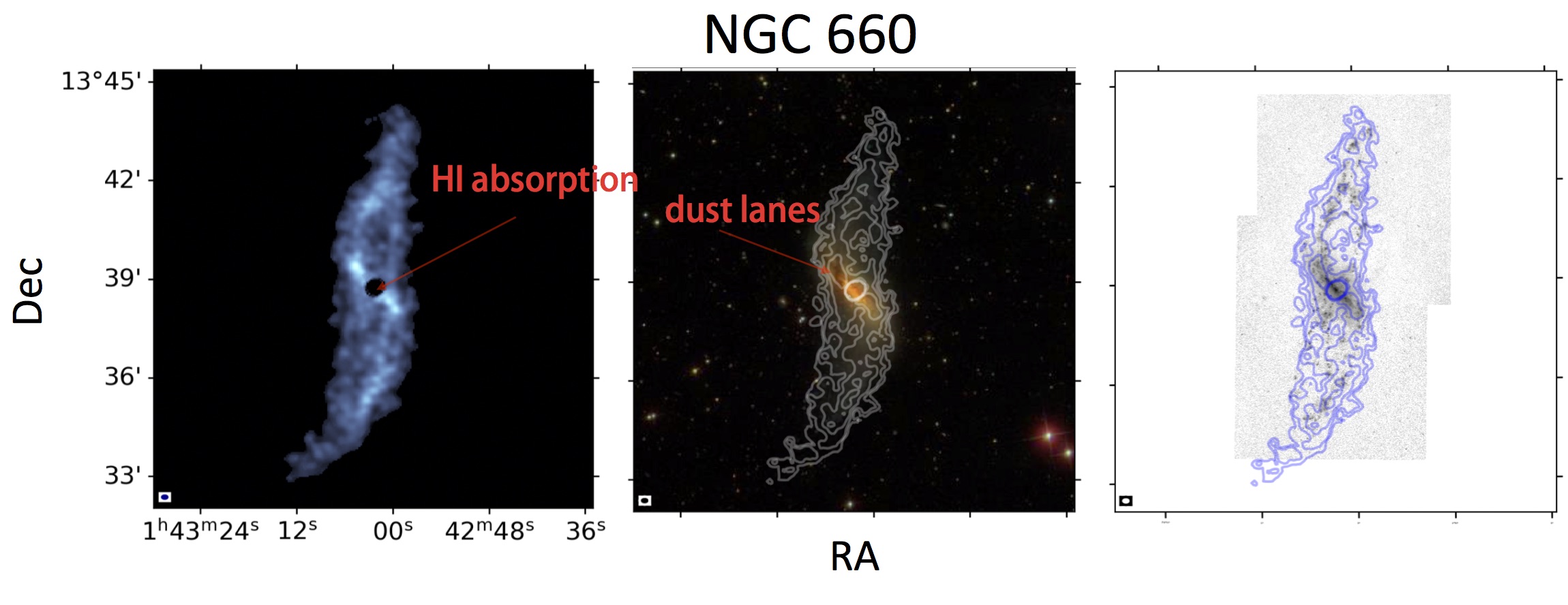}\\ \vspace{-4mm}
\includegraphics[height=0.27\textheight]{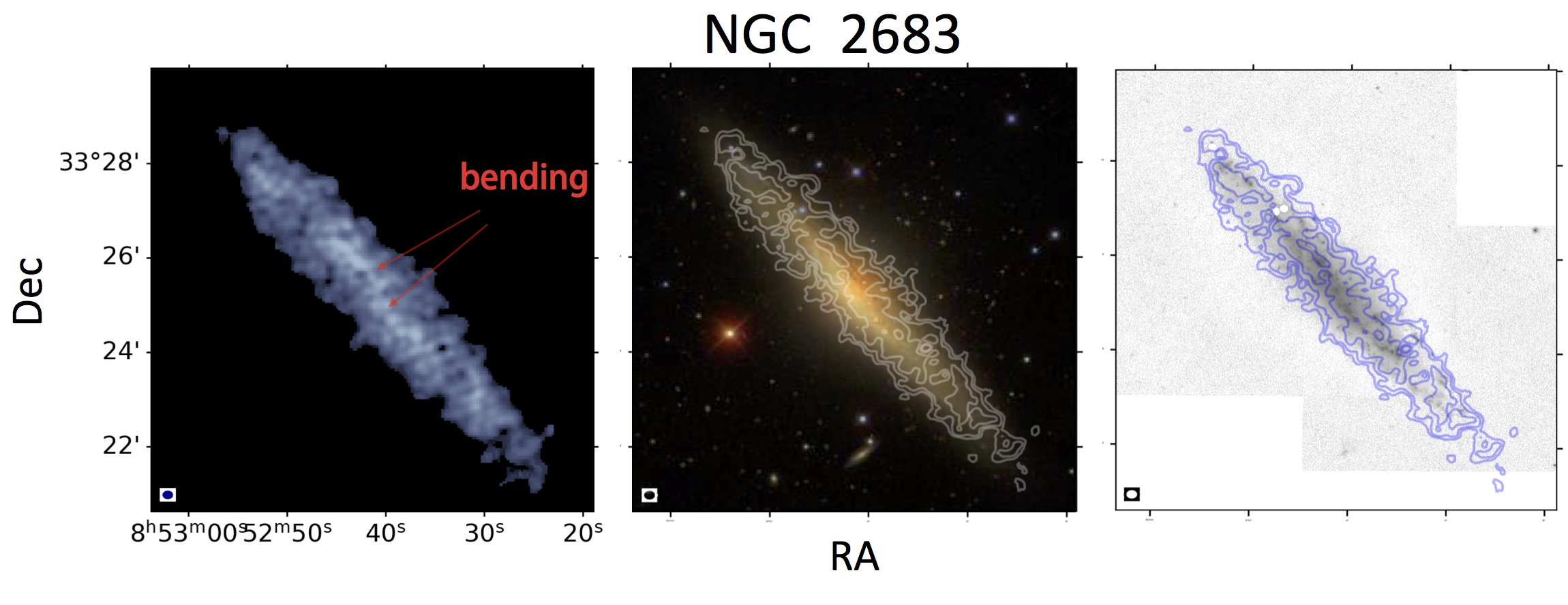}\\ \vspace{-4mm}
\includegraphics[height=0.27\textheight]{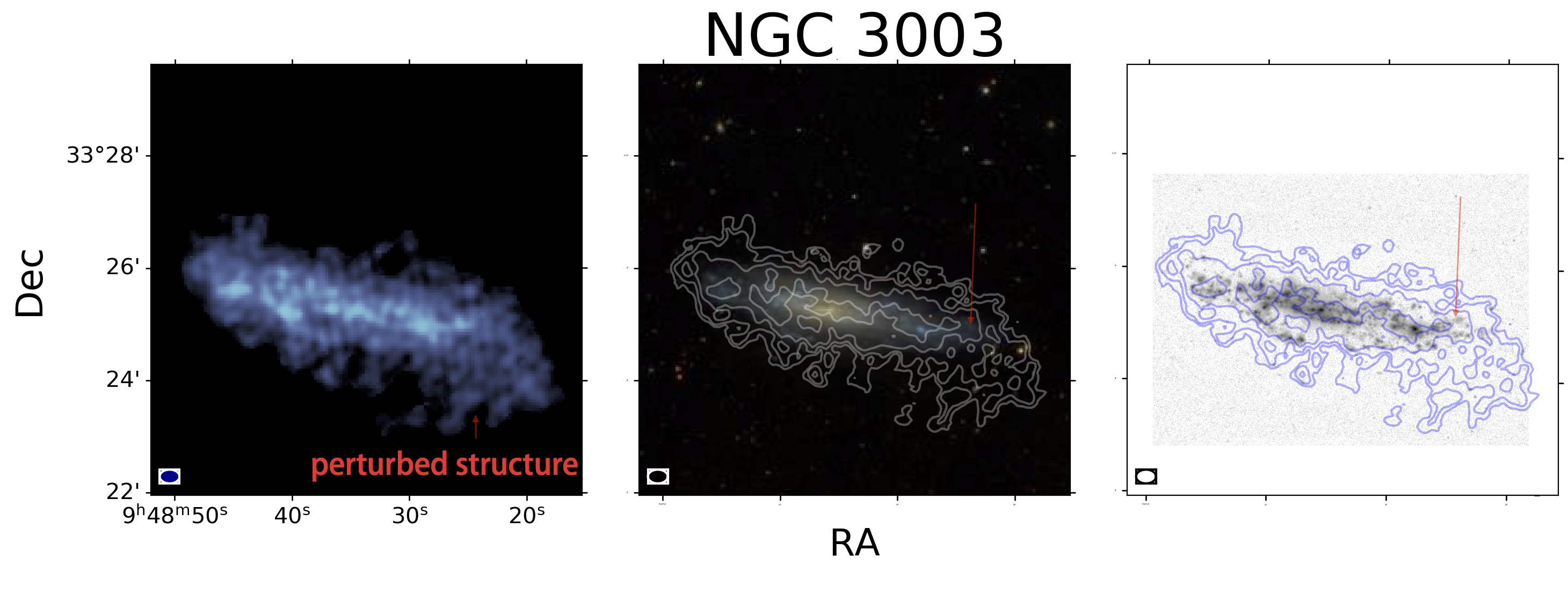}\\ \vspace{-4mm}
\caption{Atlas of our sample galaxies. The three images for each galaxy from left to right are the VLA $\hi$ intensity map, the SDSS false-color RGB ($gri$) images with $\hi$ contours, and the APO H$\alpha$ images with $\hi$ contours. The $\hi$ intensity image on the left is a composite of two displays of the moment-0 map. The log stretch, which highlights the fainter outer regions, is assigned a darker blue. The linear stretch, in lighter cyan blue highlights structure in the inner disk. In each panel, the white (or blue) contours correspond to the level of 1, 2.5, 5, 10, 20, 40 $\times$ 10$^{20}$ cm$^{-2}$. We point out salient features in red text. The two galaxies with new $\hi$ data have their names highlighted in red. The dashed red lines in NGC 3448 and NGC 5775 show how we divide the $\hi$ emission of NGC 3448 from UGC 6016 and that of NGC 5775 from NGC 5774.}
\label{atlas}
\end{figure*}

In Figure \ref{atlas}, we show $\hi$ intensity maps, SDSS false-color RGB ($gri$) images with $\hi$ contours, and H$\alpha$ images also with $\hi$ contours. 
 
Some galaxies display $\hi$ disks that are much more extended than the optical ones (like NGC 660, NGC 3003, and NGC 3448), while others appear to have truncated $\hi$ disks (like NGC 3877 and NGC 4302) due to the limit of our detection sensitivity. Extended $\hi$ disks are common in $\hi$-rich galaxies, as the $\hi$-to-optical size ratio was found to be well correlated with $\hi$ richness \citep{2013MNRAS.433..270W}.

Some galaxies manifest thin $\hi$ disks with regular morphologies (like NGC 4565 and UGC 10288), while some others show highly asymmetric outer disks (like NGC 3003, NGC 3079, NGC 3448, NGC 4631,NGC 4666, and NGC 5775).
Previous studies suggest that NGC 3079, NGC 3448, NGC 4302, NGC 4631, NGC 4565, NGC 4666 and NGC 5775 (obvious from Figure \ref{atlas}) experience tidal interactions \citep{1987ApJ...313L..91I, 1991ApJ...371..111I, 1979MNRAS.187..509R, 1984AJ.....89..350B, 1986AJ.....92.1048N, 1978A&A....63..363B, 2009AJ....138.1741C, 2015ApJ...799...61Z, 1993AJ....105.2098R, rand1994atomic, 1994ApJ...429..618I, 1998A&A...331..428D}.
As was shown more clearly in the data of \cite{1990MNRAS.246..324Z}, NGC 5084 presents a misalignment in position angle, along with a significant difference in rotation curves, between the $\hi$ and optical disks. This suggests that the $\hi$ may be recently accreted through merging a satellite galaxy and has yet to settle into the disk \citep{1997AJ....113.1585C}. NGC~5084 also harbours an active galactic nucleus (AGN) \citep{2019AJ....158...21I}.

Except for the central regions, H$\alpha$ and $\hi$ features correspond well to each other spatially, and both distributions trace spiral arms or rings evident in the optical broadband data. For example, in NGC 660, the $\hi$, optical, and H$\alpha$ all follow the outside polar ring; in NGC 4096, the emission follows the southern asymmetric spiral arm; and in NGC 5792, the inner ring and outside extended spiral arm are traced. The spatial correlation between the H$\alpha$ and the $\hi$ supports the role of $\hi$ as a reservoir of material fueling star formation \citep{bigiel2008star, leroy2008star, 2008ApJ...689..865K, 2009ApJ...699..850K, 2012ApJ...745...69K, 2019A&A...622A..64B}.

\begin{figure*}
\centering
\includegraphics[height=0.27\textheight]{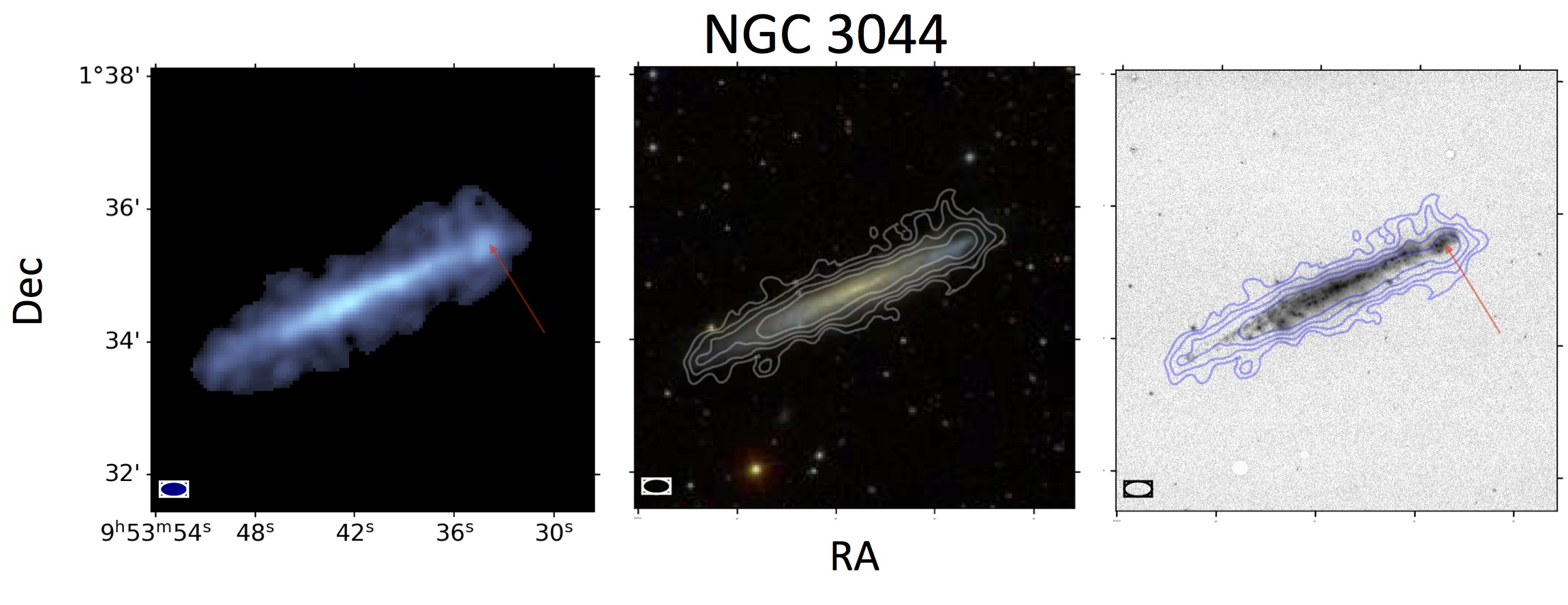}\\ \vspace{-4mm}
\includegraphics[height=0.27\textheight]{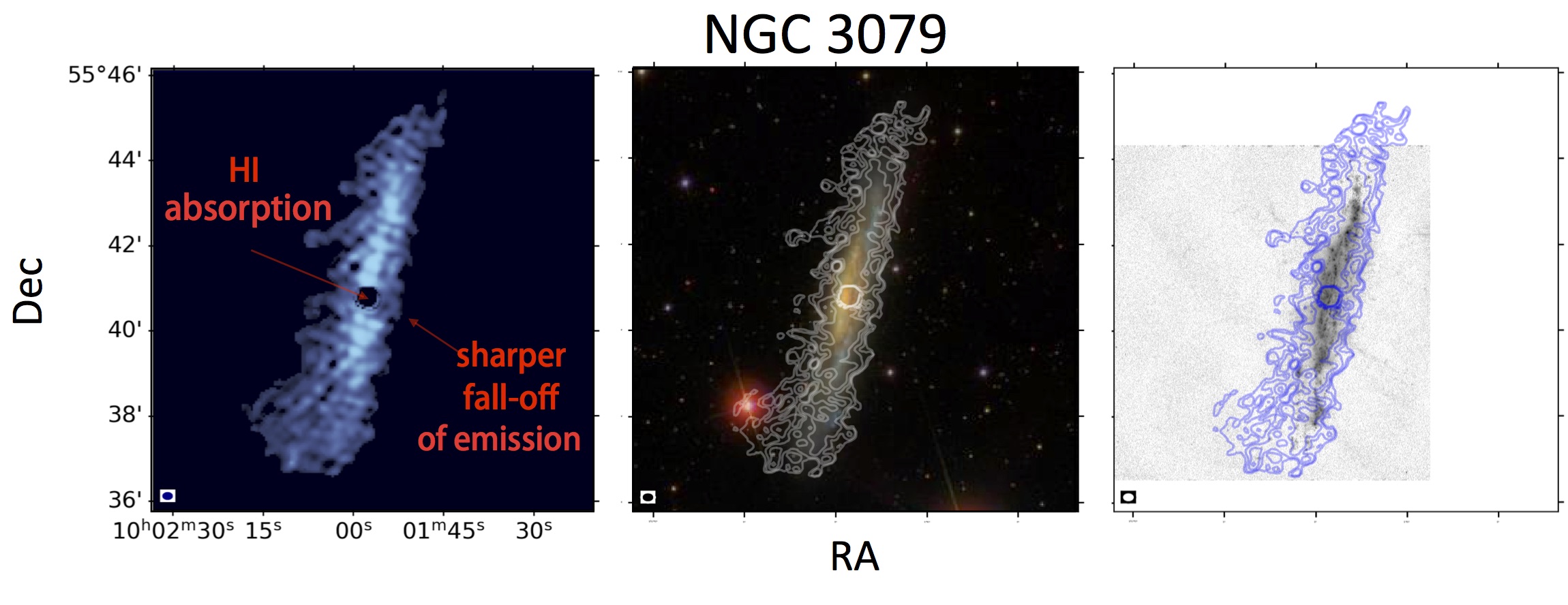}\\ \vspace{-4mm}
\includegraphics[height=0.27\textheight]{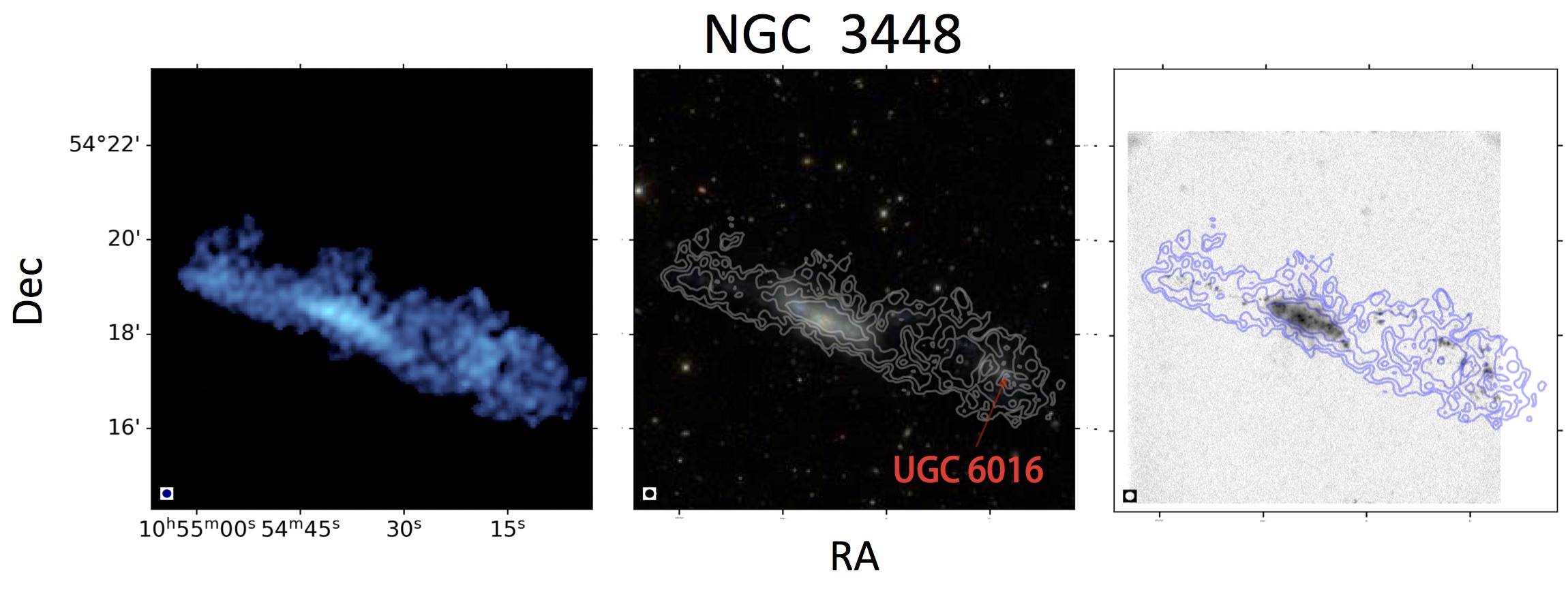}\\ \vspace{-4mm}
\includegraphics[height=0.27\textheight]{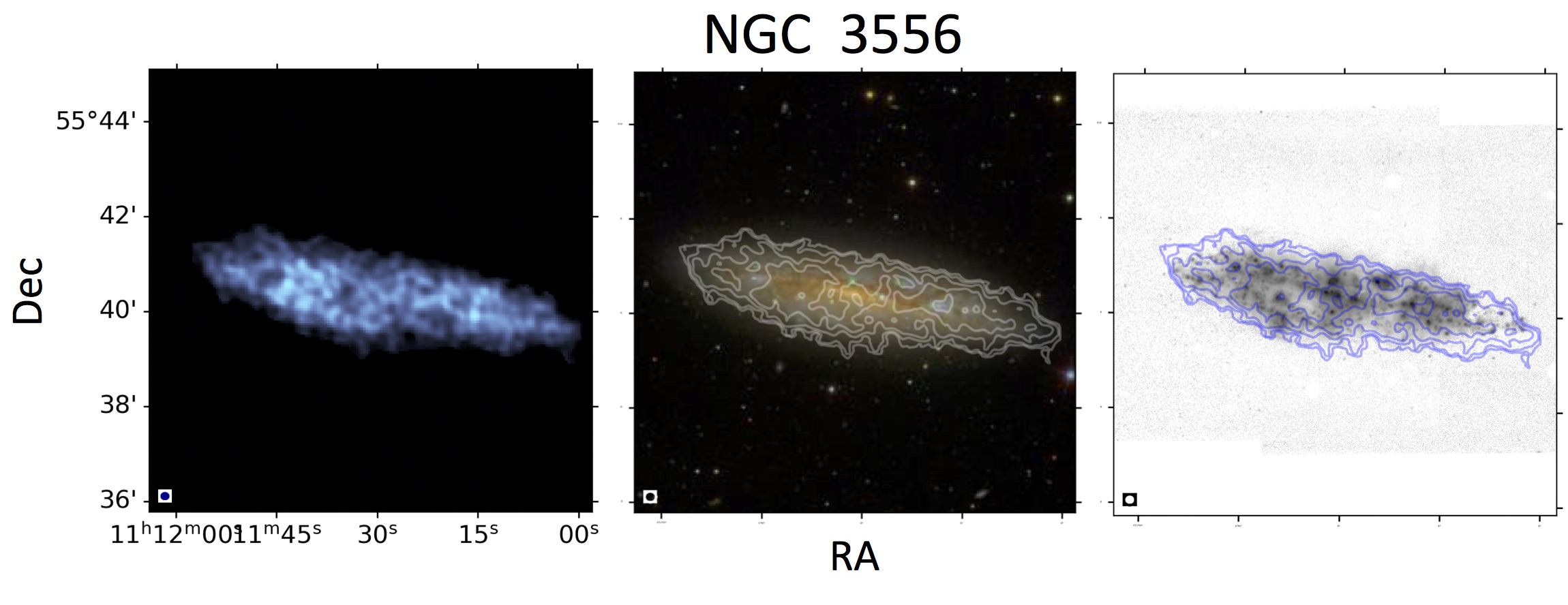}\\ \vspace{-4mm}
{\textbf{Figure 5 continued}}
\end{figure*}

\begin{figure*}
\centering
\includegraphics[height=0.27\textheight]{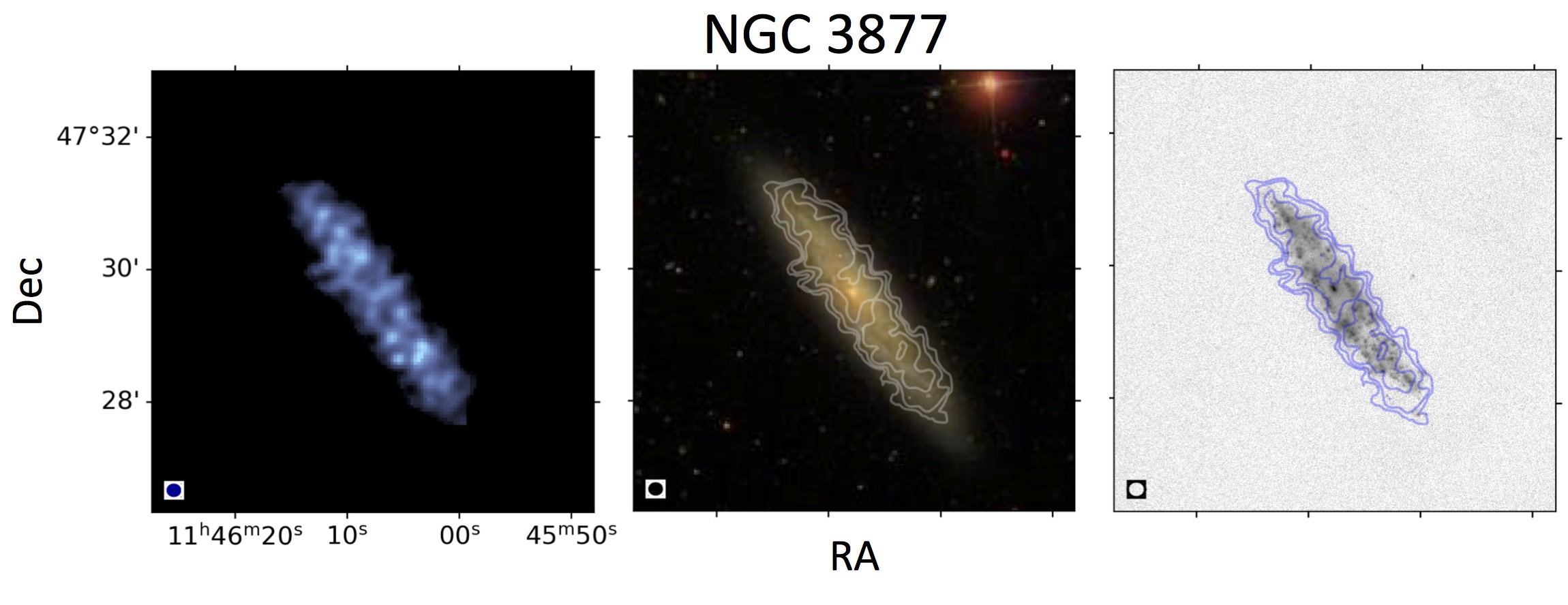}\\ \vspace{-4mm}
\includegraphics[height=0.27\textheight]{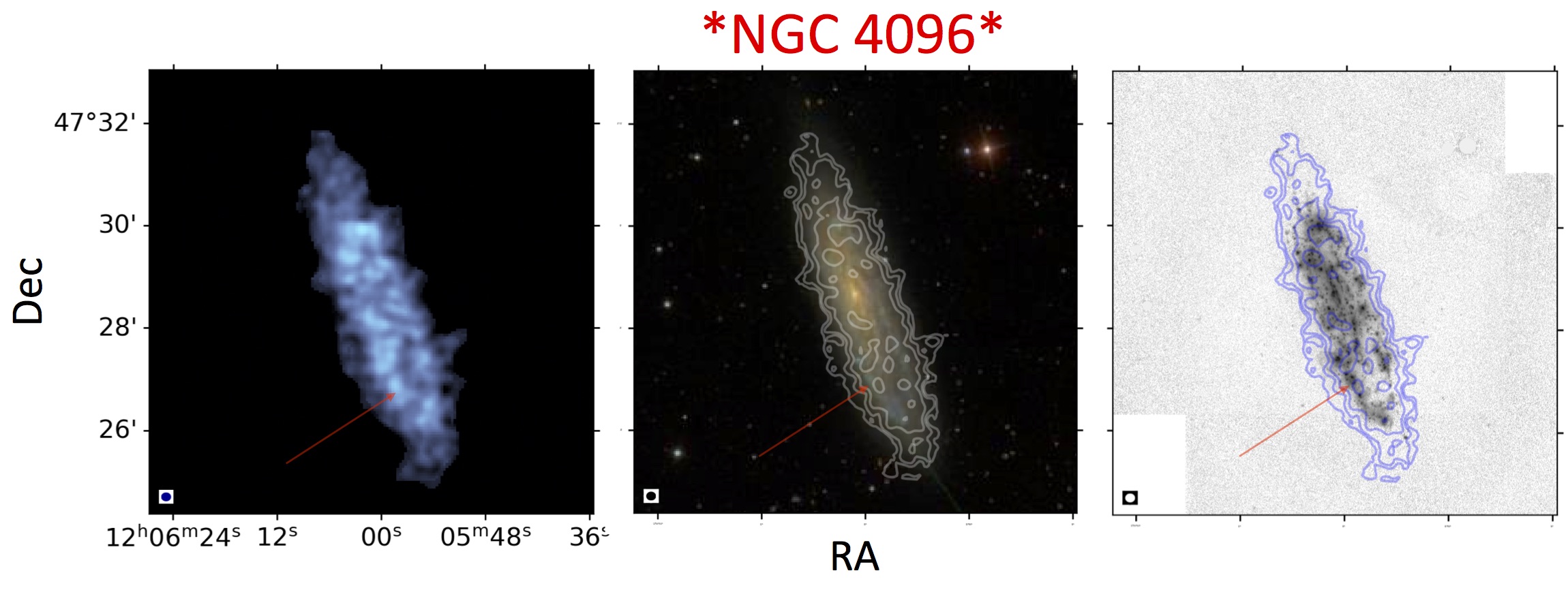}\\ \vspace{-4mm}
\includegraphics[height=0.27\textheight]{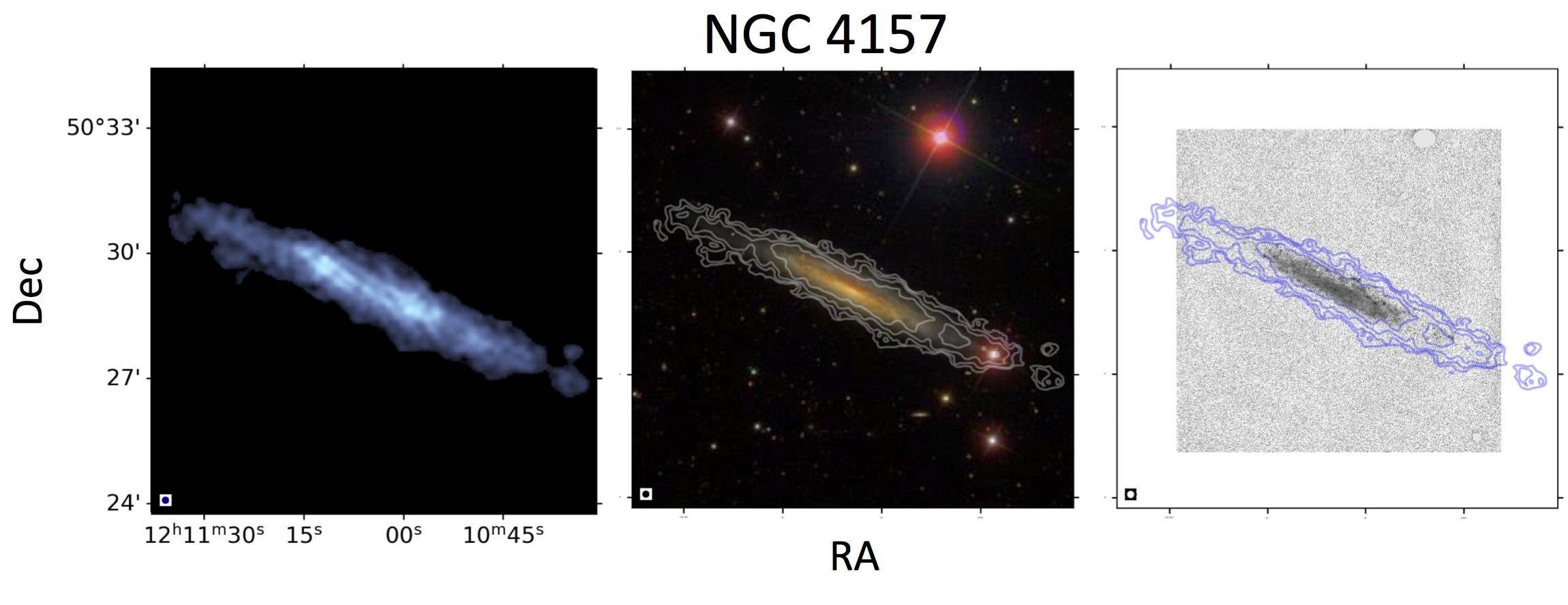}\\ \vspace{-4mm}
\includegraphics[height=0.27\textheight]{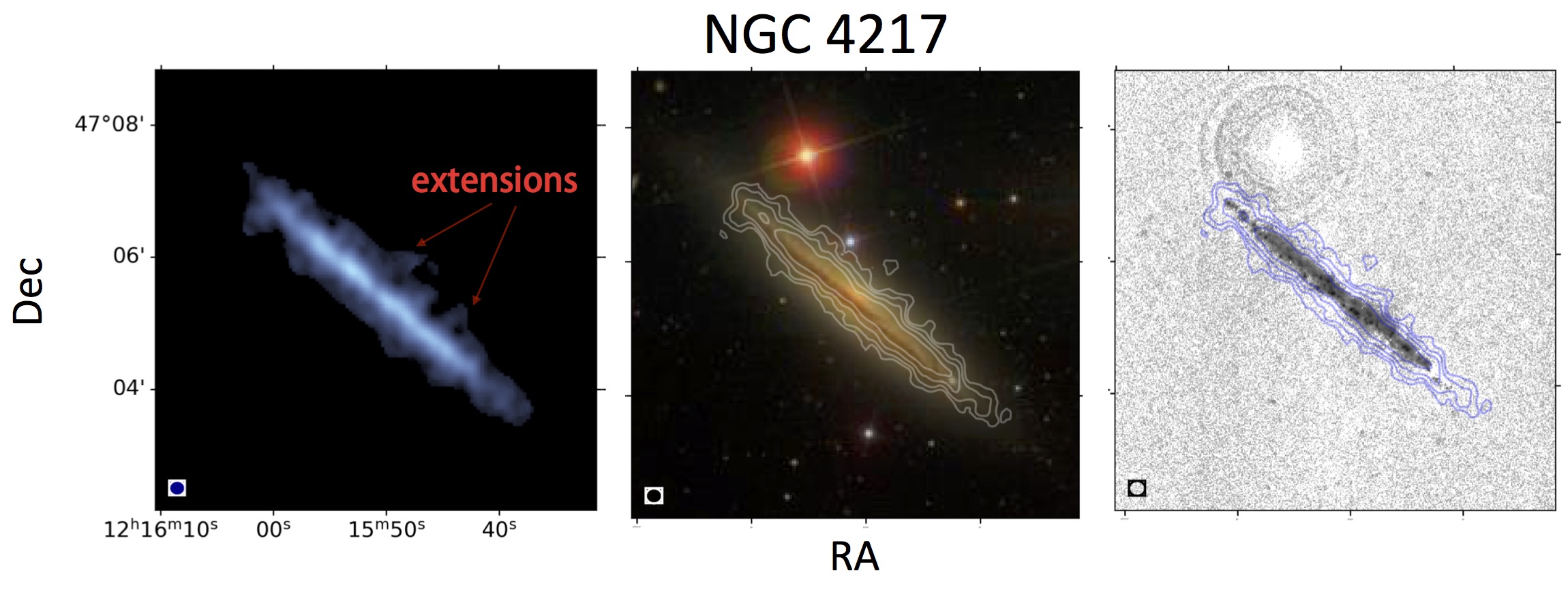}\\ \vspace{-4mm}
{\textbf{Figure 5 continued}}
\end{figure*}

\begin{figure*}
\centering
\includegraphics[height=0.27\textheight]{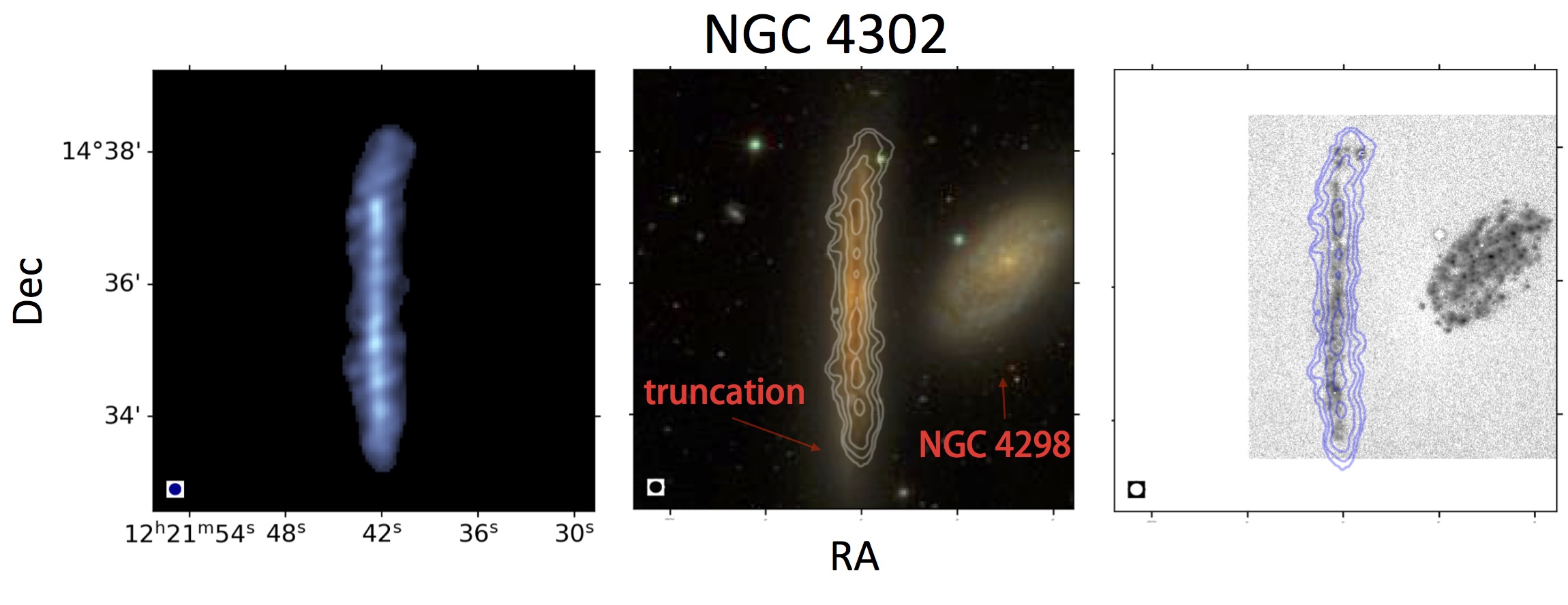}\\ \vspace{-4mm}
\includegraphics[height=0.27\textheight]{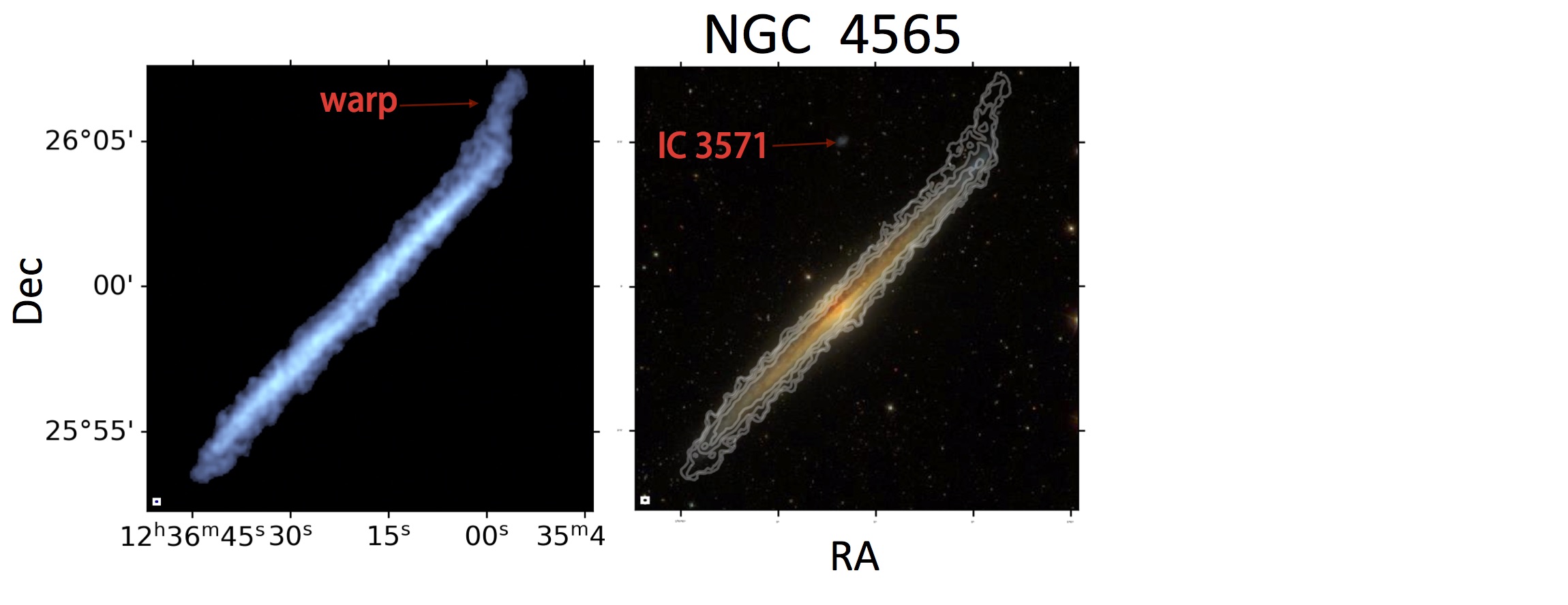}\\ \vspace{-4mm}
\includegraphics[height=0.27\textheight]{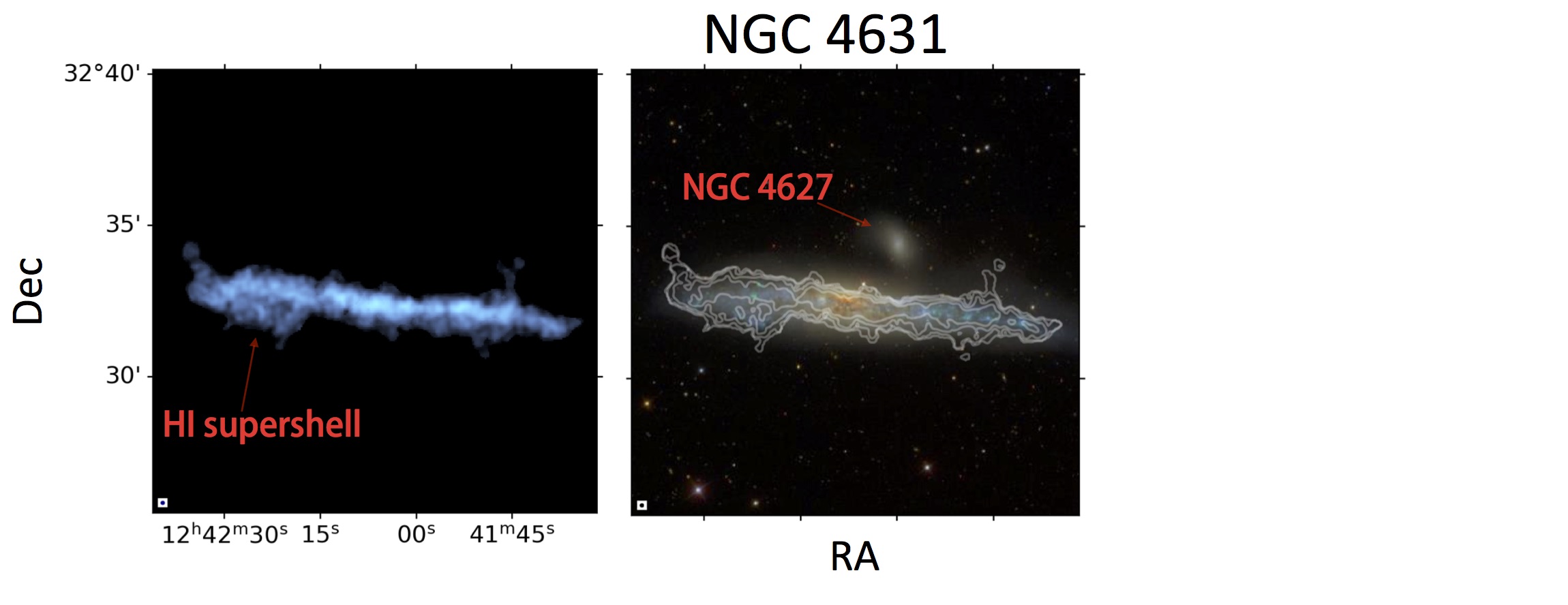}\\ \vspace{-4mm}
\includegraphics[height=0.27\textheight]{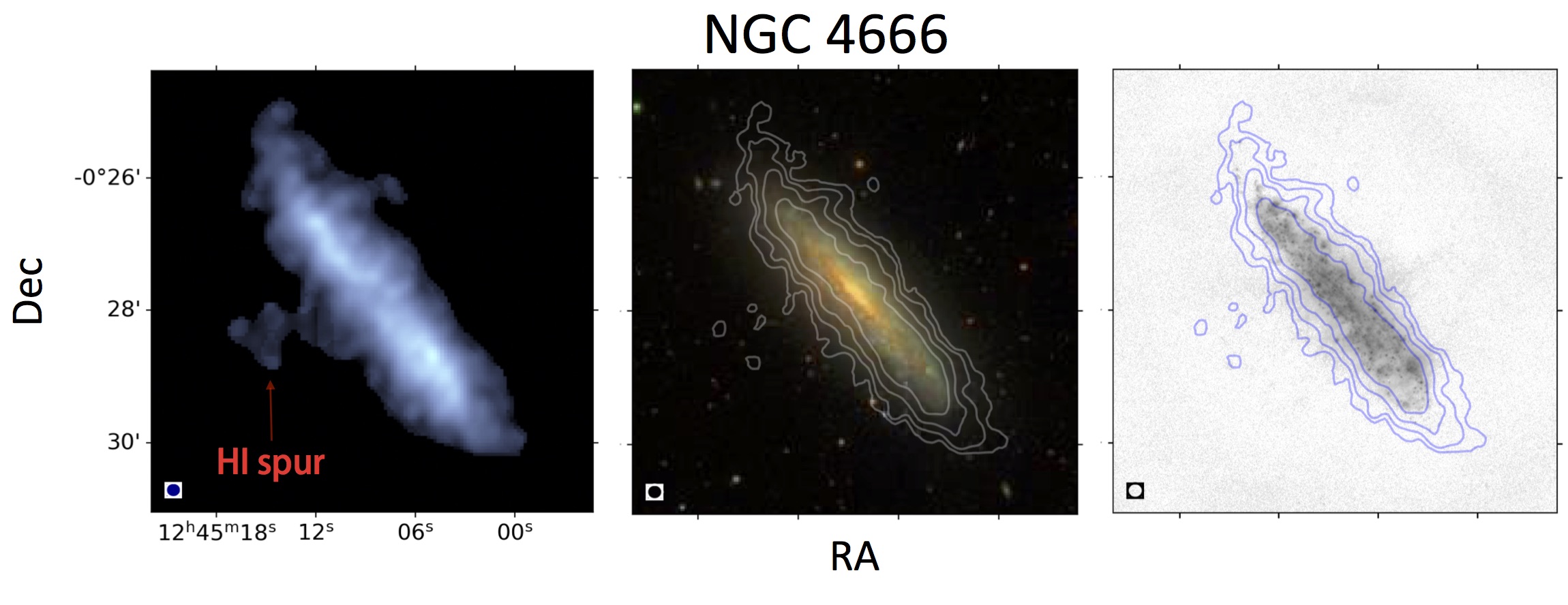}\\ \vspace{-4mm}
{\textbf{Figure 5 continued}}
\end{figure*}

\begin{figure*}
\centering
\includegraphics[height=0.27\textheight]{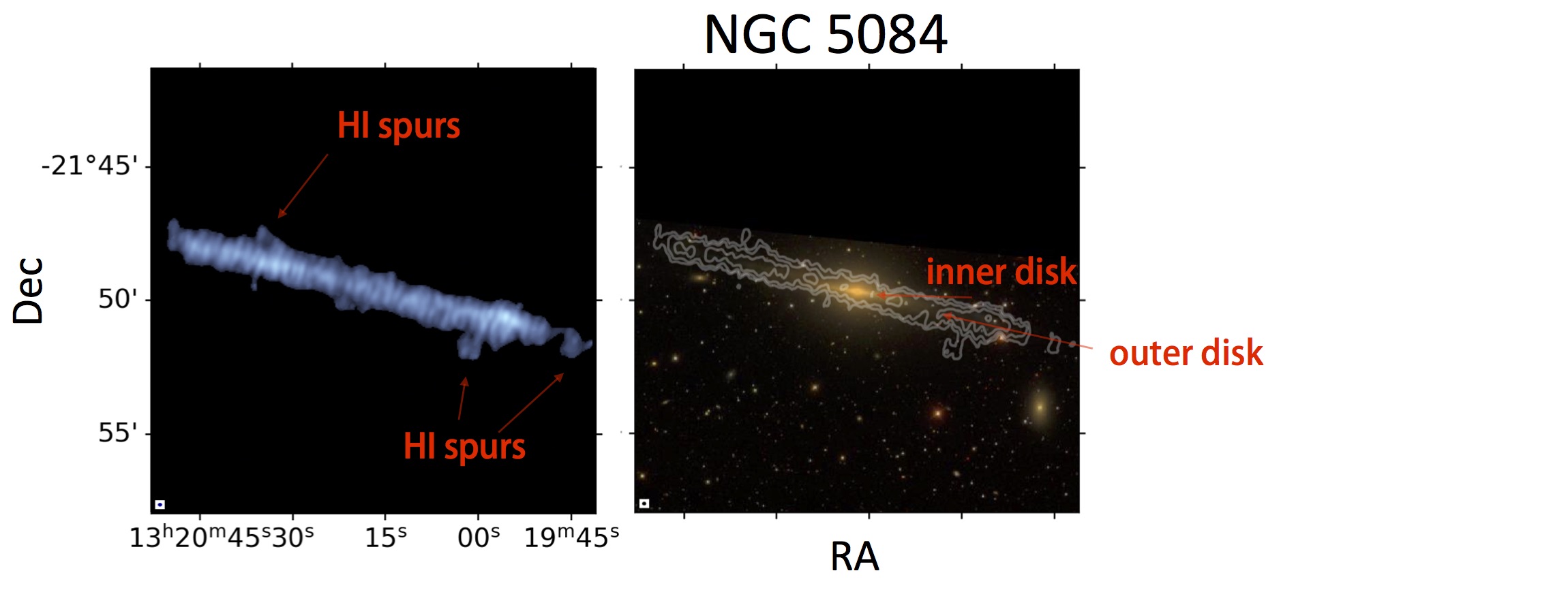}\\ \vspace{-4mm}
\includegraphics[height=0.27\textheight]{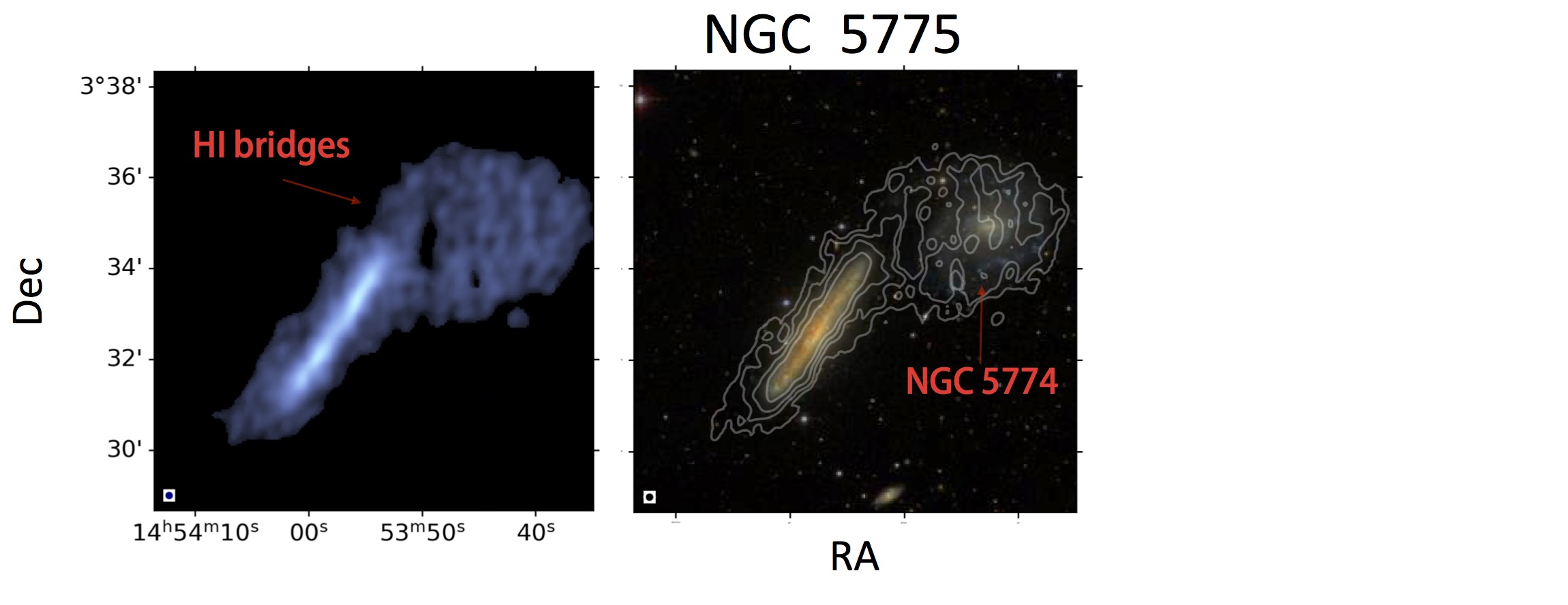}\\ \vspace{-4mm}
\includegraphics[height=0.27\textheight]{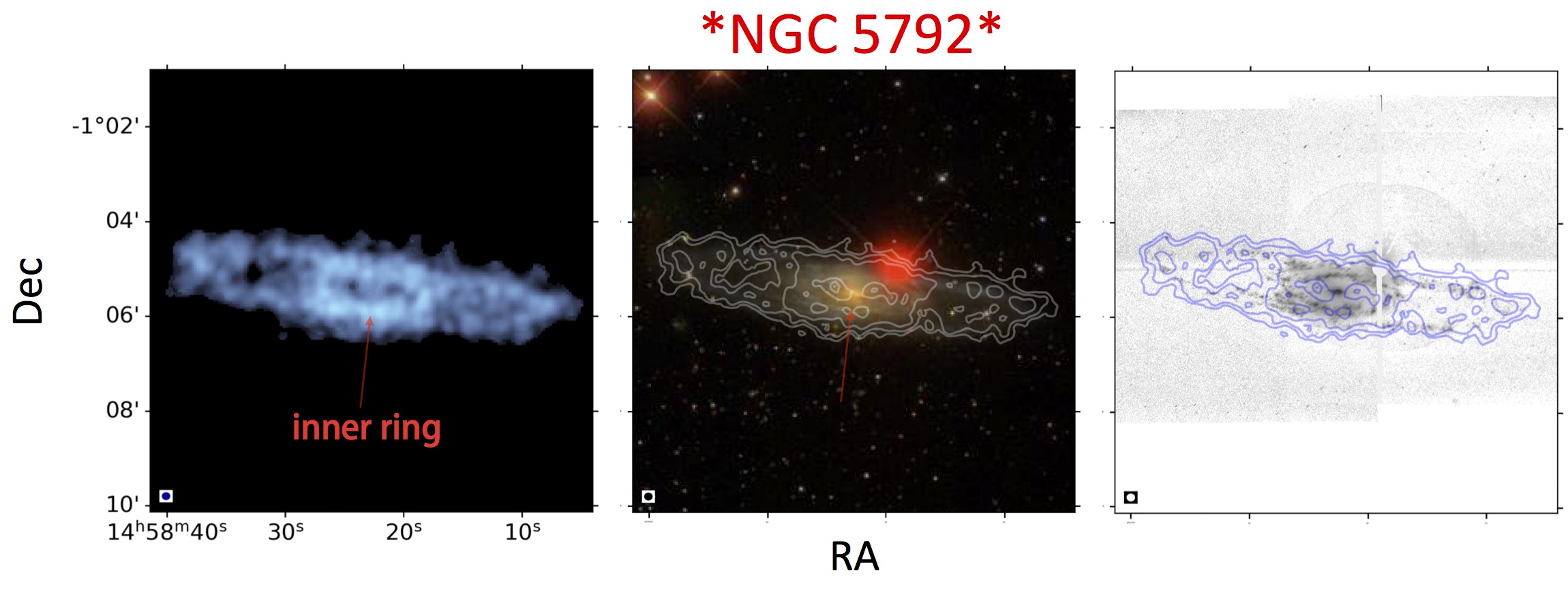}\\ \vspace{-4mm}
\includegraphics[height=0.27\textheight]{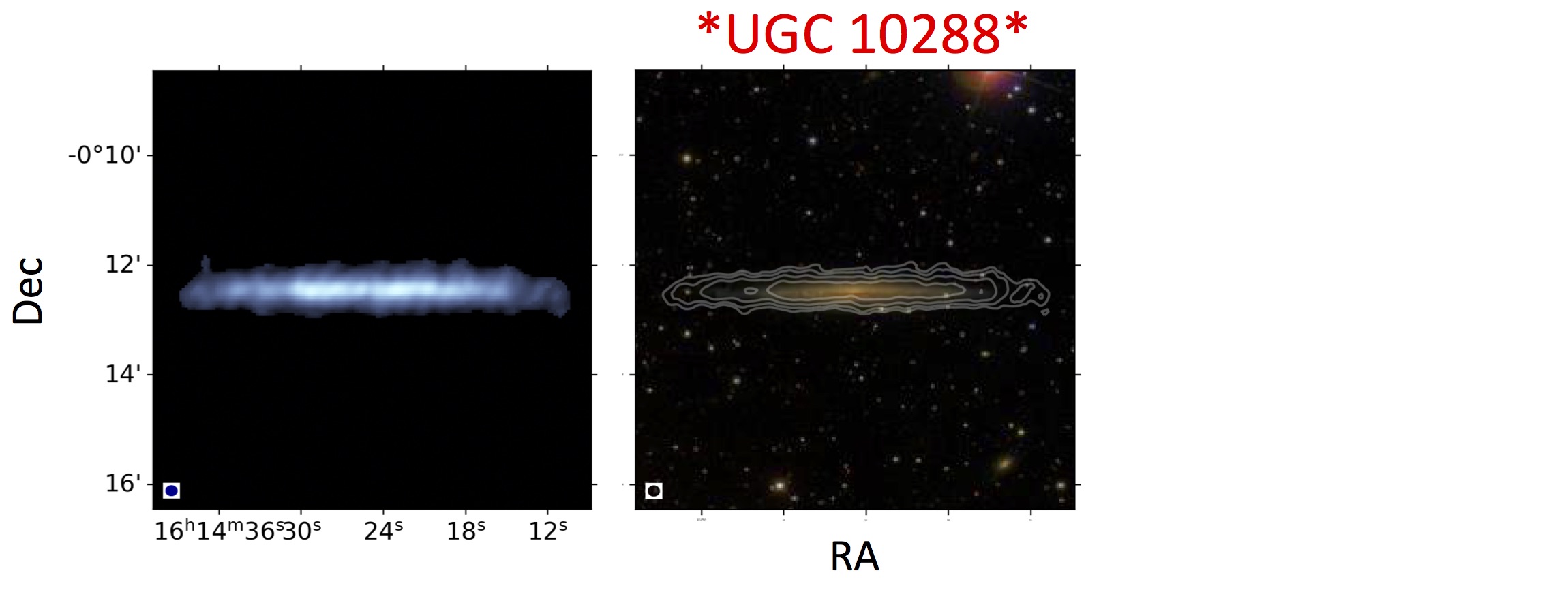}\\ \vspace{-4mm}
{\textbf{Figure 5 continued}}
\end{figure*}

\subsection{HI scale heights}
We study how $\hi$ scale heights are correlated to the scale heights in the radio continuum. We also study how $\hi$ scale heights are correlated to other galactic properties, in a close comparison to the behavior in the radio continuum studied and presented in \cite{2018AA...611A..72K}. The Pearson correlation coefficient $R$ is a measure of linear correlation with a value ranging from -1 to 1. While a value of -1 implies a perfectly linear anti-correlation between the two data sets, a value of 0 represents no correlation, and a value of 1 implies a perfect linear correlation. We use an absolute Pearson correlation coefficient of 0.4 (corresponding to a $p$ value of 0.09) as the threshold for significant correlations.

\subsubsection{HI scale heights compared to radio scale heights}
\begin{figure}
\includegraphics[width=0.45\textwidth]{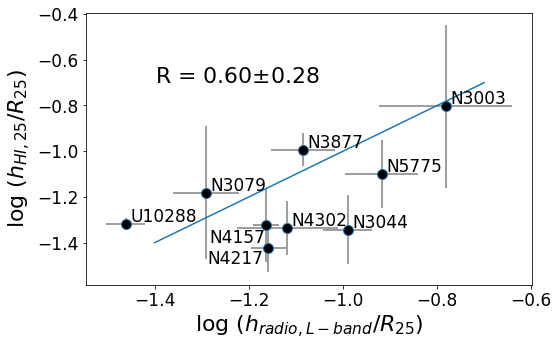}
{\textbf{(a)}}
\includegraphics[width=0.45\textwidth]{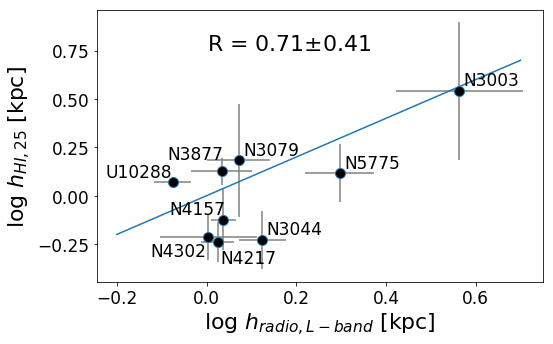}
{\textbf{(b)}}
\caption{The relation between the (a) average $\hi$ exponential scale height over optical radius ($\bar{h}_{\rm HI,25}/R_{25}$) and L-band radio exponential scale height over optical radius ($h_{\rm radio,L-band}/R_{25}$), and (b) average $\hi$ exponential scale height ($\bar{h}_{\rm HI,25}$) and L-band radio exponential scale height ($h_{\rm radio,L-band}$).
The Pearson correlation coefficient, R, of each relation is listed in the corresponding panel. The blue line show the 1:1 line.}
\label{hiradio}
\end{figure}

We present how the $\hi$ exponential scale heights $\bar{h}_{\rm HI,25}$ correlate with the radio exponential scale heights $h_{\rm radio,L-band}$ \citep{2018AA...611A..72K} in Figure \ref{hiradio}. In order to remove the effect of both types of scale heights being correlated with the optical disk sizes, we divide them by $R_{25}$, shown in Figure \ref{hiradio}(a). The Pearson correlation coefficient ($R$) is labeled in the corner of each panel.

In Figure \ref{hiradio}, there is a considerable correlation between $\hi$ scale heights and L-band radio scale heights. 
The $\hi$ scale and the L-band scale heights are related around the one to one line (i.e. close in value), supporting our speculation earlier that the $\hi$ scale heights indicate the thickness of the general $\hi$ (i.e. including the extra-planar component) rather than just the (thin) disk. This result suggests that the vertical structure of $\hi$ and L-band radio continuum might be shaped by similar physical processes.

The correlation between un-normalized scale heights (see Figure \ref{hiradio}(b)) is better than between the ones normalized by $R_{25}$. This is unexpected because in the un-normalized correlation the scatter in scale heights should be increased by the scatter in radius, which should lead to a lower R-value, not a higher one. However, as the number of points is small, the difference in R-values is probably not significant. Indeed, the trend seems to be strongly driven by the outlier NGC 3003, and the R-value has a large uncertainty. The normalization procedure does not improve the correlation between scale heights in $\hi$ and radio, so that there must other factors causing most of the scatter in both correlations (normalized and un-normalized), e.g., mass density and/or star formation density, as investigated later in Figure \ref{hiopp}.

\subsubsection{HI scale heights related to HI disk sizes}
\begin{figure}
\includegraphics[width=0.45\textwidth]{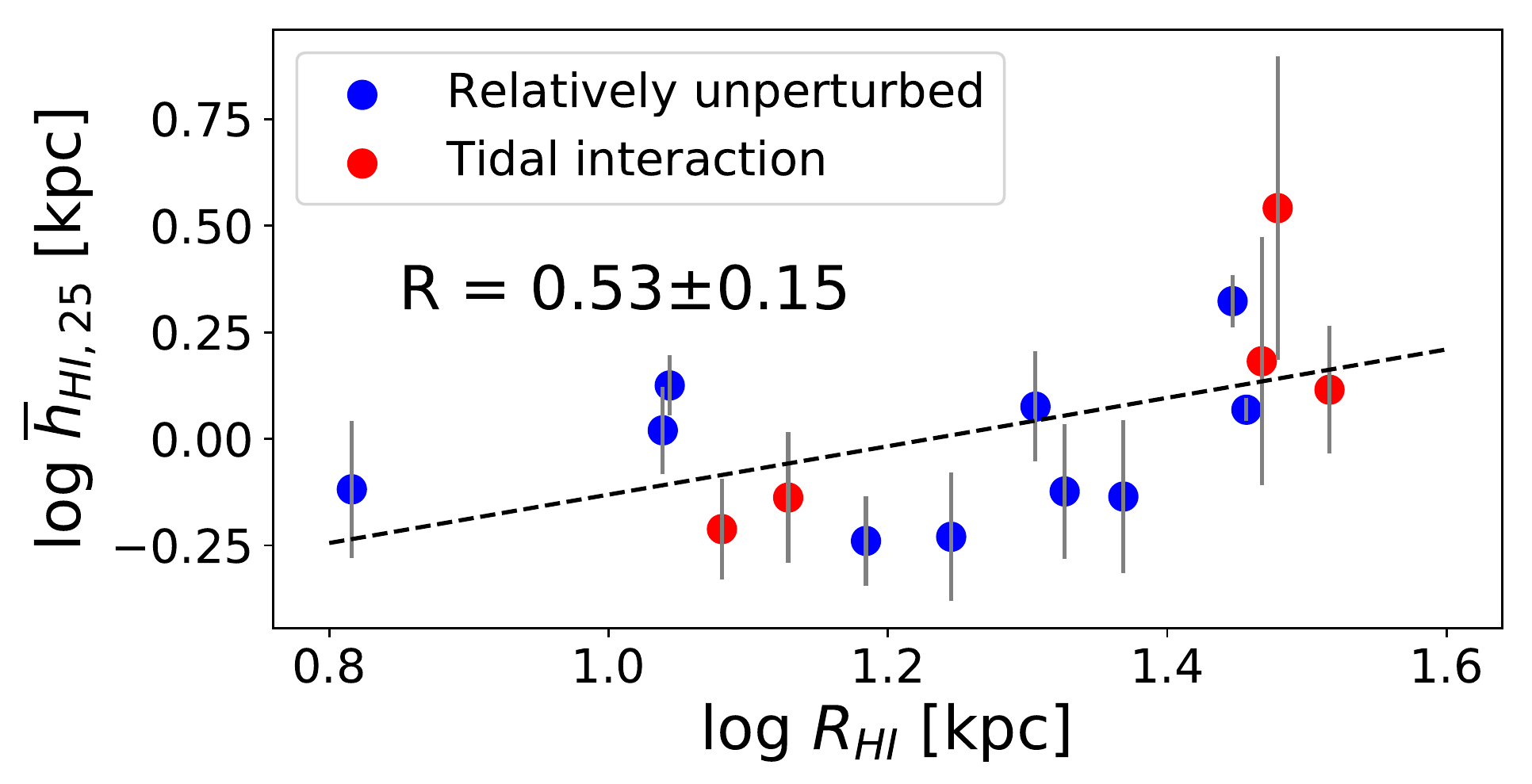}
\caption{The relation between the average $\hi$ exponential scale height ($\bar{h}_{\rm HI,25}$) and $\hi$ radius ($R_{\rm HI}$). The blue points represent the relatively unperturbed galaxies and the red points are obvious tidally interacting galaxies. The Pearson correlation coefficient, R, of each relation is listed in the corresponding panel. The black dash line presents the best-fit linear relation of ${\rm log}~ \bar{h}_{\rm HI,25} = 0.57 \times {\rm log}~ R_{\rm HI} - 0.70$.}
\label{hhiradius}
\end{figure}

We present how $\hi$ scale height $\bar{h}_{\rm HI,25}$ correlates with $\hi$ disk sizes $R_{\rm HI}$ in Figure \ref{hhiradius}, with the Pearson correlation coefficients labeled in figures. We use different colors to separate the tidally interacting from the relatively unperturbed galaxies, though these two types do not reveal much difference in any of the relations.

Figure \ref{hhiradius} shows a correlation between $\bar{h}_{\rm HI,25}$ and $R_{\rm HI}$. The black dashed line shows the best-fit linear relation ${\rm log}~ \bar{h}_{\rm HI,25} = 0.57 \times {\rm log}~ R_{\rm HI} - 0.70$. We consider using {$R_{\rm HI}$} as a proxy for $\hi$ scale length since the radial profiles of the $\hi$ surface density of galaxies show a homogeneous shape in the outer region when they are normalized radially by $R_{\rm HI}$, which can be characterized by an exponential scale length of about 0.2 $R_{\rm HI}$ \citep{2014MNRAS.441.2159W, 2016MNRAS.460.2143W}. Then the correlation in Figure \ref{hhiradius} is not totally unexpected because correlations between scale lengths and scale heights have been found for other galactic components, including the optical light \citep{1997A&A...320L..21D, 2002AstL...28..527Z}, and the radio continuum \citep{2018AA...611A..72K}. 

To remove this strong dependence on radial extension and to be consistent with the way of analysis in \cite{2018AA...611A..72K}, we define and focus on the normalized $\hi$ scale height $\bar{h}_{\rm HI,25}/R_{\rm HI}$, when explore the relation of $\hi$ thickness with surface density properties in the following. But we also show the relevant trends of the un-normalized $\bar{h}_{\rm HI,25}$ for reference.

\subsubsection{Normalized HI scale heights related to other galactic properties}

\begin{figure*}
\includegraphics[width=0.45\textwidth]{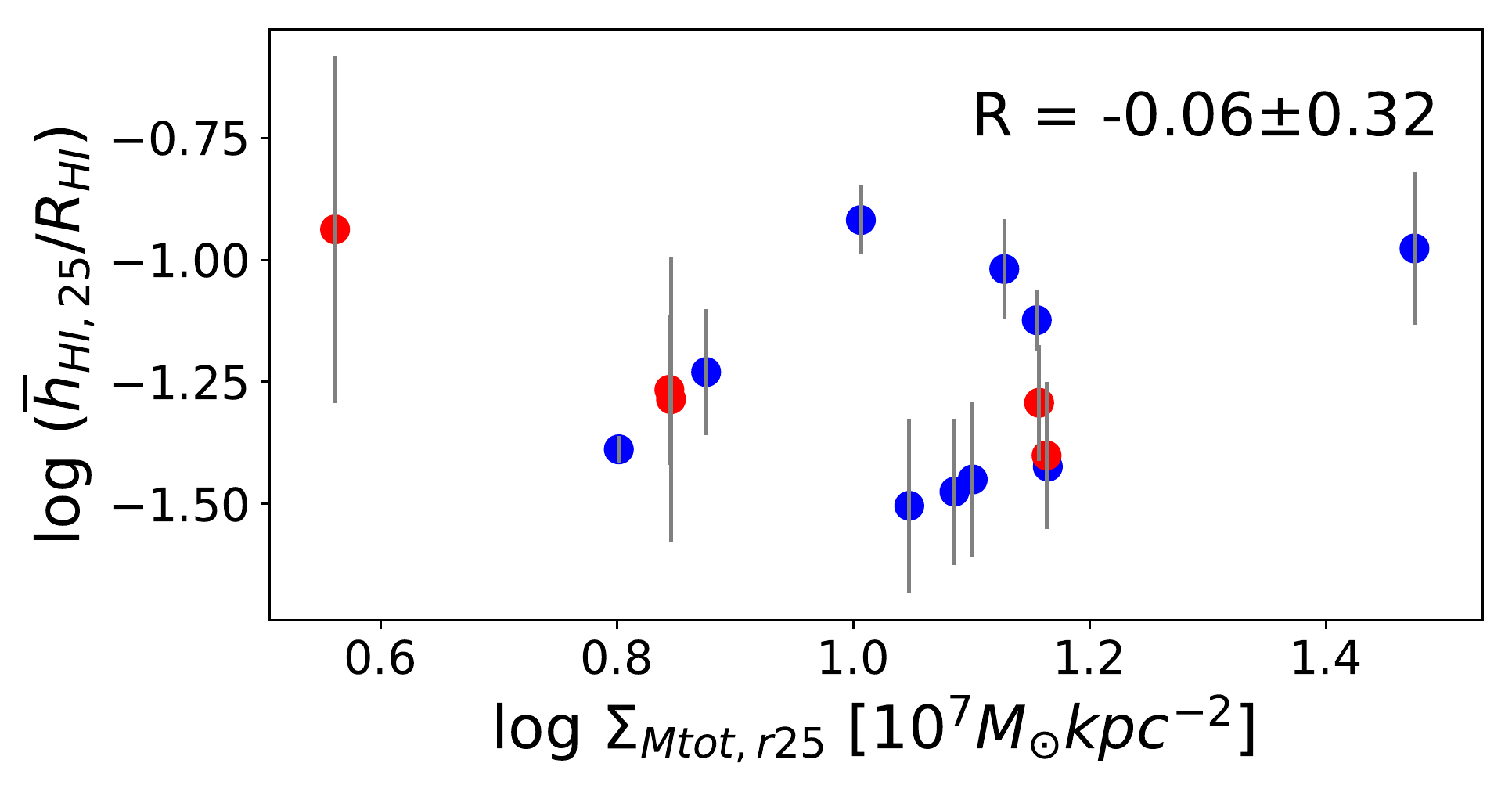}
{\textbf{(a.1)}}
\includegraphics[width=0.45\textwidth]{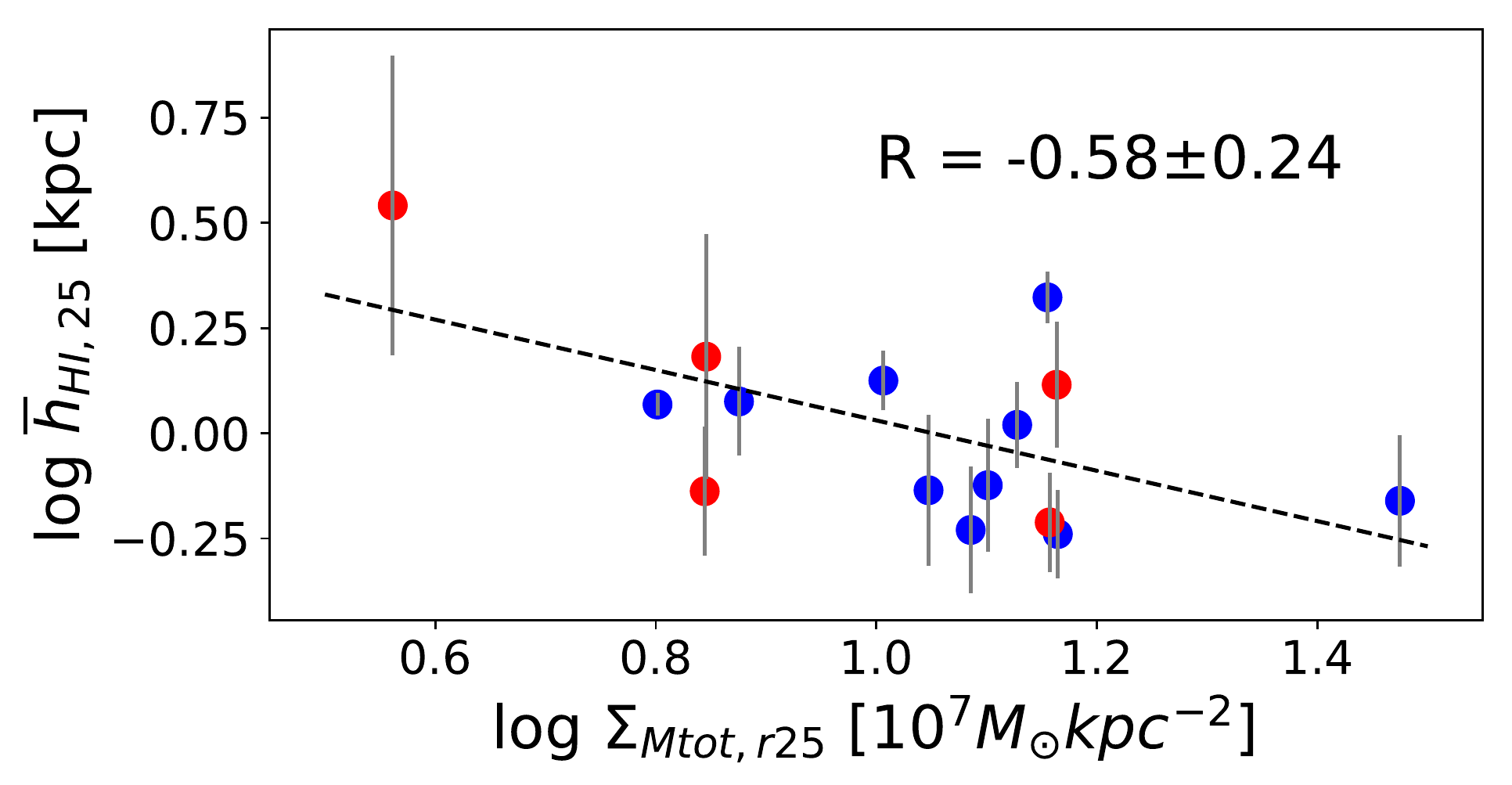}
{\textbf{(a.2)}}
\includegraphics[width=0.45\textwidth]{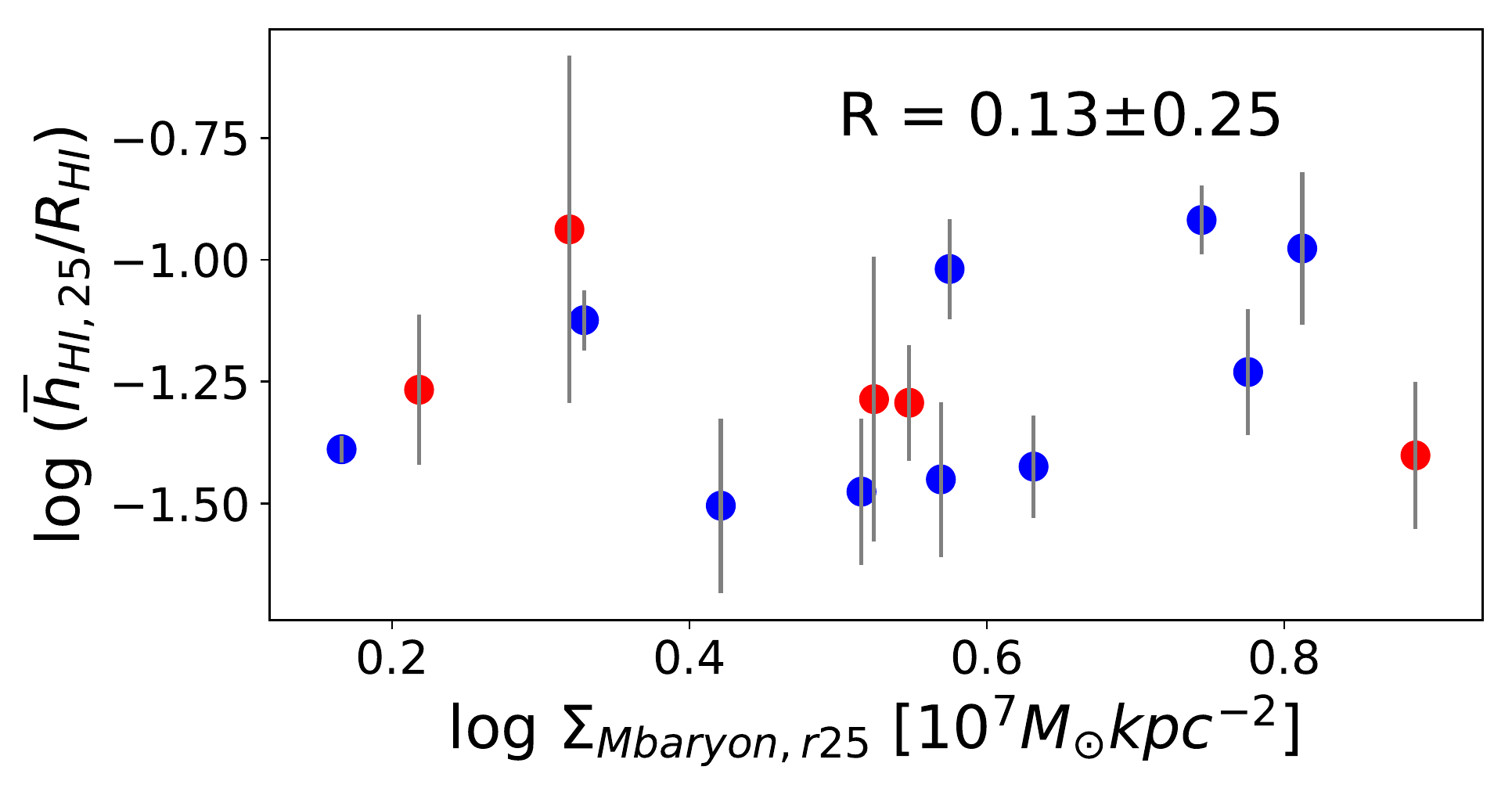}
{\textbf{(b.1)}}
\includegraphics[width=0.45\textwidth]{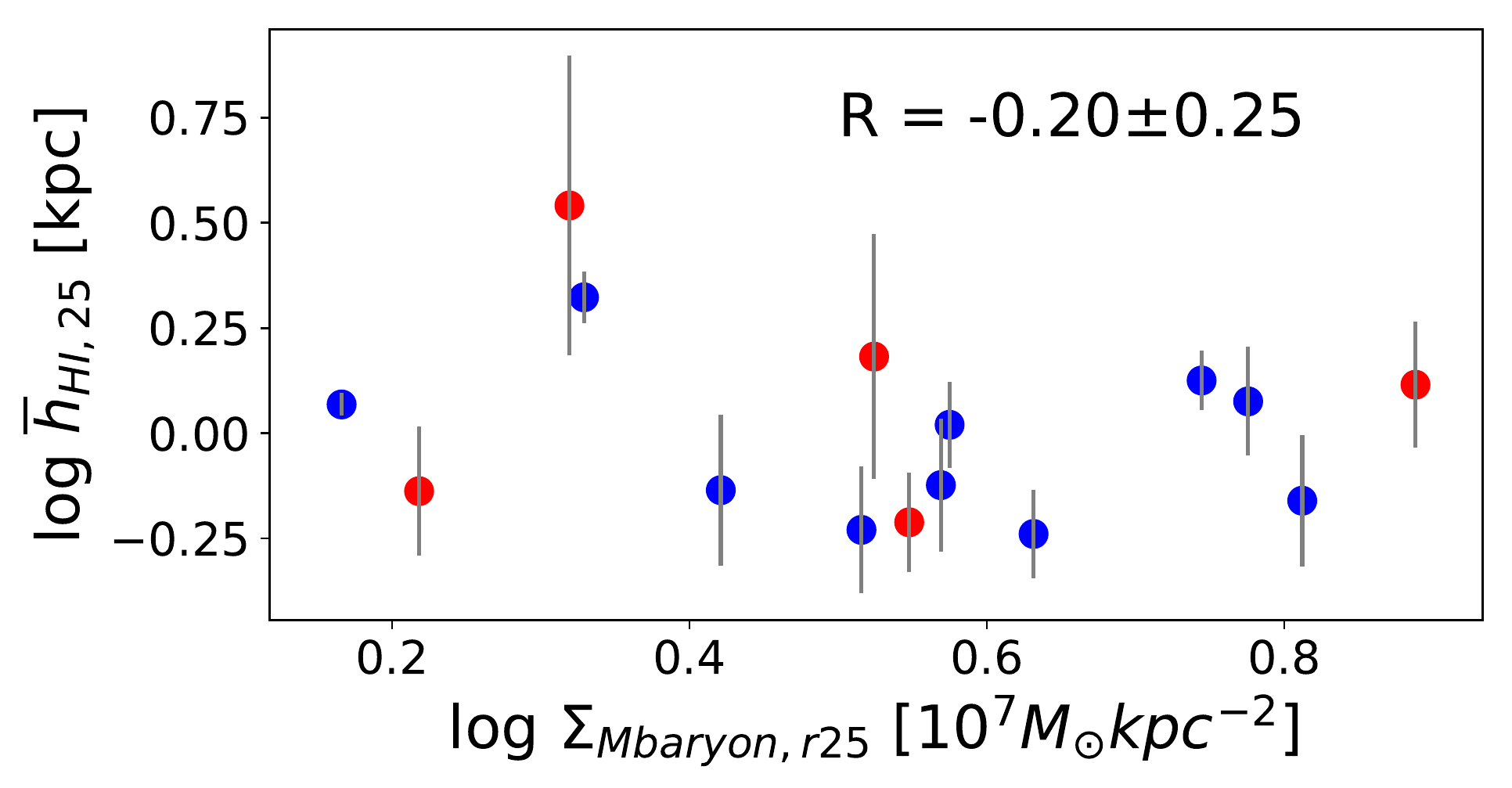}
{\textbf{(b.2)}}
\includegraphics[width=0.45\textwidth]{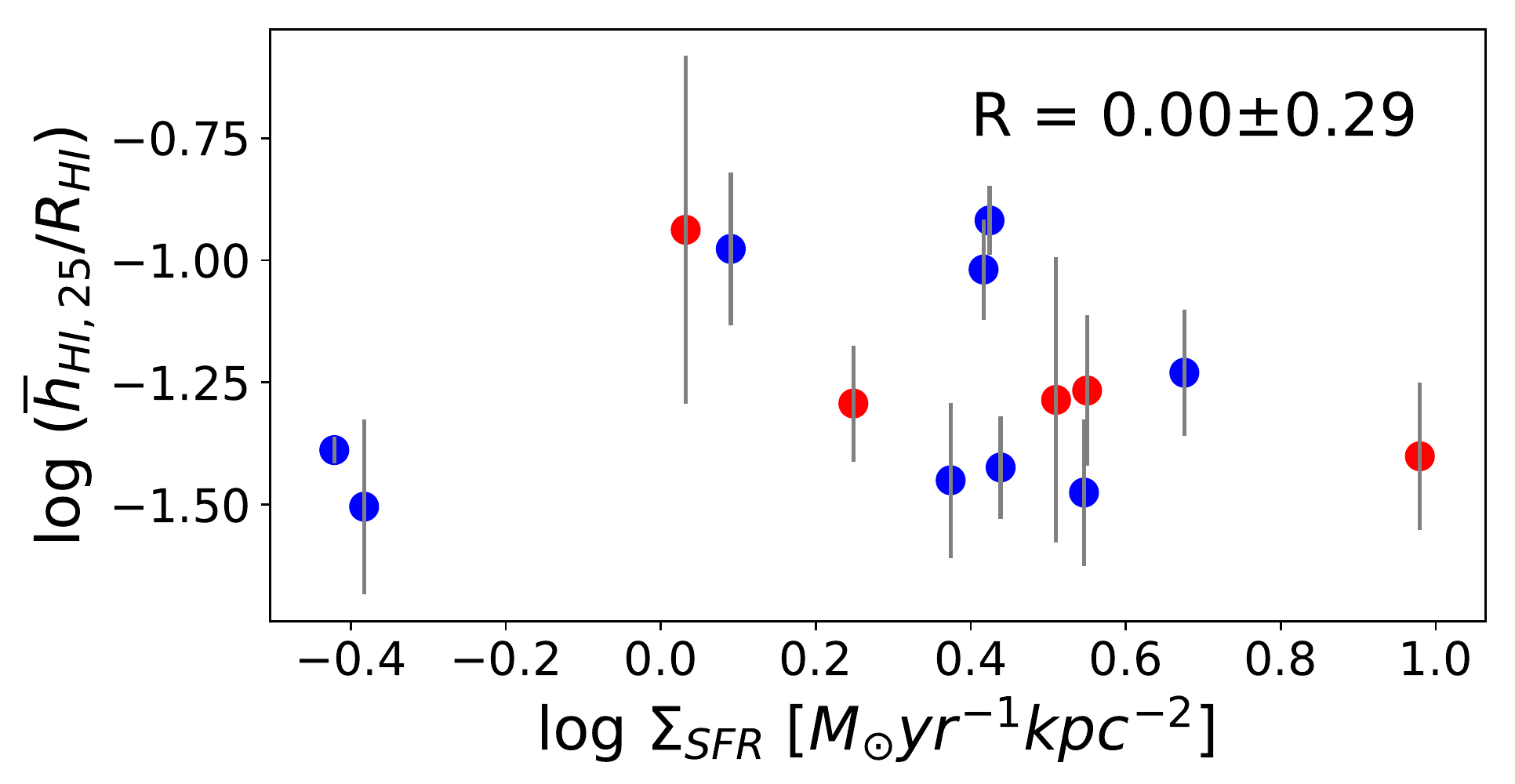}
{\textbf{(c.1)}}
\includegraphics[width=0.45\textwidth]{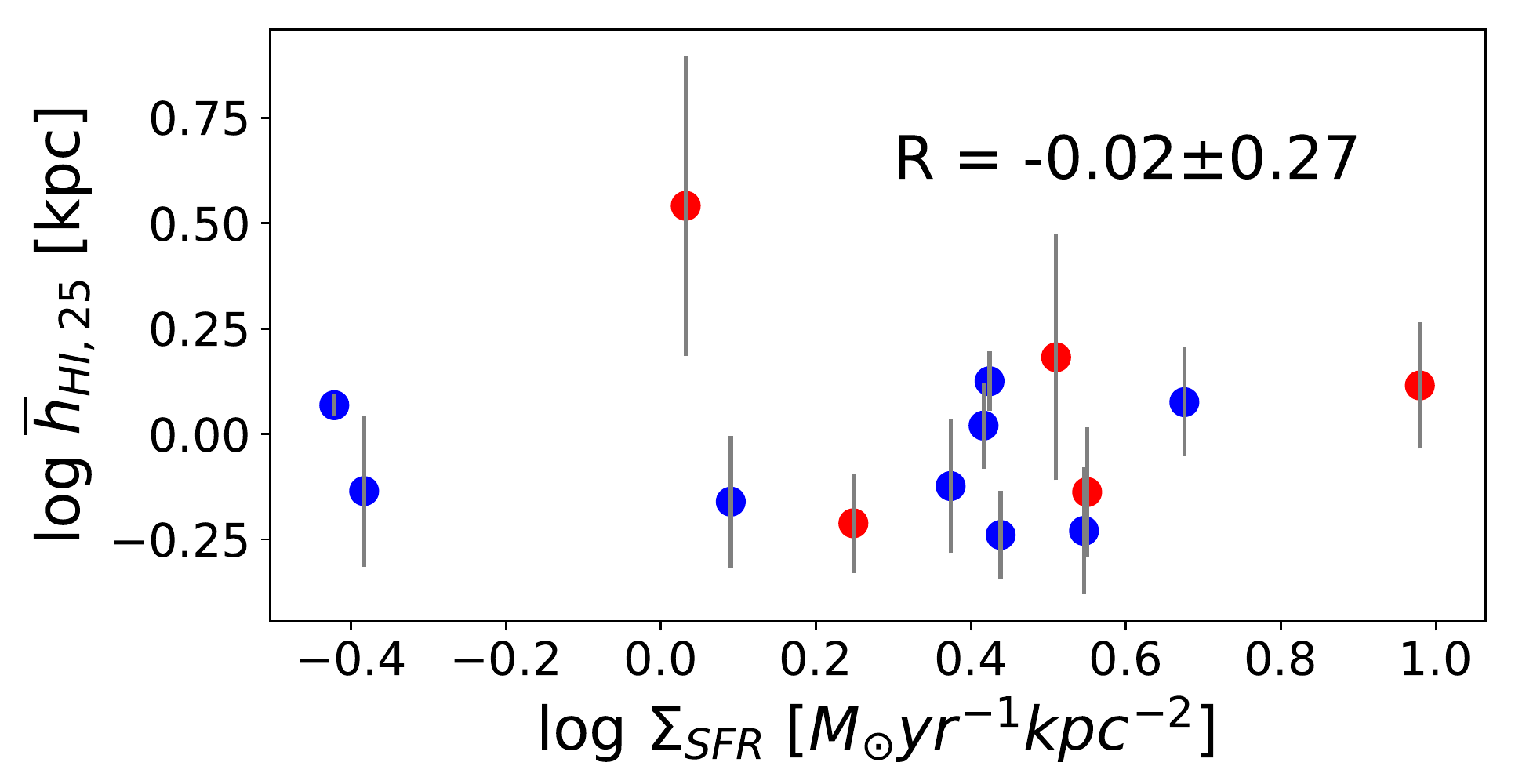}
{\textbf{(c.2)}}
\caption{The left panel shows the relation between the normalized average $\hi$ exponential scale height ($\bar{h}_{\rm HI,25}/R_{\rm HI}$) and (a.1) the total mass surface density ($\Sigma_{\rm Mtot,25}$), (b.1) the baryonic mass surface density ($\Sigma_{\rm Mbaryon,25}$), and (c.1) the star formation rate surface density ($\Sigma_{\rm SFR}$). The right panel shows how the un-normalized $\hi$ scale height ($\bar{h}_{\rm HI,25}$) correlated with those three galaxies properties in (a.2), (b.2), and (c.2).
The blue points represent the relatively unperturbed galaxies and the red points are obvious tidally interacting galaxies. The Pearson correlation coefficient, R, of each relation is listed in the corresponding panel. The black dash line in (a.2) presents the best-fit linear relation of ${\rm log }\bar{h}_{\rm HI,25} = -0.59 \times {\rm log} \Sigma_{\rm Mtot,25} + 0.63$.}
\label{hiopp}
\end{figure*}

We present how normalized $\hi$ scale height $\bar{h}_{\rm HI,25}/R_{\rm HI}$ correlates with surface density properties in Figure \ref{hiopp}. We focus on the dependence on surface densities of mass and SFR, motivated by the quasi-static model of the vertical distribution of ISM \citep{krumholz2018unified}. In the context of the model, we expect the scale heights to be anti-correlated with the mass surface densities, and correlated with the SFR surface densities. 

In Figure \ref{hiopp}(a.1), the normalized $\hi$ scale height shows no correlation with total mass surface density $\Sigma_{\rm Mtot,25}$. The total mass surface density, also called dynamical mass surface density, is taken from Paper IX \citep{2012AJ....144...43I}.
This lack of relation between $\bar{h}_{\rm HI,25}/R_{\rm HI}$ and $\Sigma_{\rm Mtot,25}$ is in contrast with the strong anti-correlation found in radio continuum \cite{2018AA...611A..72K}, suggesting different behavior of the two ISM components. We notice that when replacing $\bar{h}_{\rm HI,25}/R_{\rm HI}$ with $\bar{h}_{\rm HI,25}$, there is indeed a considerable anti-correlation with $\Sigma_{\rm Mtot,25}$ (with Pearson correlation coefficient of $-0.58\pm0.24$, see Figure \ref{hiopp}(a.2)). We also point out that, we test and do find an anti-correlation between $\bar{h}_{\rm HI,25}/R_{\rm HI}$ and $\Sigma_{\rm Mtot,25}$ when using the scale heights derived in \cite{2019A&A...622A..64B} for 12 galaxies from the $\hi$ Nearby Galaxy Survey \citep[THINGS,][]{2008AJ....136.2563W}. However, because \cite{2019A&A...622A..64B} assumed the hydrostatic equilibrium model when calculating the $\hi$ scale height, the resulted anti-correlation is somewhat by construction. Also, THINGS galaxies are on average radio fainter than CHANG-ES galaxies. We emphasize the need for larger samples of edge-on galaxies in the future to explore this possible controversy. 

Similarly, in Figure \ref{hiopp}(b.1), there is no correlation of $\bar{h}_{\rm HI,25}/R_{\rm HI}$ with baryonic mass surface density $\Sigma_{\rm Mbaryon,25}$. The baryonic mass is calculated as the sum of the stellar mass ($M_{*,25}$; See section 2.4) and gas mass (See section 2.3) within $R_{25}$, $M_{\rm baryon,25} = M_{*,25} + 1.4 M_{\rm HI,25}$, where 1.4 is the standard correction factor to account for helium and metals. Also, no correlation is found when replacing $\bar{h}_{\rm HI,25}/R_{\rm HI}$ with $\bar{h}_{\rm HI,25}$ (Figure \ref{hiopp}(b.2)). 

In Figure \ref{hiopp}(c.1), there is further no correlation of $\bar{h}_{\rm HI,25}/R_{\rm HI}$ with the SFR surface density $\Sigma_{\rm SFR}$. We additionally check for a relation when replacing $\bar{h}_{\rm HI,25}/R_{\rm HI}$ with $\bar{h}_{\rm HI,25}$ (see Figure \ref{hiopp}(c.2)) or replacing $\Sigma_{\rm SFR}$ with $\Sigma_{\rm SFR}/\Sigma_{\rm Mtot,25}$, and still find no correlations. The lack of a correlation of $\hi$ scale heights with $\Sigma_{\rm SFR}$ implies that additional energy input other than supernova feedback (e.g., from the gravitational potential energy of radially inflowing gas, \cite{krumholz2018unified}) might be present to push the $\hi$ gas up. Alternatively, the large uncertainty in the derived $\hi$ scale heights have blurred its possible dependence on SFR density. Comparing observed SFR and ISM pressure with stellar feedback models, it was indeed found that, the supernova feedback, including energy and momentum injections, may be the major source of energy to sustain the ISM pressure in active star-forming regions; However, it does not provide sufficient energy when consider the the whole optical disk, particularly in relatively passive regions \citep{tamburro2009driving,2021ApJ...909..131B,2021MNRAS.503.3643B}. Because our measure of $h_{\rm H,25}$ is averaged throughout the optical disk, and because $h_{\rm H,25}$ is pushed up by pressure, whether or not it in quasi-static equilibrium with other forces, our results based on morphology analysis are qualitatively consistent with those previous results based on dynamic analysis. A similar lack of correlation between the radio scale heights and SFR density has been also found in \cite{2018AA...611A..72K}, implying a similar complex origin of energy input sustaining the vertical structure of the two ISM components.

\section{Summary, discussion, and conclusions}

We successfully extracted $\hi$ column density images from the 1.4 GHz data obtained with the VLA in C configuration for 19 out of 35 galaxies from the CHANG-ES. The moment-0 maps from this work are available on the CHANG-ES data release website, i.e., \url{https://www.queensu.ca/changes}.

Two of the galaxies did not previously have $\hi$ interferometric images published in the literature. These galaxies are NGC 5792 and UGC 10288. For ten galaxies for which previous $\hi$ data existed, our new images have better spatial resolution, and in four cases better sensitivity in 3-$\sigma$ column density limit, as listed in Table \ref{HIcompare}. 

The whole sample is highly inclined by design, but we can still observe some spatial correlation of the $\hi$ distribution with the broad band optical emission and with the H$\alpha$ distribution in particular. We identify $\hi$ warps, spurs, and large-scale perturbed features, which are common in galaxies with extended $\hi$ disks. Almost all of the galaxies show vertical structures away from the mid-plane, as revealed in Figure \ref{atlas}. 

We have found in our sample of 19 galaxies that 42.1\% show evidence of tidal interactions while 31.6\% seem to host bars; 10.5\% have evidence of both. Very often, the galaxies show a centrally enhanced SFR and$\slash$or nuclear activity (31.6\%). 
Based on our descriptions of the $\hi$ mass and morphology, when combined with multi-wavelength data multi-wavelength properties, we can largely classify our sample into the following categories, noting that some galaxies may fall into more than one category:
\begin{itemize}
    \item Galaxies showing relatively featureless $\hi$ disks (except for spurs) and likely evolving in a secular way: NGC 2683, NGC 3044, NGC 3556, NGC 3877, NGC 4096, NGC 4157, NGC 4217, UGC 10288.
    \item Galaxies with an tidally enhanced SFR (0.5-$\sigma$ above the SFMS), superwind, and numerous $\hi$ spurs: NGC 3079, NGC 3448, NGC 4631, NGC 4666, NGC 5775.
    \item Galaxies that appear to have tidal features but low SF:  NGC 4302, NGC 4565.
    \item Galaxies with outer rings: NGC 660, NGC 5084, NGC 5792.
    \item Galaxies with possibly bar driven strong AGN and/or star forming activities (with SFR 0.5-$\sigma$ above the SFMS): NGC 3044, NGC 3079, NGC 3556, NGC 5792.
    \item Galaxies with deep $\hi$ absorption in the centre produced by strong radio AGN: NGC 660, NGC 3079.
    \item Galaxies with significant $\hi$ shells which cannot be supported by energy injection from star formation alone: NGC 3044, NGC 3556, NGC 4631.
\end{itemize}

The sample presented in this paper enriches the literature of $\hi$ atlas of edge-on galaxies. As demonstrated in this paper, the edge-on view of $\hi$ can illuminate characteristics such as the amount of fuel available for star formation. Comparing such characteristics with an individual galaxy's placement relative to the SFMS, and with datasets at other wavelengths, can lend support for proposed scenarios that produce the observed morphology. Thus the data may contribute to a statistical study of these mechanisms in the future when more $\hi$ images of edge-on galaxies are compiled. 

As a first application of the data, we measure exponential $\hi$ scale heights using a similar procedure of \cite{2018AA...611A..72K} for the radio continuum, and thus conveniently compare behaviors of the two ISM components. Both components have scale heights strongly related with the characteristic radius, and hardly correlated with the SFR surface densities. Those consistencies imply the possible similar complex energy inputs and outflow mechanisms that build the gas vertical structure. But the normalized $\hi$ scale height is not anti-correlated with the mass surface density while the normalized scale height in the radio continuum is. These results suggest that the $\hi$ and radio continuum vertical distribution may be governed by similar fundamental physics of gravity, energy and pressure, but there can be subtle differences possibly due to different origin, temperature, density, and sensitivity to internal and external physical processes.

This study serves as an exploration of the feasibility of extracting and using $\hi$ data from broadband continuum observations. Our method is also applicable to $\hi$ data cubes whose velocity resolution should be significantly degraded in order to improve sensitivity. Such data may be abundant in the near future from pilot surveys with the SKA pathfinders \citep{2007PASA...24..174J, 2008ExA....22..151J, 2009IEEEP..97.1507D, 2020Ap&SS.365..118K} and the ngVLA \citep{2017arXiv171109960B}.

\section*{Acknowledgements}

We thank an anonymous referee for constructive suggestions. We thank J. Shangguan, Gregory J. Herczeg, X. Feng, Y. Fu, S. Yu, especially thank H. Xu, Y. Yang, N. Yu, Z. Liang. JW thank support from National Science Foundation of China (12073002, 11721303), and the science research grants from the China Manned Space Project with NO. CMS-CSST-2021-B02. Parts of this research were supported by High-performance Computing Platform of Peking University. Funding for the SDSS and SDSS-II has been provided by the Alfred P. Sloan Foundation, the Participating Institutions, the National Science Foundation, the U.S. Department of Energy, the National Aeronautics and Space Administration, the Japanese Monbukagakusho, the Max Planck Society, and the Higher Education Funding Council for England. The SDSS Web Site is \url{http://www.sdss.org/}. This research made use of Photutils, an Astropy package for detection and photometry of astronomical sources \citep{bradley2020astropy}. CASA \citep{2007ASPC..376..127M}, SoFiA \citep{2015MNRAS.448.1922S}, NOD3 \citep{2017A&A...606A..41M}, Photutils \citep{bradley2020astropy}

\section*{Data Availability}

The data underlying this article are available in the CHANG-ES webpage at \url{https://www.queensu.ca/changes}. The datasets were derived from sources in the public domain:
\begin{itemize}
\item The optical data are available in SDSS website at \url{http://www.sdss.org/}.
\item The H$\alpha$ data are available in the CHANG-ES webpage at \url{https://www.queensu.ca/changes}.
\end{itemize}

\bibliographystyle{mnras}
\bibliography{main}{}

\appendix
\section{HI Scale heights}

We only measure the $\hi$ scale heights in galaxies with inclinations larger than $80^{\circ}$. The same criteria was adopted by \citet{2018AA...611A..72K}, who investigated the scale height of radio continuum halos from the CHANGE-ES sample. In total, 15 galaxies in our sample meet this criterion. We derive the $\hi$ scale height $h_{\rm HI}$ largely following the method described in \citet{2018AA...611A..72K}. 

\begin{figure*}
    \centering
    \includegraphics[width=0.7\textwidth]{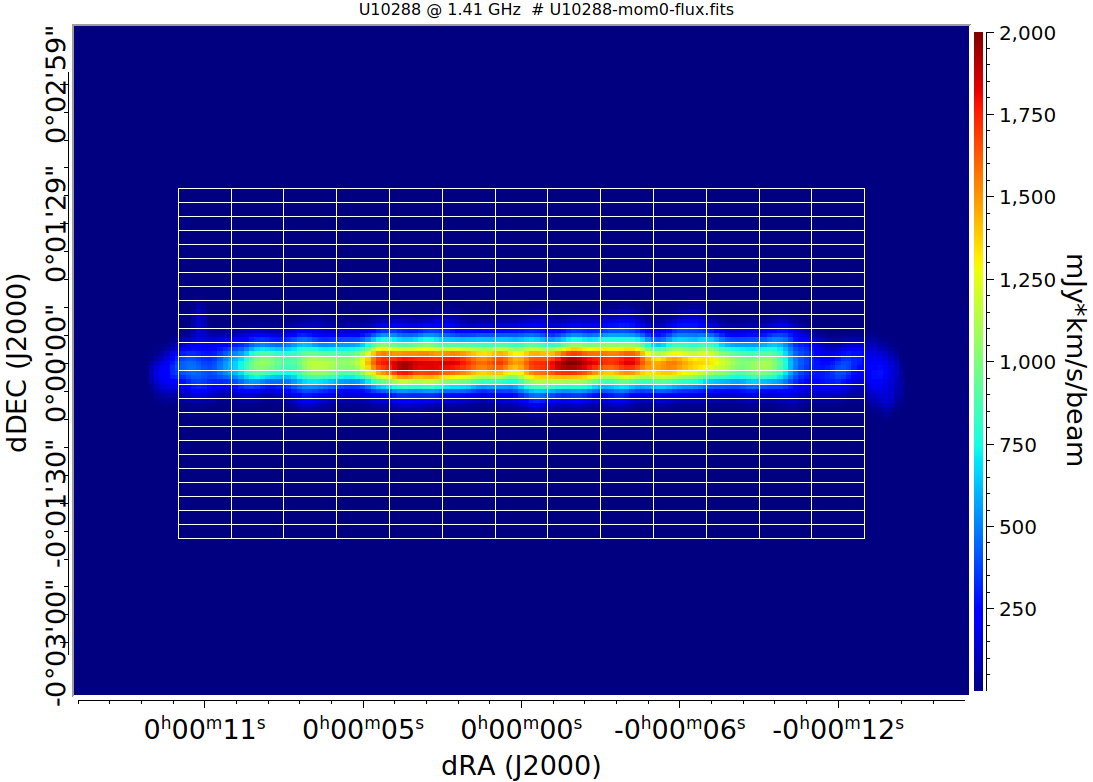}
    \caption{Illustrated strips (in the vertical direction) of UGC 10288 used for the $\hi$ scale height measurement. The color map shows the rotated $\hi$ intensity distribution of UGC 10288, where the white vertical lines divide the galaxy into 13 strips; each is 30 arcsec wide and is further divided into 25 9-arcsec high segments as marked by the white horizontal lines.}
    \label{nod3}
\end{figure*}

\citet{2018AA...611A..72K} used the “BoxModels” task in the NOD3 program package \citep{2017A&A...606A..41M} to fit the vertical distribution of surface densities. The moment-0 image, inclination, and position angle are required as input for each galaxy. 
The $\hi$ moment-0 images are convolved to have a circular beam shape, with FWHM equivalent to the major axis FWHM of the initial beam. Then the images are rotated for the galaxies to have horizontally oriented major axes. For each galaxy, strips are set perpendicular to the major axis (along the z-direction), which are further divided into boxes with a width slightly larger than the beam FWHM (30 arcsec in our sample) and a height about half the beam FWHM (9 arcsec in our sample).
An example of sampling the $\hi$ image of galaxy UGC 10288 is shown in Figure \ref{nod3}.
The mean intensity and the standard deviation are derived within each box, and the vertical profile of surface density is derived for each strip. These strip profiles serve as the data to fit the vertical models.

For galaxies with an inclination lower than $90^{\circ}$ the flux distribution away from the mid-plane is considered to be a mixture of the intrinsic vertical component and the projected planar component of fluxes from the disk. In the observed data, such a mixture is further convolved with the beam PSF. 
A exponential model is assumed for the intrinsic vertical distribution, with the exponential $\sigma$ indicating $h_{\rm HI}$. This exponential model is consistent with \cite{2018AA...611A..72K} and prepare for the comparison between $\hi$ and radio continuum.
The contamination of the planar projection is accounted for as a pseudo-increment of the PSF's FWHM along the z-direction in a radius $r$ dependent way: $\Delta FWHM = R\  {\rm cos}(r/R* \pi/2)\  {\rm cos}(i)$, where $R$ is the galaxy disk size and $i$ is the inclination. The effective PSF then has an effective FWHM along the z direction: $FWHM_{\rm eff,z} = \sqrt{FWHM^2+\Delta FWHM^2}$. 
In this way, the planar contamination is stronger at a small radius than at a large radius. The vertical model is then convolved with this effective PSF before being compared with the observed profile of a strip, and the best-fit model is determined with the least-square algorithm. 

We largely follow the procedure of \citet{2018AA...611A..72K} and “BoxModels” as described above, but make two additional adjustments.
Firstly, we do not directly use the position angles from the CHANG-ES main paper or \citet{2018AA...611A..72K} for the strip fitting, but compensate for the fact that the $\hi$ disk may be misaligned with the disk orientation observed at other wavelengths. We use the previous CHANG-ES inclinations and position angles only as initial guess, to produce an initial set of strip samplings. For most galaxies, we then determine the vertical, i.e., `z-coordinate', centre of each strip at the peak surface density. By doing so, we re-define the mid-plane direction of each disk. Exceptions are NGC 3556 and NGC 4096, in which we apply the surface density weighted mean z-coordinates for the mid-plane centre, because most of their strip profiles are singly peaked. For the remainder of the analysis we set the inclination to the values in \cite{2018AA...611A..72K} or \cite{2012AJ....144...43I}. Exceptions are galaxies with inclinations from kinematic $\hi$ analysis in the literature, e.g. NGC 4565 \citep{2019A&A...632A..12S} and NGC 4631 \citep{2019A&A...632A..10M}.

Secondly, the NOD3 procedure “BoxModels” fits both sides of the strip profile simultaneously. However, $\hi$ disks can be much more asymmetric to either side of the mid-plane than radio halos. So we developed an additional procedure which also fits the two sides independently but generates two individual profile fittings emanating from the mid-plane's central z-coordinate.

\begin{figure*}
\vspace{-5mm}
\includegraphics[height=0.49\textheight]{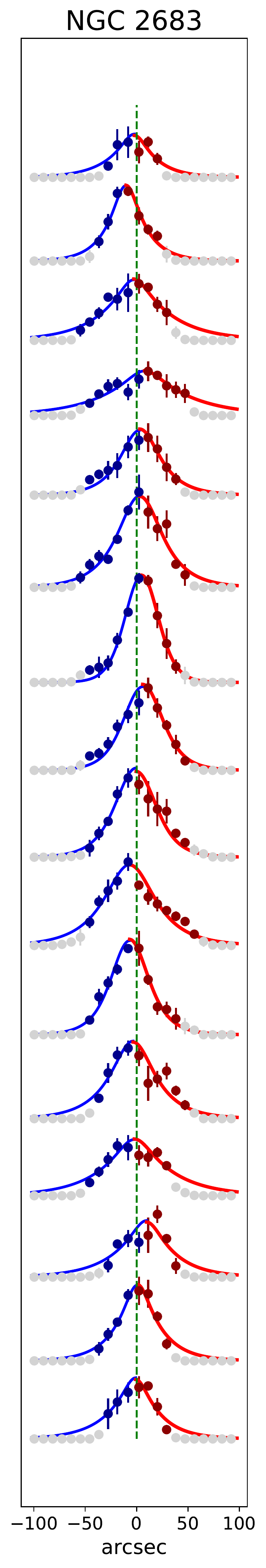}
\includegraphics[height=0.49\textheight]{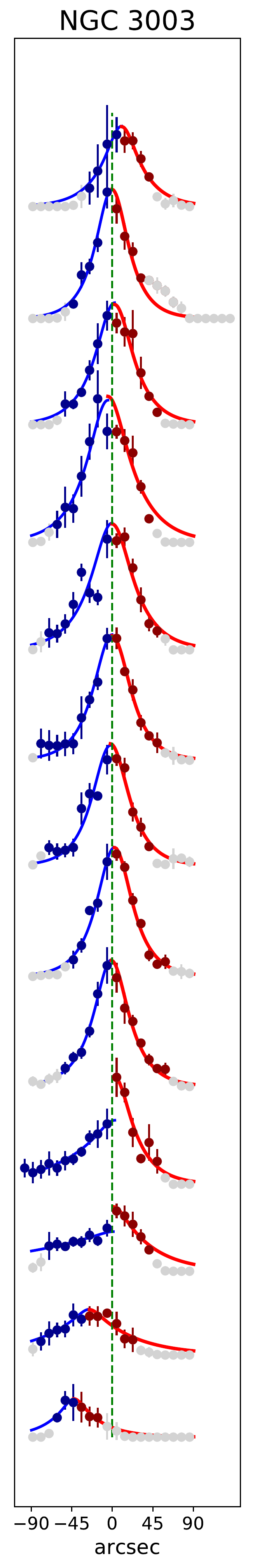}
\includegraphics[height=0.49\textheight]{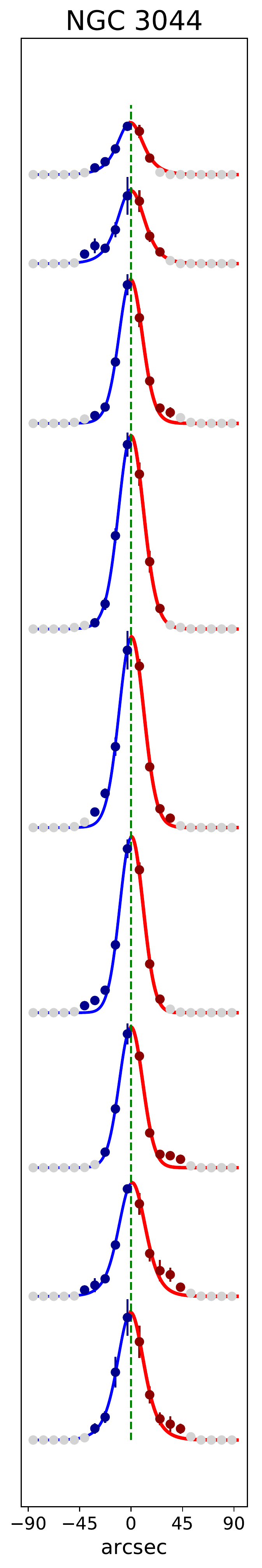}
\includegraphics[height=0.49\textheight]{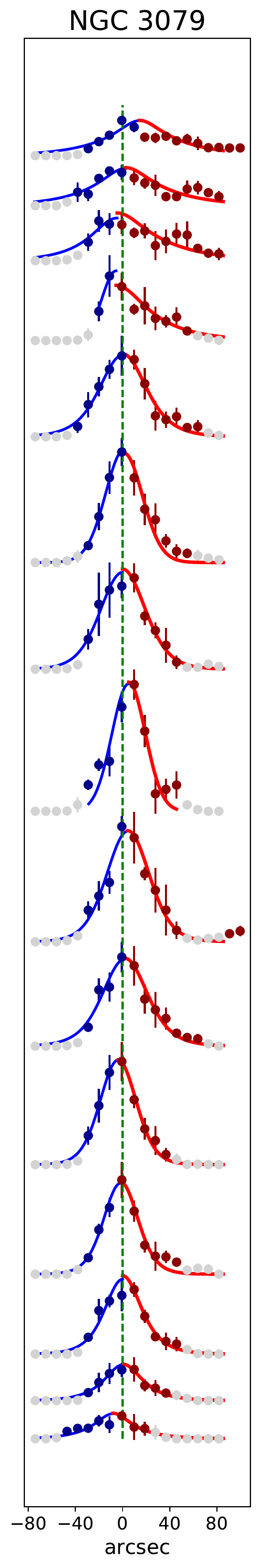}
\includegraphics[height=0.49\textheight]{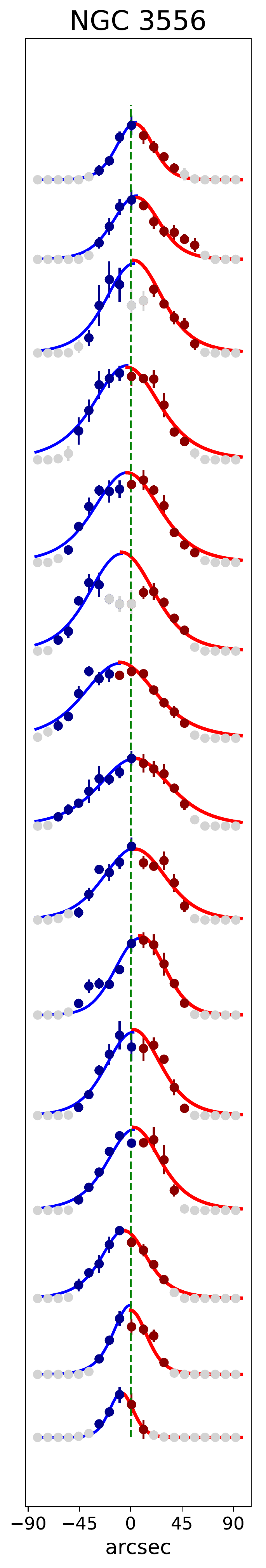}
\includegraphics[height=0.49\textheight]{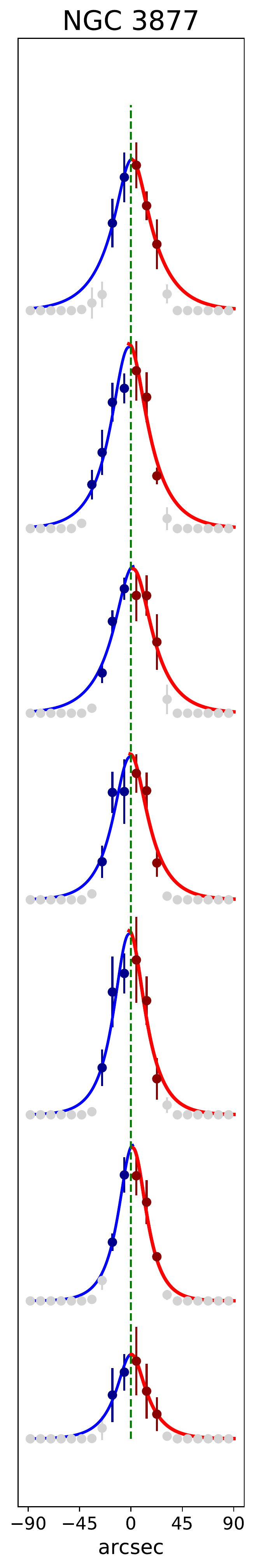}
\includegraphics[height=0.49\textheight]{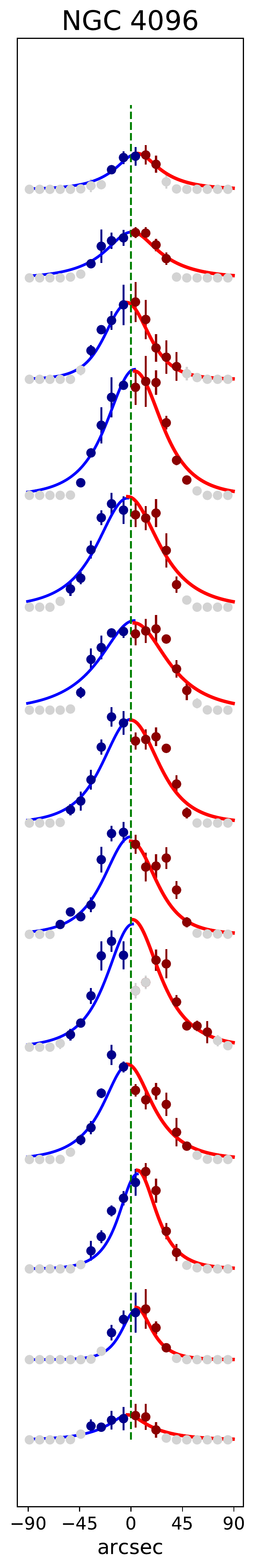}
\includegraphics[height=0.49\textheight]{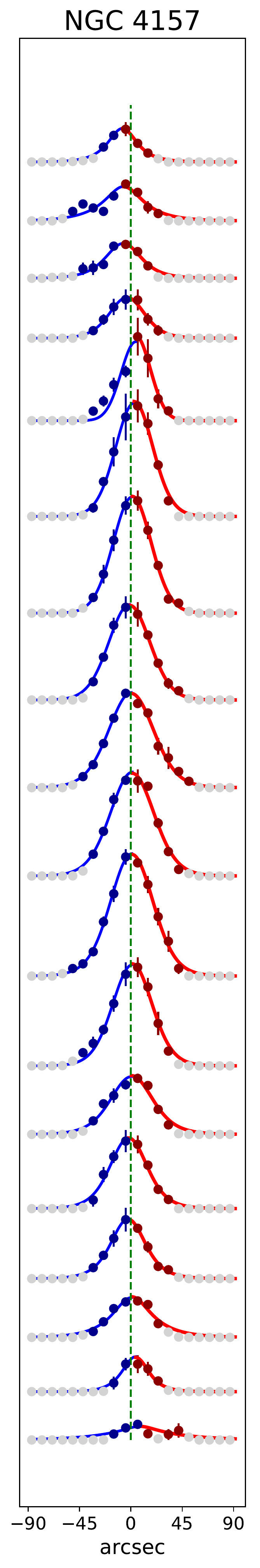}

\includegraphics[height=0.49\textheight]{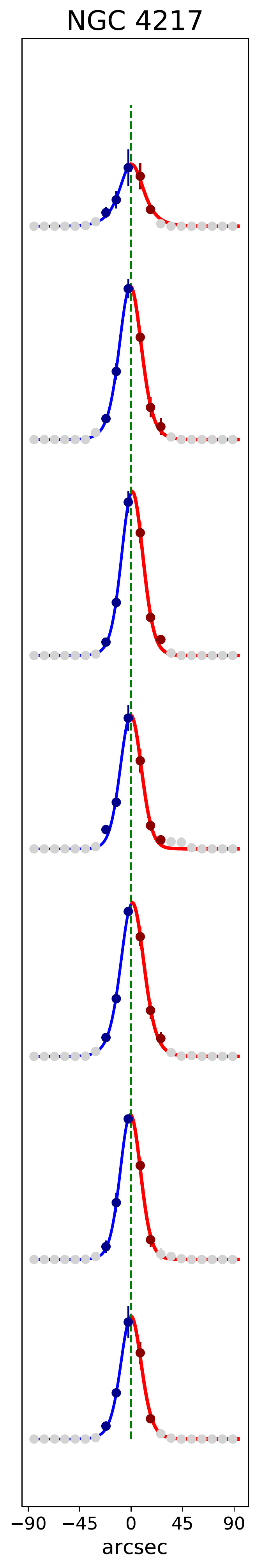}
\includegraphics[height=0.49\textheight]{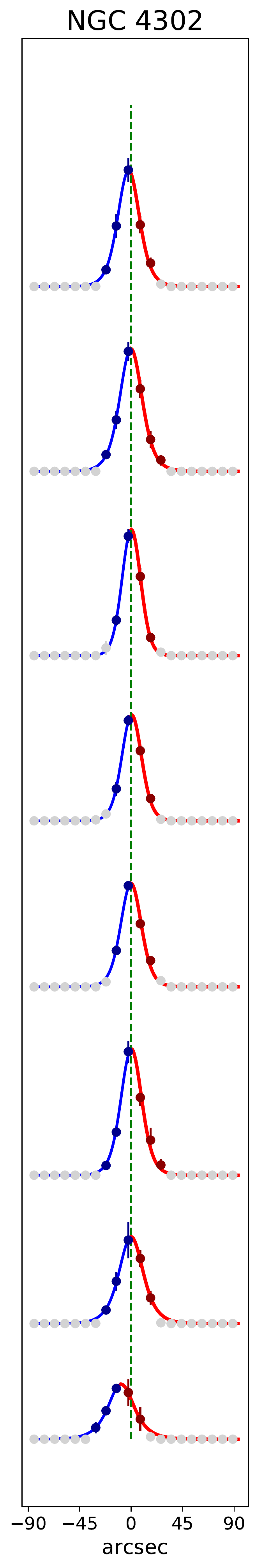}
\includegraphics[height=0.49\textheight]{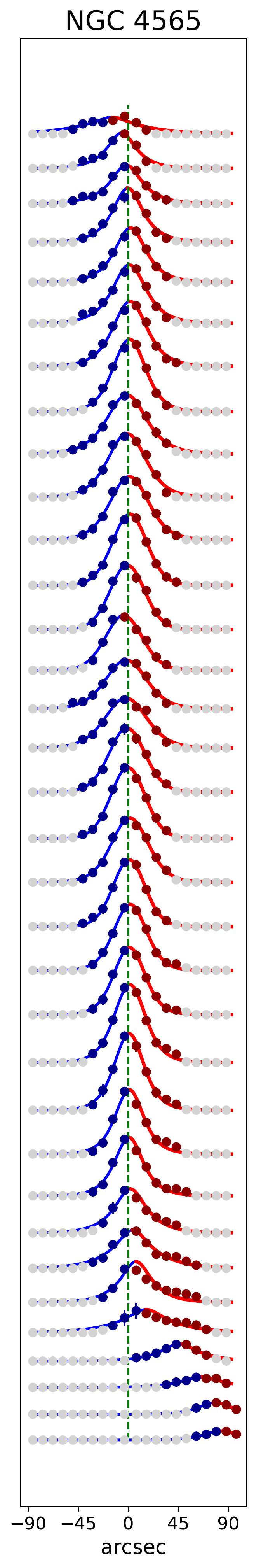}
\includegraphics[height=0.49\textheight]{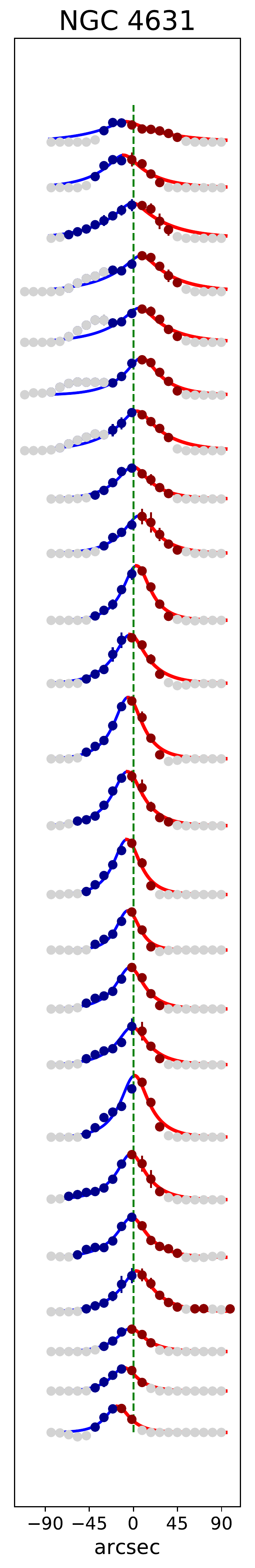}
\includegraphics[height=0.49\textheight]{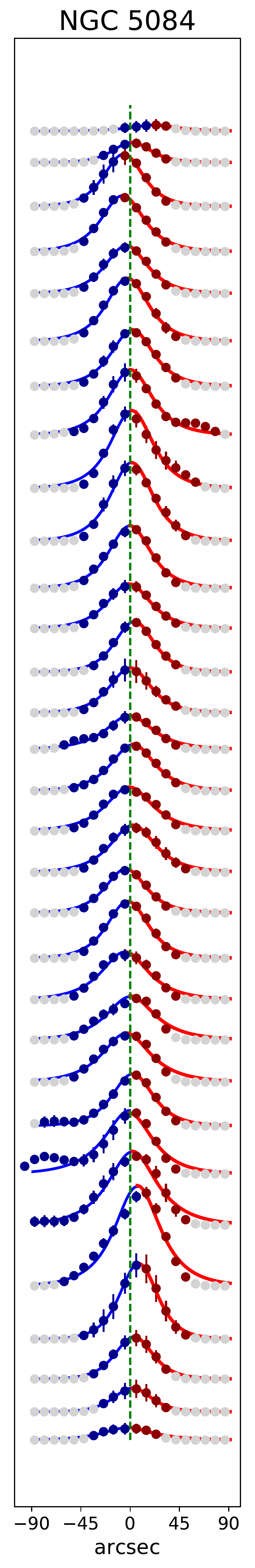}
\includegraphics[height=0.49\textheight]{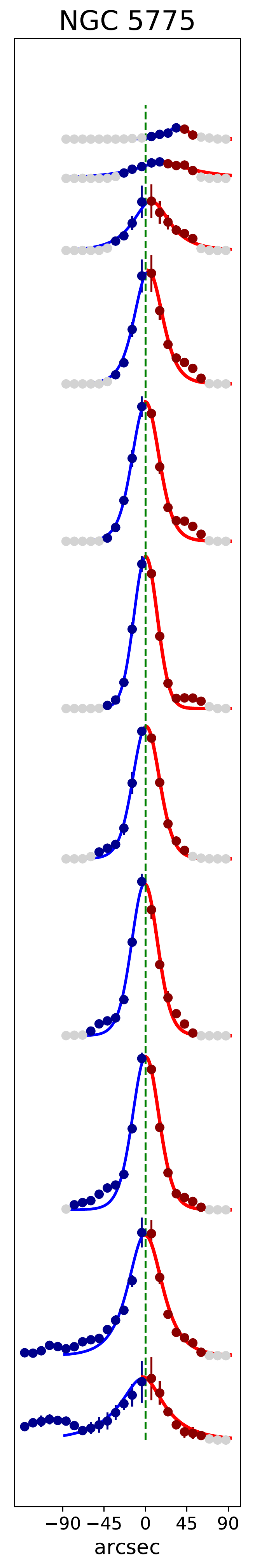}
\includegraphics[height=0.49\textheight]{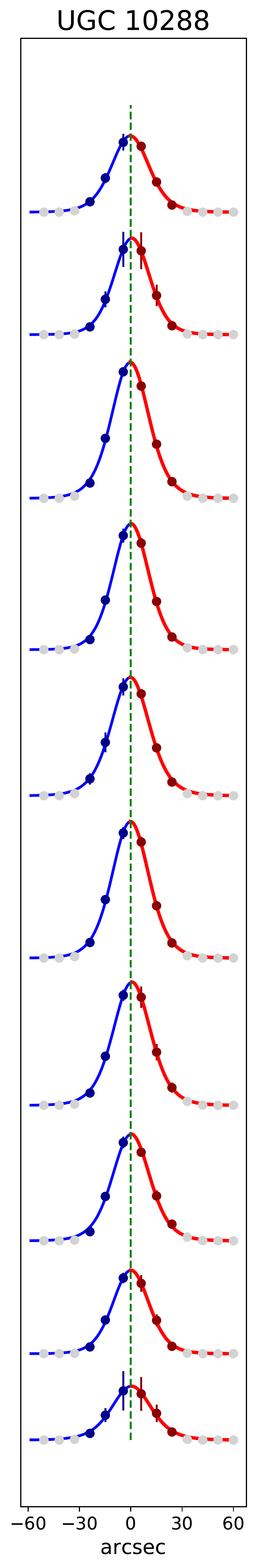}
\caption{The results of fitting exponential profiles to each galaxy. The grey points represent the masked data. The dark blue and dark red points represent data that contribute to the fitting of the galaxy's upper and lower sides, respectively. The blue and red curves represent the exponential fitting results of upper and lower sides, respectively. The green dashed line shows the x=0 line, which is the redefined mid-plane axis.}
\label{fitting}
\end{figure*}

The strip profiles, as well as best-fit models for each galaxy, are shown in Figure \ref{fitting}. The observational data that contribute to the fitting process are represented by blue and red points with errorbars, while the grey points represent data with values lower than 3$\times$RMS or have beem masked. Every strip profile has two independently fitted sides (blue and red curves) with respect to the separately determined profile centre, and the vertical green dashed lines represent x=0. Due to rotation to the horizontal position, z=0 is redefined as x=0. We note that there are strip profiles that do not strictly follow a exponential shape (as can be found in, e.g, NGC 2683, NGC 3556, and NGC 4096). These irregularities are likely associated with disk instability structures like spiral arms and bars, or external perturbations \citep{2009AJ....138.1082K, 2015AstBu..70..379K, 2015ApJ...805..144K, vollmer2016flaring, 2020Ap&SS.365..111S, 1997NewA....2..251K, 2000A&A...361..888G}. We still fit exponential profiles to them to conserve the uniformity of the $h_{\rm HI}$ parameter. Despite the poor matches between the model and data, the best-fit $\sigma$ is still a good measure of the width of the strip profile.

Since the data lack sufficiently resolved kinematic information for properly modelling the possible flaring of the disk, we only use the radially averaged $h_{\rm HI}$ within the optical radius $R_{25}$ ($\bar{h}_{\rm HI,25}$) as an indicator of the disk thickness in our analysis (listed in Table \ref{HIsample}). Due to our limited sensitivity we are unable to detect much $\hi$ beyond $R_{25}$ (median $R_{\rm HI}/R_{25} = 1.14 \pm 0.33$), i.e., the radius beyond which most warps arise \citep{1990ApJ...352...15B}. Fortunately, this may reduce a putative warp's contribution to our photometric determination of $h_{\rm HI}$, regardless of projection effects. The error used for the radially averaged $h_{\rm HI}$ combines the standard deviation of the values of a galaxy's individual $h_{\rm HI}$ strips and the uncertainty from the model fitting. 
Our method involves calculating the radially averaged $h_{\rm HI}$ for the upper and lower halves with respect to the mid-plane of each galaxy separately ($\bar{h}_{\rm HI,up}$, $\bar{h}_{\rm HI,low}$). Except for the severely perturbed galaxies NGC 3003 and NGC 3079, all other edge-on galaxies have similar value between $\bar{h}_{\rm HI,up}$ and $\bar{h}_{\rm HI,low}$. We thus use the mean of $\bar{h}_{\rm HI,up}$ and $\bar{h}_{\rm HI,low}$ as the final measure of $\bar{h}_{\rm HI,25}$ for each galaxy.

\section{Galaxies with new HI images}

In Appendix B and C, we discuss $\hi$ features in combination with other known galaxy properties. The descriptions that we present, galaxy-by-galaxy, are meant to describe our own results, and also introduce the reader to some previously known information about the galaxies. The references to earlier work are not exhaustive, but we focus on information that is most relevant for comparisons to our $\hi$ results. We report a galaxy's position in a scaling relation diagram using the number of $\sigma$ it differs from the relation. We use the SFMS of \cite{2016MNRAS.462.1749S} in the following discussion, as the measurement methods for SFR are relatively close. We point out a caveat that other SFMSs may result in different $\sigma$ values, particularly at the high stellar mass end.

$\hi$ interferometric images for two galaxies in the sample (NGC 5792 and UGC 10288) have not previously been presented in the literature.

\subsection{NGC 5792}
NGC 5792 is a `SBb' barred spiral galaxy. The $\hi$ disk has a central hole, where there is strong H$\alpha$ emission. This indicates active star formation in the centre. There is an inner $\hi$ ring (labeled in Figure \ref{atlas}), which can also be seen in the optical and the H$\alpha$. This ring structure is possibly related to the dynamics of the bar \citep{1979ApJ...227..714K}. There also appears to be an outer ring in $\hi$. The optical disk is faint and appears to have an outer ring, or spiral arms, as well at the same spatial location. NGC 5792 is the brightest in a small galaxy group of six galaxies or less \citep{2011MNRAS.412.2498M, 2017ApJ...843...16K}. Thus the environment may be contributing to the formation of the putative outer $\hi$ ring.

\cite{1980A&AS...41..189R} observed the $\hi$ line profile, with a velocity resolution of 22 km/s and a RMS of 350 mJy/beam, with the National Radio Astronomy Observatory (NRAO) 300-ft single-dish telescope. Using the drift scan mode of this single disk telescope they also constructed an $\hi$ distribution contour map. The main beam of this antenna is 10 arcmin \citep{1973AJ.....78..565H} and scans were separated by 5 arcmin. Since this image is not a result of interferometry, we include NGC~5792 in this section.

There is an evidence that this galaxy hosts an AGN, which possibly contributes to the H$\alpha$ emission and may be related to the $\hi$ hole (Yang et al. in prep). 

Figure \ref{xGASS} shows that, NGC 5792 has slightly high $M_{\rm HI}$ (0.77-$\sigma$ above the HIMS) and SFR (0.92-$\sigma$ above the SFMS) for its stellar mass. This is consistent with the rings and$\slash$or spiral arms observed in the $\hi$, which are associated with strong disk instabilities that tend to trigger star formations \citep{1969ApJ...158..123R, 2005ApJ...633L.137V}. 

\subsection{UGC 10288}
UGC 10288 is an `Sc' spiral galaxy. The $\hi$ in UGC 10288 forms a highly regular, flat, and thin disk, with signs of a very weak warp on the western side. While these features imply that in the past few gigayears the galaxy is unlikely to have been strongly perturbed by gravitational or hydrodynamic effects, we note that UGC 10288 may be in a small group of 2-3 galaxies, e.g. \cite{2011MNRAS.412.2498M}; Lyon Group of Galaxies \citep{1993A&AS..100...47G}.

Despite the featureless appearance in the $\hi$, \cite{1996ApJ...462..712R} and \cite{1984MNRAS.209...93S, 1981ApJ...247...77S} found spurs in the H$\alpha$ disk which extend toward the east side of the disk. Some of the spurs appear as shell-like structures. They also found a bulge and dust lanes in the near infrared.
\cite{2001ApJ...551...57C} presented deep long-slit optical spectra at two positions in the halo of the diffuse ionized gas (DIG), and found that the line ratios are best described by a composite photo-ionization model. UGC 10288 has discrete high-latitude radio continuum features; it lacks a global radio continuum halo (exponential scale heights are typically $\sim$1 kpc) \citep{2013AJ....146..164I}.

The H$\alpha$ spurs, the line ratio of the DIG in the halo indicate outflows from the galaxy disk. It is interesting to point out that, as shown in Figure \ref{xGASS}, UGC 10288 has an sSFR lower than expected from the $\hi$ mass fraction. Indeed it is the most extreme case of three galaxies with this feature; the other two are NGC 3003 and NGC 4565. The low star forming efficiency of UGC 10288 may be the result of either the lack of a gravitational perturbation (to seed an instability) or quenching by the outflow that creates the warm gas halo. 

\section{Previously imaged galaxies}
We discuss in detail the $\hi$ morphology of the galaxies for which $\hi$ images have previously appeared in the literature.  We note that although these galaxies have been presented in $\hi$ before, they are shown here for the first time as part of a uniform data-set.

\subsection{NGC 660}
NGC 660 is identified as a polar ring spiral galaxy with a LINER-type nucleus. It has a morphologically perturbed companion galaxy UGC 01195 which is outside the field of view of the $\hi$ image shown in Figure \ref{atlas}. It belongs to the Messier 74 Group which has six members \footnote{\href{http://www.messier.seds.org/more/m074gr.html}{http://www.messier.seds.org/more/m074gr.html}}. In Figure \ref{atlas}, both optical and H$\alpha$ images show that this galaxy contains two distinct components; roughly measured from these data the almost edge-on disk has a position angle of $\sim45^{\circ}$ and the strongly inclined ring has a mean position angle about $\sim170^{\circ}$. In the optical image, two prominent dust lanes cross each other, one in the northern part of the central disk and the other associated with the polar ring (labeled in Figure \ref{atlas}).
The $\hi$ distribution follows these two structures and has an $\hi$ absorption feature in the core (labeled in Figure \ref{atlas}) associated with strong radio continuum.

The disk and the polar ring have been detected before in $\hi$ and robustly confirmed by position-velocity analysis \citep{1990NASCP3098..209G, 1995AJ....109..942V}. \cite{1995AJ....109..942V} estimated the polar ring contains 75\% of $\hi$ the total mass and displays regular rotation. 

In the H$\alpha$ image, the $\rm H_{II}$ regions appear distributed throughout the disk and rings, indicating that a large number of stars continue to form \citep{2019ApJ...881...26V}. \cite{2004A&A...421..833K} used data from the Wide Field Camera-2 of the Hubble Space Telescope (HST) to study the stellar content of the ring based on the color-magnitude diagram. They illustrated that the star formation in the ring is continuous and that the youngest stellar age is about 7~Myr. \cite{2017OAst...26...88S} showed that in NGC 660 the SFR in the star forming regions in the disk is normal compared to other disk galaxies, but that the SFR of star forming regions in the ring is much lower. 

In the galaxy core, $\hi$ absorption is observed instead of $\hi$ emission, which implies the existence of nuclear activity. \cite{2015MNRAS.452.1081A} observed a spectacular radio outburst based on radio continuum and X-ray data. These showed a compact radio core, a northeastward jet-like component, and possibly a weaker westward counter-jet. Further evidence of activity is provided by \cite{2009A&A...506..689I} who found that the CO emission has a strong central peak and follows solid-body, rapid rotation. By fitting CO ratios, they concluded that most of the molecular gas is very hot ($T_{kin}$=150 K) and tenuous in the central regions of NGC 660. Activity in the core may be the result of a recent major merger of two spiral galaxies, which built the $\hi$ polar ring at the same time. \cite{2019ApJ...878...76Y} simulated NGC 660 as the result of merger of two galaxies a few billion years ago. 

In summary, NGC 660 is a polar ring galaxy and appears to have strong nuclear activity. Taken together these features suggest that a merger occurred within a few billion years. In Figure \ref{xGASS}, NGC 660 has normal $M_{\rm HI}$ (0.32-$\sigma$ above the HIMS) for its stellar mass, but significantly high SFR (1.35-$\sigma$ above the SFMS). The enhanced SFR is consistent with the strong nuclear activity, both of which can be triggered by strong gas inflows driven by a major merger \citep{1996ApJ...464..641M}.

\subsection{NGC 2683}

NGC 2683 is a `Sb' spiral galaxy with an $\hi$ disk that is slightly more radially extended than the optical one. A significant S-shape structure, starting from the galaxy centre, is observed in the $\hi$ disk (labeled `bending' in Figure \ref{atlas}). 

\cite{1991AJ....101.1231C} and \cite{vollmer2016flaring} also observed this bent structure in $\hi$ starting from a galactocentric radius of 3.5' out to 8'.   \cite{2009AJ....138.1082K} argued that NGC 2683 hosts a bar, using evidence from kinematic and photometric features in long-slit spectra, integral field unit spectroscopy, and multi-band images.

Based on higher sensitivity data than our own, \cite{vollmer2016flaring} detected a flared and much more radially extended $\hi$ disk on both sides of the nucleus, with  a diameter of 26.5'. Unlike normal spiral galaxies, the flaring in NGC 2683 does not continue, but suddenly halts at an intermediate radius ($\sim$15 kpc) and then fades. They suggest that it may be caused by external accretion \citep{vollmer2016flaring}.
NGC 2683 may be interacting with its companion dwarf galaxies, KK 69, KK 70, dw1, and dw2 \citep{2015AstBu..70..379K,2015ApJ...805..144K,2020Ap&SS.365..111S} and one of these could be the source for external accretion of $\hi$ gas.

In summary, NGC 2683 may host a bar and be experiencing external accretion at the same time. Our $\hi$ image confirms the inner distribution of $\hi$, which may possibly be associated with the bar, but misses the extended outer disk associated with gas accretion, probably due to our limited sensitivity. NGC 2683 appears to be the most significant $\hi$-poor galaxy and passive outlier in the scaling relations of $M_{\rm HI}$ mass versus stellar mass (2.09-$\sigma$ below the HIMS), and SFR versus stellar mass (1.51-$\sigma$ below the SFMS) in Figure \ref{xGASS}. 

\subsection{NGC 3003}

NGC 3003 is a `Sbc' spiral galaxy and has a nearby companion NGC 3021 which is outside of the field of view in Figure \ref{atlas}. The morphology of the $\hi$ looks asymmetric and perturbed, with the south-western corner displaying a complex structure (labeled in Figure \ref{atlas}). Many $\hi$ spurs are observed on each side of the disk plane. The scaling relations in Figure \ref{xGASS} show that NGC 3003 has normal a SFR (0.32-$\sigma$ above the SFMS) but significantly high $M_{\rm HI}$ (1.36-$\sigma$ above the HIMS) for its stellar mass. 

NGC 3003 has been observed, and catalogued as UGC~5251, in the Westerbork $\hi$ survey of spiral and irregular galaxies (WHISP) \footnote{ \href{https://www.astro.rug.nl/~whisp/}{https://www.astro.rug.nl/~whisp/}}. The WHISP catalogue includes intensity moment-0 and velocity moment-1 maps for up to 1000 galaxies at three different resolutions: 30'',  60'' and a full resolution that is listed in Table \ref{HIcompare} for five of our galaxies.
In the 30'' resolution moment-0 map of NGC 3003, the curved $\hi$ structure extending to the southwest is consistent with the extension in our observations.

The asymmetry is also noticeable in H$\alpha$ \citep{1999ApJ...522..669H, 2019ApJ...881...26V} and optical light (SDSS) (labeled in Figure \ref{atlas}). They point out that the asymmetric feature indicated with an arrow in the H$\alpha$ image may be associated with a dwarf galaxy merger, or tidal stripping causing the relocation to this position of a gaseous star forming region, or a tidally induced asymmetric spiral arm \citep{2019ApJ...881...26V}. 
It also shows an unusually large scale height in the L-band radio continuum halo \citep{2018AA...611A..72K}. 

We notice that the dynamical mass surface density of NGC 3003, derived in Paper IX \citep{2012AJ....144...43I} is among the lowest values of our sample, which implies that this galaxy has a low gravitational potential. Thus, the extended morphology of the $\hi$ disk could relatively easily be shaped by tidal perturbations from its companion NGC 3021.

\subsection{NGC 3044} 

NGC 3044, a 'SBc' galaxy. While it is identified as an isolated in many galaxy group catalogues, there is some evidence in the HIPASS observations that this galaxy has companions (HIRF 0948+0043; \cite{2011A&A...533A.122P}). Figure \ref{atlas} shows that it has a radially and vertically extended $\hi$ disk compared to the optical one.
The $\hi$ disk reveals a slightly asymmetric morphology and clear $\hi$ spurs above and below the disk's midplane. On the northwest end of the disk, exists a dense region in $\hi$ (indicated in Figure \ref{atlas} using an arrow), with counterparts in the H$\alpha$ emission but not in the optical. This northwest end of the $\hi$ disk is also slightly broader vertically than the southeast end. 

\cite{1997ApJ...490..247L} found clear $\hi$ supershells in NGC 3044, mostly on the upper side of the disk plane and near the west end of the $\hi$ disk. We do not see these features in Figure \ref{atlas} due to the poorer sensitivity. \cite{1997ApJ...490..247L} built kinematic models for the $\hi$ distribution of supershells. They concluded that the $\hi$ supershells originate from internal perturbations, but need to be sustained by additional energy, like that of the magnetic fields. Following these studies with data of better resolution and sensitivity, \cite{2015ApJ...799...61Z} considered vertical gradients in the rotational velocity in their modeling, and found that NGC 3044 can be described as a single, thick $\hi$ disk with a warp component along the line of sight. They postulate that the complicated kinematics may suggest a previous minor merger or possibly internal feedback processes. They also found a decrease in rotation speed with height.

The DIG of NGC 3044 has also been well-studied and shows very similar structure to the neutral hydrogen \citep{2000A&A...359..433R,2000ApJ...536..645C, 2000A&A...362..119T}. In general, NGC 3044 shows an extended radio halo out of the galaxy plane \citep{1995ApJ...444..119D, 1996ApJ...467..551C, 2013MNRAS.433.2958I}. The radio continuum emission is also highly asymmetric, but is more truncated on the south-eastern (left-bottom) side.  
\cite{2006A&A...448...43T} presented the first X-ray observation of NGC 3044 and showed a spatial correlation between DIG and soft X-ray band (0.5–1.0 keV) emission.

In short, NGC 3044 has a slightly asymmetric disk and many spurs that are seen in gases of different phases. On the other hand, in Figure \ref{xGASS}, its slightly high SFR (0.60-$\sigma$ above the SFMS) 
and normal $M_{\rm HI}$ (0.26-$\sigma$ above the HIMS) imply that the galaxy is close to a typical, actively star-forming galaxy. Thus, it may serve as a good laboratory to study disk-halo connections in a typical star formation galaxy in which minor interactions may play a very modest role.

\subsection{NGC 3079}

NGC 3079 is a `SBcd' barred spiral galaxy with two companions, NGC 3073 and MCG 9-17-9, both of which are outside the field of view of the $\hi$ image shown in Figure \ref{atlas}. Its $\hi$ disk is more radially extended than the optical disk and shows clear asymmetry indicative of external perturbation. More specifically, its southern end is much broader vertically than the northern side. NGC 3079 also has a strong radio core \citep{2019AJ....158...21I} which results in an absorption feature in the $\hi$ image (labeled in Figure \ref{atlas}).

\cite{1987ApJ...313L..91I} and \cite{1991ApJ...371..111I} studied the $\hi$ of NGC 3079 and also observed its two companions. They found an elongated $\hi$ tail from NGC 3073 which they propose is caused by ram pressure stripping from the nuclear outflow of NGC 3079. For NGC 3079 itself, their observations and ours show a sharper fall-off of emission on one side of the mid-plane (the west (right) side; also see Figure \ref{atlas}). They suggest that this may also be related to ram pressure stripping, although observations of hot gas in the environment are required to confirm this scenario.

Evidence of a starburst-driven superwind is provided by X-shaped filaments in H$\alpha$ \citep{1990ApJS...74..833H} and X-ray observations \citep{2004ApJS..151..193S,2004ApJ...606..829S,2005astro.ph..4237A}. Additionally, \cite{2001ApJ...555..338C} showed a superbubble, in HST optical emission lines and VLA radio continuum data, in the inner $\sim$19 kpc region.  On the other hand, NGC 3079 is classified as a Seyfert galaxy, with bright $H_2O$-maser emission \citep{1984A&A...141L...1H, 2002A&A...387L..29H}. \cite{2003MNRAS.346..977I} using the GMRT showed a radio halo at least 4.8 kpc in height as well as peculiar nuclear radio lobes. They propose these lobes are a result of both the superwinds and the active galactic nucleus jet. 

Kinematic analysis of CO observations indicate that this galaxy has both a nuclear ring aligned with the jet \citep{1992PASJ...44..353S,1992ApJ...396L..75I} and a bar \citep{2002ApJ...573..105K, 1999AJ....118.2108V}.

In summary, tidal interactions and the bar may have driven strong inflows of gas in NGC 3079, which can simultaneously fuel both the starburst and AGN activity. In Figure \ref{xGASS}, its $\hi$ mass is slightly higher than expected for its stellar mass (0.83-$\sigma$ above the HISM), but its SFR is significantly high (1.28-$\sigma$ above the SFMS), which are consistent with this scenario.

\subsection{NGC 3448}

NGC 3448 is a `S0-a' galaxy and interacting with its dwarf irregular companion UGC 6016 (labeled in Figure \ref{atlas}). The $\hi$ disk is highly radially extended compared with the optical one, and shows a highly perturbed morphology due to the tidal interaction. Beyond the main disk region, there are sporadic H$\alpha$ emission features in the extended $\hi$ tail to the east and the $\hi$ `bridge' region between NGC 3448 and UGC 6016, which may imply the interaction caused new star formation.
The $\hi$ disk shows a sharp $\hi$ gradient on the southern side and extensions on the northern side. This is morphologically consistent with ram pressure stripping for a galaxy moving southwards, which however awaits confirmation based on properties of the surrounding hot gas halo in the future.

The tidally perturbed structure of the $\hi$ disk was noticed previously and modelled extensively in the literature. 
\cite{1978A&A....63..363B} analyzed the $\hi$ emission line of NGC 3448 observed with the Nancay Radio telescope and found a double velocity profile in the inner region. 
\cite{1984AJ.....89..350B} compared the optical and $\hi$ velocity curves of NGC 3448. They presented a “hump” structure in the velocity curve, which was recognized as a high-velocity component corresponding to the double velocity feature reported by \cite{1978A&A....63..363B}. They concluded that the high-velocity component probably originated from a tidal interaction.
\cite{1986AJ.....92.1048N} presented a high resolution $\hi$ image using the VLA in C configuration and successfully simulated the complex large-scale structure with a three-body interaction model. They found no continuous stream of gas from NGC 3448 to UGC 6016 (also confirmed in our $\hi$ data cube) and managed to separate the two galaxies in the $\hi$ data cube. Most of the $\hi$ mass in NGC 3448 resides in the tail. The inner part of its $\hi$ disk rotates like a solid body, while the $\hi$ tail does not participate in the inner solid-body rotation. They concluded that before the interaction with UGC 6016, NGC 3448 was probably a late-type spiral galaxy.

Bright radio and H$\alpha$ fluxes found throughout the central disks, indicative of a starburst, are Likely driven by the tidal interactions \citep{1987AJ.....93.1045N}. 

To sum up, NGC 3448 is a tidal-interacting galaxy with strong starbursts. The morphology in our $\hi$ observation strongly supports the tidal interaction scenario. In Figure \ref{xGASS}, the SFR of NGC 3448 is close to the upper envelop of the SFMS (1.11-$\sigma$ above the SFMS). This enhanced SFR is consistent with the scenario of a tidally driven starburst. Interestingly, it has the highest $M_{\rm HI}/M_*$ in our sample. Considering the $\hi$ bridge, the amount of $\hi$ gas flowing from UGC 6016 to NGC 3448 might be replenishing the cold gas consumed by active star formation in this galaxy.

\subsection{NGC 3556}
NGC 3556 is classified as an isolated `SBc' barred spiral. Figure \ref{atlas} shows that its $\hi$ disk is slightly more extended than the optical disk, and does not show much distortion in the outskirts. Like in many other spiral galaxies, the densest region of $\hi$ gas is not distributed in the centre but follows a spiral-arm-like structure with clear counterparts in the H$\alpha$.

A prominent $\hi$ feature of this galaxy is its large eastern loop which resembles an expanding half-shell \citep{1997NewA....2..251K}. This feature can not be directly seen in Figure \ref{atlas} due to the limited sensitivity.
Given its isolated environment, the super shell is most likely of internal origin. Like in NGC 3044, the current star formation energy is insufficient to support the supershell \citep{1997NewA....2..251K}. 
It is possible that some additional energy (possibly magnetic fields) are necessary to help explain this supershell. \cite{2000A&A...361..888G} raised one possible mechanism, that the $\hi$ supershell is caused by radio lobes produced by radio jets. The radio jets were ejected 10 Myr ago and led to a localized flaring. However, neither radio continuum nor DIG counterparts are found around the $\hi$ supershell \citep{2000ApJ...536..645C}.

The large-scale radio continuum halo of this galaxy was initially identified by \cite{1979A&A....73..196D, 1984A&A...133...19K}. The X-ray emission of the galaxy was studied by \cite{2003ApJ...598..969W}, and shows an extended extraplanar distribution which can be explained by supernova blast waves. The hot gas halo in NGC 3556 is less extended compared with interacting galaxies, but the isolated environment makes it more pristine for studying the disk-halo interaction.

Overall, NGC 3556 has a strong disk-halo interaction. It is a pity that the limited sensitivity has prohibited us from investigating the $\hi$ supershell in detail. But its regular morphology helps confirm there is little perturbation from the external environment. Figure \ref{xGASS} shows that its SFR is slightly high (0.90-$\sigma$ above the SFMS) and $M_{\rm HI}$ is close to expectations (0.06-$\sigma$ below the HIMS) based on the stellar mass and average scaling relations, which supports the relatively secular nature of NGC 3556.

\subsection{NGC 3877}
NGC 3877 is identified as a `Sc' spiral in group environment \citep{2011MNRAS.412.2498M}. Its $\hi$ disk is well aligned and almost similar in size to the optical disk both radially and vertically. As can be seen in Figure \ref{xGASS}, NGC 3877 has a normal SFR (0.01-$\sigma$ below the SFMS) but significantly low $M_{\rm HI}$ (1.27-$\sigma$ below the HIMS) compared to star-forming galaxies with similar $M_*$.

Figure \ref{atlas} shows that, except for the spurs typically observed in star-forming galaxies, NGC 3877 presents a regular $\hi$ disk with very few signatures of perturbation. Similar to NGC 3556, the dense $\hi$ regions are distributed along the spiral arm which is clear in the H$\alpha$ image, while the strongest H$\alpha$ emission is in the centre. 

A previous $\hi$ observation of NGC 3877 was presented in \cite{verheijen2001ursa}. Despite the apparent regular morphology in the intensity image, \cite{verheijen2001ursa} found a kinematic asymmetry that the rotation curve on one side rises more steeply than the other side. The total mass in NGC 3877 was estimated to be $4.76 \times 10^{10} M_{\odot}$ \citep{2010MNRAS.404..468G}. 

Early studies based on H$\alpha$ and X-ray data found NGC 3877 lacked a hot gas halo \citep{2006A&A...448...43T}. The H$\alpha$ shows some vertical structures consistent with the $\hi$ spurs we observe in Figure \ref{atlas}, but they do not reach into the extended halo \citep{2006A&A...448...43T}. The diffuse X-rays are only detectable in the soft band around the central region \citep{2006A&A...448...43T}. In spite of the apparent lack of an H$\alpha$ or X-ray halo, CHANG-ES results have revealed a significant radio halo around this galaxy \citep{wiegert2015chang}.

The $\hi$ has a symmetric intensity distribution but perturbed kinematics, and the hot gas halo seems to show complex structures. These features imply that NGC 3877 may be recovering from a past environmental perturbation.

\subsection{NGC 4096}

NGC 4096 is a `SABc' spiral galaxy. Figure \ref{atlas} shows that its $\hi$ gas, even with our limited sensitivity, is radially and vertically at least slightly extended compared to its optical disk.
The $\hi$ disk shows lopsided structure in the south associated with a spiral arm (labeled with an arrow in Figure \ref{atlas}). The H$\alpha$ light also follows this structure. 

NGC 4096 is presented in WHISP and catalogued as UGC~7090 \citep{garcia2002neutral}. The south lopsided $\hi$ structure is presented in the full resolution intensity image. \citep{garcia2002neutral} showed there is no warp in NGC 4096.

Figure \ref{xGASS} has shown that, NGC 4096 has a normal SFR (0.19-$\sigma$ below the SFMS) and a slightly low $M_{\rm HI}$ (0.67-$\sigma$ below the HIMS) for its stellar mass, which is consistent with the largely unperturbed morphology and kinematics of the inner region of the gas disk. The slightly asymmetric morphology and kinematics in the outer region may be the result of its environment. This galaxy is in the Canes II group and has two companions (NGC 4144 and PC~1200+4755) \citep{garcia2002neutral}.

\subsection{NGC 4157}
NGC 4157 is a `SABb' barred spiral galaxy with a companion galaxy UGC 7176. Its $\hi$ is two times more extended than the optical disk and appears reasonably symmetric (Figure \ref{atlas}). The outskirt of the $\hi$ disk beyond the optical disk is slightly warped. The H$\alpha$ emissions are concentrated on the main disk, and are spatially correlated with the $\hi$ contours at high intensity levels.

$\hi$ observations of NGC 4157 were presented by \cite{verheijen2001ursa} and WHISP. Its scale height was kinematically modelled, and its $\hi$ gas and SFR were found to follow the volumetric star formation law of general disk galaxies by \cite{2014AJ....148..127Y}. Thus, the kinematics of $\hi$ in this galaxy are unlikely to be strongly influenced by its companion. 

\cite{2013MNRAS.433.2958I} showed the radio continuum image of NGC 4157. The morphology in the 617 MHz is consistent with the morphology in the 1.4 GHz for the entire disk, except for an extraplanar structure in the north-east which is only detectable in 617 MHz but not in 1.4 GHz. With much deeper data from CHANG-ES, a radio halo around this galaxy was revealed in \cite{wiegert2015chang}.

In short, NGC 4157 has a slightly warped but symmetric $\hi$ disk, which supports the scenario that this galaxy is currently not strongly effected by its companion and could be largely evolving secularly. Such a secularly evolving scenario is also supported by the featureless $\hi$ kinematics, and multi-wavelength properties including the normal SFR (0.36-$\sigma$ above the SFMS) and $M_{\rm HI}$ (0.40-$\sigma$ above the HIMS) for its $M_*$ as shown in Figure \ref{xGASS}. 

\subsection{NGC 4217}
NGC 4217 is a `Sb' spiral galaxy, which is a possible companion of the galaxy, M106. Its $\hi$ disk is similar in size with the optical disk and has quite a few spurs on both sides of the disk (labeled in Figure \ref{atlas}). In Figure \ref{xGASS}, we have shown that NGC 4217 has slightly low $M_{\rm HI}$ (0.66-$\sigma$ below HIMS) but normal SFR (0.25-$\sigma$ above the SFMS) for its stellar mass, which are consistent with these morphological features.
Although it may be a companion of M106, we do not observe clear evidence for $\hi$ warps.

Based on the $\hi$ image taken with Westerbork Synthesis Radio Telescope (WRST), \cite{verheijen2001ursa} noticed two blobs of $\hi$ gas located at each end of the disk, and interpreted them as an outer ring with no optical counterpart. \cite{verheijen2001ursa} pointed out that the outer ring appears to be spatially and kinematically separate from the main disk, thus may have an external origin. This ring is not observed in our moment-0 image in Figure \ref{atlas}. \cite{2015A&A...582A..18A} used tilted-ring modeling to show that the $\hi$ distribution is best described by a flat geometry without a warp and flare, consistent with our impression from the moment-0 map in Figure \ref{atlas}.

The stellar disk is reddish in color which suggests an old stellar population \citep{1982A&A...110...61V}. The outer disk is truncated and the truncation radius has a wavelength dependence, indicating the effect of secular stellar migration \citep{2001A&A...378...82F}.
\cite{2004AJ....128..662T} used HST Wide Field Camera 2 to study extraplanar dust structures which have an extensive distribution on both sides of NGC 4217's midplane, consistent with the spurs we observe from the $\hi$ image in Figure \ref{atlas}. Supernova-driven galactic chimneys may be responsible for these features. 

To sum up, since this galaxy has a high density of $\hi$ spurs on both sides of the mid-plane, the scenario that star formation produces them is more likely than ram pressure stripping. Additionally it seems that NGC 4217 is not strongly affected by environmental effects since the $\hi$ disk flat.

\subsection{NGC 4302}

NGC 4302 is a `Sc' galaxy in the Virgo Cluster, with a relatively face-on companion NGC 4298. Its $\hi$ shows a little truncation on the south side. Likely because of the tidal force from NGC 4298, the outer contours of the $\hi$ disk bend to the west with respect to the faint stellar halo in the optical. The bending is most clear on the north. We do not detect any $\hi$ between the two galaxies (see Figure \ref{atlas}). 

NGC 4302 and its companion NGC 4298 are also in the sample of the VLA Imaging survey of Virgo galaxies in Atomic gas (VIVA) \citep{2009AJ....138.1741C}. \cite{2009AJ....138.1741C} discussed the possible orbit of NGC 4302 falling into Virgo, and concluded that the lopsided morphology in the $\hi$ disk of NGC 4302 cannot be solely attributed to tidal interaction with NGC 4298; it should be partly or mainly explained by ram pressure stripping. The absence of $\hi$ bridges between NGC 4302 and NGC 4298 supports this explanation. Future simulations of the system could help assess whether this is more likely than the galaxies having retrograde spins with respect to their orbits, which can also be responsible for the absence of strong tidal features.

\cite{2015ApJ...799...61Z} presented a kinematic model of the $\hi$ distribution, which comprises a thin disk with a 300 pc scale height and a thick disk with 1.8 kpc scale height.
The thick $\hi$ disk seems to have a counterpart in the DIG. The extraplanar DIG in the inner disk is stronger than in the outer disk \citep{1996ApJ...462..712R}. A radial variation in the velocity gradient of the DIG perpendicular to the disk (i.e. a lag in rotation) was found and is possibly related to the low level of global SFR \citep{2007ApJ...663..933H}.

We have shown in Figure \ref{xGASS}, that it has a normal SFR (0.42-$\sigma$ below the SFMS) and a significantly low $M_{\rm HI}$ (1.05-$\sigma$ below the HIMS) when compared to normal galaxies of similar stellar masses. If its $M_{\rm HI}$ dropped prior to a decrease in SFR then its $\hi$ mass fraction would be low compared to the average expectation from the sSFR. These global properties are consistent with a scenario in which NGC 4302 is stripped of $\hi$, possibly by the combined effect of ram pressure and tidal forces.

\subsection{NGC 4565}
NGC 4565 is a `Sbc' galaxy with two known companion galaxies, IC 3571 (labeled in Figure \ref{atlas}) and NGC 4562. Its $\hi$ disk is radially more extended than the optical one, and shows a clear warp on the northwest end bending toward the north (labeled in Figure \ref{atlas}), possibly caused by the interaction with IC 3571. 

The distribution and kinematics of $\hi$ in NGC 4565 were extensively studied in the literature. 
Previous studies estimated that the warp contains one-fifth of the total $\hi$ gas \citep{1991AJ....102...48R}. Using $\hi$ data of higher resolution and based on kinematical modeling, \cite{2012ApJ...760...37Z} found evidence for the existence of a bar and a possible ring. The kinematical modeling also reveals a flaring of the disk. The scale height of the $\hi$ disk was also derived through kinematical modeling in \cite{2014AJ....148..127Y}. Extraplanar $\hi$ is found at small radii, and has a lag in its rotation which suggests an association with the gas halo \citep{2012ApJ...760...37Z}. Additional diffuse extraplanar $\hi$ along the minor axis of NGC 4565 was detected by the Green Bank Telescope (GBT) recently by \cite{2020MNRAS.494.4558Y}. 

\cite{2019MNRAS.483..664M} found evidence for disk truncations in the NUV, optical, and 3.6 $\mu m$, and found that the truncation radius is independent of wavelength. Asymmetry in the truncation on the two sides of the optical disk was identified recently, implying that the disk is growing by stripping gas from its companion \citep{2020ApJ...897..108G}. However, NGC 4565 has weak H$\alpha$ emission and shows no chimney structures \citep{1992ApJ...396...97R}, suggesting a low level of SFR. The diffuse radio halo and extraplanar X-rays are possibly remnants of enhanced star forming activity in the past \citep{2013MNRAS.428.2085L}. The radio halo also shows a warp structure \citep{1984A&A...133....1H, 1991ApJ...382..100S, 2019A&A...628L...3H}. CHANG-ES radio and polarization data of this galaxy have been extensively studied by \cite{2019A&A...632A..12S}.

Overall, NGC 4565 is growing the northern side of its disk mainly through stripping material from its companion IC 3571. The $\hi$ warp can be clearly seen in Figure \ref{atlas}. 
From Figure \ref{xGASS}, NGC 4565 seems to have normal $M_{\rm HI}$ (0.13-$\sigma$ above the HIMS) and slightly low SFR (0.60-$\sigma$ below the SFMS) for its stellar mass. Thus, if there is tidally driven disk growth, it does not seem to enhance the global stellar mass growth of the galaxy. This is very different from other tidally interacting systems in this sample (e.g., NGC 4631, NGC 4666, to be discussed soon).

\subsection{NGC 4631}
NGC 4631 is a `SBcd' spiral galaxy with a dwarf elliptical companion, NGC 4627 (about 3' to the northwest), and an edge-on companion NGC 4656, (about 30' to the southeast and outside the field of view of our $\hi$ image). From Figure \ref{xGASS}, we can see that its sSFR is much higher than the  average expected for its $\hi$ mass fraction, indicating significantly enhanced star forming efficiency. In Figure \ref{atlas}, its $\hi$ disk shows many $\hi$ spurs which may be the result of its interactions with the companions and the enhanced star formation. The optical disk is blue in color and the dust is mainly concentrated in the centre.

\cite{1993AJ....105.2098R} and \cite{rand1994atomic} discovered two $\hi$ supershells in this galaxy. The larger one has a diameter of about 3 kpc, while the smaller one has a diameter of about 1.8 kpc. The two shells are broken in several places. The bright X-ray counterparts of the two $\hi$ shells can be seen in ROSAT PSPC images \citep{1996A&A...311...35V}. And the larger shell corresponds to an H$\alpha$ perturbation feature in the eastern half of the disk \citep{1992ApJ...396...97R}. An internal origin for the $\hi$ shells are excluded, and the shells are likely produced by the oblique impact of a high velocity cloud \citep{1996AJ....111..190R}. The large supershell can be seen in Figure \ref{atlas} on the eastern side close to the major axis of the galaxy. \cite{2015AJ....150..116M} discovered a giant stellar tidal stream between NGC 4631 and NGC 4656, spatially correlated with the larger $\hi$ supershell. Both modeling and inference from the morphology indicate the streams originate from a relatively recent, gas-rich merger event rather than current tidal interactions with the relatively distant companions. It further supports the high velocity cloud origin of the $\hi$ shells. 

NGC 4631 shows a double-peaked central CO emission  \citep{1994A&A...286..733G, 2011MNRAS.410.1423I}, with a similar structure in the dust emission \citep{2006ApJ...652..283B}. \cite{2011MNRAS.410.1423I} suggested that the interaction between NGC 4631 and its companion galaxies agitates the disk and also initiates star formation. Subsequent outflows from the star forming regions are necessary for the formation of a prominent halo.

Analysis of the thickness and asymmetries of the NGC 4631 radio halo suggests that the halo is influenced by the star formation in the disk \citep{1995ApJ...444..119D}, the gravitational interaction with companions \citep{1988A&A...197L..29H,1990A&A...236...33H}, and interaction with the magnetic field \citep{1988A&A...197L..29H,1991A&A...248...23H,1994A&A...284..777G}. 

The X-ray morphology \citep{1995ApJ...439..176W,2001ApJ...555L..99W,2009PASJ...61S.291Y} resembles the radio halo, suggesting a close connection between hot gas outflows, CRs, and the magnetic field from the galactic disk. The radio continuum and polarization structure of NGC 4631 have been extensively studied by \cite{2019A&A...632A..10M} and the Faraday rotation measure behaviour has been modelled by \cite{woo19}.

To sum up, the tidal interaction with companions and the tidal induced star formation seem to influence the multi-phase gas in and around NGC 4631. The $\hi$ supershells and spurs observed in our $\hi$ image support this scenario.

\subsection{NGC 4666}
NGC 4666 is a `SABc' spiral galaxy interacting with NGC 4668, which is located southeast of NGC 4666 and is outside the field of view of our $\hi$ image shown in Figure \ref{atlas}. The $\hi$ disk is more extended than the optical disk and there are extended $\hi$ spurs on the south-east side (labeled in Figure \ref{atlas}). Figure \ref{xGASS} shows that, similar to NGC 4631, it has a significantly enhanced sSFR for its $\hi$ mass fraction. It has the highest SFR (1.91-$\sigma$ above the SFMS) and a slightly high $M_{\rm HI}$ (0.61-$\sigma$ above the HIMS) for its stellar mass, indicating it is a starburst galaxy. 

Based on deeper $\hi$ images, \cite{2004ApJ...606..258W} found that NGC 4666 is obviously interacting, and the tidal arms and tidal tails cover an area of at least $5 \times 10^3$ kpc$^2$.
NGC 4666 has a central concentration of CO gas and starburst possibly due to this interaction \citep{2004ApJ...606..258W}.

NGC 4666 is identified as a “superwind” galaxy. \cite{1997A&A...320..731D} presented evidence for a starburst driven galactic superwind, which is visible in optical emission line images, radio continuum maps, and in soft X-ray images. Later observation and analysis of images of the DIG \citep{2013A&A...554A.133V}, radio continuum \citep{1988MNRAS.231..765S,1995ApJ...444..119D}, and X-ray \citep{1998ApJS..118..401D,2004MmSAI..75..515E}, robustly confirm outflows in NGC 4666. Especially, the H$\alpha$ emission data presents a clear “X”-shaped structure near the galaxy centre \citep{1997A&A...320..731D}. 
\cite{2019A&A...623A..33S} further revealed an “X”-type magnetic field in this galaxy using CHANG-ES data.

Thus, the $\hi$ spurs in NGC 4666 are possibly associated with both the tidal interaction and the superwind.

\subsection{NGC 5084}
NGC 5084 is a lenticular galaxy and is also the central galaxy of the NGC 5084 Group. The $\hi$ disk is much more extended than the optical disk, as can be seen in Figure \ref{atlas}. Thanks to the improved sensitivity of our data, there are clear $\hi$ spurs on both sides of the disk which was not observed before (labeled in Figure \ref{atlas}).
The position angles of the $\hi$ disk and the inner disk (or pseudo bulge) of stars are not perfectly aligned. The majority of the $\hi$ is aligned with an outer optical ring (labeled in Figure \ref{atlas}, with arrows pointing to the direction of the position angle of the ring). 

The inner disk and outer ring in the optical are more clearly seen in Figure 1 of \cite{1990MNRAS.246..324Z}. \cite{1990MNRAS.246..324Z} confirmed the $\sim$5 degrees tilted between the bright inner stellar disk and the fainter outer stellar structure based on spectroscopic and photometric observations. 
The association of the $\hi$ with the optical outer ring was observed by \cite{1986MNRAS.219..759G}. 
They estimated a high mass-to-light ratio compared to normal lenticular type galaxies, and suggested the possibility that this galaxy experienced a recent merger with a dwarf galaxy.

In summary, our observations of the $\hi$ and optical (mis)alignment in NGC 5084 is consistent with those presented in the literature. The possibility of a recent merger suggested by the (mis)alignment is consistent with the group environment of NGC 5084. Found for the first time in Figure \ref{atlas}, are $\hi$ spurs.

\subsection{NGC 5775}
NGC 5775 is a `Sbc' spiral galaxy. The $\hi$ image in Figure \ref{atlas} shows that its $\hi$ disk is much more extended than the optical one, and is highly asymmetric. The $\hi$ disk is likely tidally interacting with a nearly face-on companion galaxy NGC 5774, with a broad $\hi$ tail on the northwest side. The tidal features are not so clearly seen in the SDSS optical image. From Figure \ref{xGASS}, NGC 5775 has significantly high SFR (1.60-$\sigma$ above the SFMS) and slightly high $M_{\rm HI}$ (0.58-$\sigma$ above the HIMS) for its stellar mass, which is likely caused by the tidal interaction discussed above. 

The $\hi$ bridge connecting NGC 5775 to NGC 5774 was first detected by \cite{1994ApJ...429..618I}. Based on the velocity images, \cite{1994ApJ...429..618I} pointed out that the $\hi$ mass is likely transferred from NGC 5774 to NGC 5775, and suggested this system may be in the early stage of merger. Due to the improved sensitivity, the $\hi$ bridges in Figure \ref{atlas} are clearer and broader than those presented in \cite{1994ApJ...429..618I}. It is worth pointing out that the $\hi$ bridge has a counterpart in the radio continuum \citep{1998A&A...331..428D}, providing further confirmation that the pair are interacting.

Possibly due to the interaction, NGC 5775 may have a complex history of star formation and mass loading. The halo hot gas reveals two characteristic components through X-ray spectral analysis, with temperatures of about 0.2 and 0.6 keV \citep{2008MNRAS.390...59L}. Shell features of soft X-rays may represent outflow wind bubbles from the central region of galaxy. Current star formation with higher mass loading may be responsible for the lower temperature hot gas component, while recent and past star formation with lower mass loading may be responsible for the higher temperature component \citep{2008MNRAS.390...59L}. 

The warm and hot gas halo of NGC 5775 reaches high latitudes, possibly due to strong outflows which are caused by the tidally triggered active star formation. \cite{2003MNRAS.340..269S} found a strong correlation between 850 $\mu$m and 617 MHz emission distributions in the disk and at high latitudes of NGC 5775, suggesting a fundamental relationship between the cold dust and synchrotron radiation. DIG can be traced out to $\sim$ 9 kpc above the mid-plane and present a vertical gradient (i.e., slow-down) in the rotational velocity \citep{2000A&A...364L..36T,2006ApJ...636..181H}.
CHANG-ES radio continuum emission has recently been extensively studied by \cite{2021MNRAS.tmp.2538H} who concluded that the mass outflow rate is of the same order as the SFR.

In summary, NGC~5775 is a very active system, showing many vertical extensions and/or halos in many wavebands. The $\hi$ morphology in our image, including the bridges due to the tidal interaction of the NGC~5775/5774 system, provides important clues to understanding the origin of the star formation activity in this galaxy.

\section{Comparison of data quality with previous HI images}

Table \ref{HIcompare} presents information about previous images produced using interferometry. For most of the galaxies (NGC 2683, NGC 3003, NGC 3079, NGC 3877, NGC 4096, NGC 4217, NGC 4302, NGC 4565, NGC 4631, NGC 4666), the CHANG-ES $\hi$ data have better spatial resolution than the observations in the literature.  We calculate the sensitivities of the intensity maps of CHANG-ES data which are typically several $10^{20} \rm cm^{-2}$ and better than the previous observations of NGC 4096, NGC 4217, NGC 5084, and NGC 5775. 

\begin{table*}
\caption{Previous $\hi$ Imaging Data}
\resizebox{\textwidth}{40mm}{
\begin{tabular}{lcccccccc}
  \hline
Galaxies & Date & Observatory & Beamsize & velocity resolution & RMS Sensitivity & 3-$\sigma$ column density limit & Ref$^a$\\
 & [year] & & [''] & [km/s] & [mJy/Beam] & [10$^{20}$ cm$^{-2}$] & \\
  \hline
NGC 660 & 1986-1987 & VLA & 14.89 $\times$ 13.5 & 21.0 & 0.76 & 2.63 & 1\\
        &      1990 & WRST & 13.2 $\times$ 60.2 & 20.0 & 1.9 & 1.58 & 2\\
NGC 2683 & 2004;2009 & VLA &     21 $\times$ 20 & 10.3 & 0.3 & 0.47 & 3\\
NGC 3003 (UGC 5251) &  1999 & WRST & 20.6 $\times$ 9.8 & 4.15 & 3.4 & 2.31 & WHISP website$^b$\\
NGC 3044 & 2010-2011 & VLA & 12.27 $\times$ 11.14 & 6.7 & 0.28 & 1.36 & 4\\
NGC 3079 & 1984;1986 & VLA &  16.9 $\times$ 16.7 & 41.4 & 0.35 & 1.70 & 5\\
NGC 3448 (UGC 6024) & 2010-2011 & VLA & 12.27 $\times$ 11.14 & 6.7 & 0.28 & 1.36 & 6\\
         &      1999 & WRST & 12.9 $\times$ 10.8 & 16.6 & 1.7 & 6.71 & WHISP website\\
NGC 3556 (UGC 6225) & 1994;1995 & VLA &  12.8 $\times$ 11.1 & 20.71 & 0.4 & 1.93 & 7\\
         &      1999 & WRST & 13.6 $\times$ 11.9 & 16.5 & 1.9 & 6.42 & WHISP website\\
NGC 3877&      1990 & WRST & 12.2 $\times$ 17.6 & 33.18 & 0.7 & 3.57 & 8\\
NGC 4096 (UGC 7090) &  1997 & WRST & 12.48 $\times$ 17.15 & 19.9 & 1.7 & 5.26 & 9 \& WHISP website\\
NGC 4157 (UGC 7183) &   - & VLA & 15.69 $\times$ 14.87 & 20.0 & 0.26 & 0.73 & 10\\
         &      1993 & WRST & 11.8 $\times$ 16.7 & 19.88 & 1.6 & 5.37 & 8\\
         &      1999 & WRST & 13.9 $\times$ 8.6 & 16.5 & 1.7 & 7.77 & WHISP website\\
NGC 4217 &      1990 & WRST & 18.6 $\times$ 13.2 & 33.2 & 1.0 & 4.48 & 8\\
NGC 4302 &      2005 & VLA & 16.85 $\times$ 15.72 & 10.0 & 0.35 & 0.88 & 11\\
NGC 4565 &         - & VLA &   6.26 $\times$ 5.59 & 20.0 & 0.16 & 2.99 & 12\\
        & 2003;2009 & WRST & 24.27 $\times$ 11.5 & 4.12 & 0.31 & 0.74 & 13\\
NGC 4631 &      1991 & WRST & 21.5 $\times$ 11.5 & 10.0 & 1.17 & 3.13 & 14\\
NGC 4666 & 1996-1997 & VLA &  29.6 $\times$ 22.2 & 5.2 & 0.39 & 0.39 & 15\\
NGC 5084 &      1989 & VLA &  13.6 $\times$ 13.4 & 41.67 & 1.44 & 2.04 & 16\\
NGC 5775 &     1989 & VLA &   13.6 $\times$ 13.4 & 41.67 & 0.5 & 3.79 & 17\\
  \hline
\end{tabular}}
\begin{tablenotes}
\item For comparison the CHANG-ES beam size is 14.5 arcsec on average.
\item $^a$ Ref.--- Reference to $\hi$ observation.
\item 1. \cite{1990NASCP3098..209G} 2. \cite{1995AJ....109..942V} 3. \cite{1991AJ....101.1231C, vollmer2016flaring} 4. \cite{2015ApJ...799...61Z} 5. \cite{1987ApJ...313L..91I, 1991ApJ...371..111I} 6. \cite{1986AJ.....92.1048N} 7. \cite{1997NewA....2..251K} 8. \cite{verheijen2001ursa} 9. \cite{garcia2002neutral} 10. \cite{2014AJ....148..127Y} 11. \cite{2009AJ....138.1741C} 12. \cite{1991AJ....102...48R} 13. \cite{2012ApJ...760...37Z} 14. \cite{1993AJ....105.2098R, rand1994atomic} 15. \cite{2004ApJ...606..258W} 16. \cite{1986MNRAS.219..759G} 17. \cite{1994ApJ...429..618I}
\item $^b$ The Westerbork $\hi$ survey of spiral and irregular galaxies (WHISP) website:
\item \href{https://www.astro.rug.nl/~whisp/Database/OverviewCatalog/ListByName/list_by_name.html}{https://www.astro.rug.nl/$\sim$whisp/Database/OverviewCatalog/ListByName/list\_by\_name.html}
\end{tablenotes}
\label{HIcompare}
\end{table*}

\end{document}